\documentclass[a4paper,11pt]{article}
\pdfoutput=1

\usepackage[height=23.6truecm,width=16truecm,top=3.05truecm]{geometry}

\usepackage[skip=6pt,indent]{parskip}

\usepackage[utf8]{inputenc}
\usepackage{graphicx}
\usepackage{amsmath}
\usepackage{amssymb}
\usepackage{hyperref}
\hypersetup{colorlinks,citecolor=TokiwaIro,linkcolor=RuriIro,urlcolor=RuriIro,linktocpage}
\usepackage{cite}
\usepackage{braket}
\usepackage{bm}
\usepackage{bbm}
\usepackage{color}
\usepackage[svgnames,table]{xcolor}
\usepackage{float}
\usepackage{hhline}
\usepackage{multirow}
\usepackage{indentfirst}
\usepackage{latexsym}
\usepackage{mathtools}
\usepackage{cancel}
\usepackage{slashed}
\usepackage{colortbl}
\usepackage[mathscr]{euscript}
\usepackage{enumitem}
\usepackage{subcaption}
\usepackage{comment}
\usepackage{amsthm}

\usepackage{soul}

\numberwithin{equation}{section}

\allowdisplaybreaks

\definecolor{RuriIro}{rgb}{0.,0.28,0.60}
\definecolor{TokiwaIro}{rgb}{0.,0.39,0.16}
\definecolor{AkaneIro}{rgb}{0.72,0.16,0.18}

\definecolor{kblue}{rgb}{0,0.48,0.73}
\definecolor{kred}{rgb}{0.73,0.25,0}
\definecolor{kgreen}{rgb}{0.48,0.73,0}
\definecolor{dgreen}{rgb}{0.2,0.51,0.19}
\definecolor{dred}{rgb}{0.7,0.15,0.09}

\theoremstyle{plain}

\theoremstyle{remark}

\newcommand{\nn}{\nonumber}

\newcommand{\mc}{\mathcal}
\newcommand{\mr}{\mathrm}
\newcommand{\ms}{\mathsf}

\newcommand{\del}{\partial}

\newcommand{\Tr}{\mathop{\mathrm{Tr}}\nolimits}

\newcommand{\EH}{\mathrm{EH}}
\newcommand{\GHY}{\mathrm{GHY}}
\newcommand{\Maxwell}{\mathrm{Maxwell}}

\newcommand{\U}{\mathrm{U}}

\newcommand{\eff}{\mathrm{eff}}
\newcommand{\bulk}{\mathrm{bulk}}

\definecolor{murasaki}{rgb}{0.42,0.19,0.61}

\usepackage{ulem}

\begin{document}

\begin{titlepage}

\begin{flushright}
\end{flushright}

\vspace{1cm}

\begin{center}

{\LARGE \bfseries
Black hole thermodynamics and KK photon quantum corrections in 2D effective dilaton gravity
}

\vspace{1cm}

\renewcommand{\thefootnote}{\fnsymbol{footnote}}
{%
\hypersetup{linkcolor=black}
Yoshihiko Abe$^{1,2,3}$\footnote[1]{yabe3@keio.jp},
\ 
Tetsutaro Higaki$^{4}$\footnote[2]{thigaki@rk.phys.keio.ac.jp},
\ and
Yu Miyauchi$^{4}$\footnote[3]{mu\_19990710@keio.jp}
}%
\vspace{8mm}

{\itshape%
$^1${Graduate School of Science and Technology, Keio University, Yokohama, Kanagawa, Japan}\\
$^2${Keio University Sustainable Quantum Artificial Intelligence Center (KSQAIC), Keio University, Tokyo 108-8345, Japan}\\
$^3${Quantum Computing Center, Keio University, 3-14-1 Hiyoshi, Kohoku-ku, Yokohama, Kanagawa 223-8522, Japan}\\
$^4${Department of Physics, Keio University, Yokohama 223-8533, Japan}
}%

\vspace{8mm}

\end{center}

\abstract{
We study black hole thermodynamics using a two-dimensional effective theory obtained by dimensional reduction of four-dimensional Einstein--Maxwell theory. 
For spherically symmetric charged black holes, the resulting dilaton gravity has a nonlinear potential that reproduces the semiclassical phase structure of four-dimensional AdS black holes, including the Hawking--Page transition and the small/large Reissner--Nordstr\"{o}m--AdS black hole transition. 
This shows that the two-dimensional theory before taking the near-horizon and near-extremal limits captures non-extremal thermodynamics beyond the Jackiw--Teitelboim gravity regime.
We also include electromagnetic Kaluza--Klein modes on the internal sphere and integrate them out to derive the one-loop effective dilaton gravity.
We consider only the local one-loop contributions, which induce constant shifts in the black hole entropy and in the effective charge parameter of the dilaton potential.
Within this approximation, the semiclassical phase structure remains qualitatively unchanged.
}

\end{titlepage}

\renewcommand{\thefootnote}{\arabic{footnote}}
\setcounter{footnote}{0}
\setcounter{page}{1}

\tableofcontents

\section{Introduction}

Understanding black holes in terms of quantum mechanics is a fundamental problem in modern physics.
One of the key ideas is black hole thermodynamics, which arises from semiclassical analysis yet yields many important insights into quantum gravity.
Typical examples include Hawking radiation, the thermal radiation emitted by black holes~\cite{Hawking:1974rv,Hawking:1975vcx}, and the Hawking--Page transition, a phase transition of anti-de Sitter (AdS) black holes~\cite{Hawking:1982dh,Chamblin:1999hg,Chamblin:1999tk,Caldarelli:1999xj}.
In particular, low-temperature or near-extremal black holes are often discussed in the context of quantum gravity consistency, for example via the weak gravity conjecture~\cite{Arkani-Hamed:2006emk,Harlow:2022ich}, as well as in studies of the connection between black hole entropy and the degeneracy of microstates~\cite{Strominger:1996sh,Sen:2008vm}.

A central issue is how to incorporate quantum corrections into black hole thermodynamics. An effective field theory framework offers a systematic way to analyze such effects in gravitational systems. In this context, dimensional reduction from higher dimensions provides a powerful framework. In particular, reduction to two-dimensional theories is computationally tractable since the gravitational dynamics simplify significantly due to the reduced number of degrees of freedom, enabling explicit and often exact treatments of quantum effects. Among these, two-dimensional dilaton gravity has often been studied as a model arising from this construction~\cite{Achucarro:1993fd,Grumiller:2002nm,Iliesiu:2020qvm,Banerjee:2021vjy,Gukov:2022oed}.
For example, by imposing spherical symmetry on four-dimensional Einstein--Maxwell theory, one can explicitly integrate over the $S^2$ directions, yielding a tractable two-dimensional theory for the reduced gravitational and electromagnetic degrees of freedom. In this framework, the dilaton emerges as a scalar field controlling the size of the $S^2$. Since the degrees of freedom differ between four and two dimensions, the reduced description is not fully equivalent to the original theory. To capture the essential features of the higher-dimensional system, one must include contributions from Kaluza--Klein (KK) modes. These correspond to non-spherically symmetric fluctuations around the background, and integrating them out induces quantum corrections in the two-dimensional effective theory.
In the near-extremal and near-horizon limits, massive KK modes are typically neglected at sufficiently low temperatures, and the effective dilaton gravity reduces to a simpler model \cite{Iliesiu:2020qvm,Banerjee:2021vjy,Gukov:2022oed}, known as Jackiw--Teitelboim (JT) gravity~\cite{Teitelboim:1983ux,Jackiw:1984je}.
Since its partition function can be computed exactly~\cite{Saad:2019lba,Iliesiu:2019xuh}, JT gravity has become an important model for studying quantum effects in near-extremal black hole physics.
Motivated by these developments, two-dimensional dilaton gravity has attracted considerable attention in recent years, particularly in the study of black hole thermodynamics.

However, the standard JT description primarily captures the near-extremal and near-horizon regimes and thus does not fully describe non-extremal thermodynamics, such as the Hawking--Page transition. To study the phase structure of non-extremal black holes within a two-dimensional effective framework, it is useful to consider effective dilaton gravity before taking the near-extremal and near-horizon limits. In general, two-dimensional dilaton gravity exhibits nontrivial phase structures~\cite{Grumiller:2007ju,Witten:2020ert}, suggesting that it can capture essential features of four-dimensional black hole thermodynamics. Furthermore, in the non-extremal regime, massive KK modes couple to the effective gravity theory and can contribute nontrivially to the effective potential. It is, therefore, important to take these modes into account and evaluate their effects on thermodynamic quantities. In particular, understanding how the KK tower modifies the effective dilaton potential away from the near-extremal regime is of central importance. The aim of this paper is to derive an effective action describing non-extremal black holes and to estimate the quantum corrections to black hole thermodynamics arising from KK modes.

In this paper, we first derive a semiclassical two-dimensional dilaton gravity describing non-extremal spherically symmetric charged black holes by performing a dimensional reduction of four-dimensional Einstein--Maxwell theory and analyze its phase structure. We find that the resulting phase structure is consistent with the known four-dimensional semiclassical results~\cite{Hawking:1982dh,Chamblin:1999hg,Chamblin:1999tk}. We then construct an effective theory including KK modes of the electromagnetic field and derive the one-loop effective dilaton gravity by integrating out the massive KK modes. 
Keeping only the leading local heat-kernel terms (up to two bulk derivatives and one boundary derivative), while omitting dilaton-dependent anomaly terms and additional massless-sector contributions, yields shifts in the entropy and the effective charge parameter. The semiclassical phase structure remains qualitatively unchanged only within this approximation. 
We do not compute the full one-loop determinants of the KK modes. 
In particular, the massless sector is treated only partially, and for the lightest massive KK level we do not establish that the retained local terms reliably approximate the full determinant.

This paper is organized as follows.
In Sec.~\ref{Sec:Reduced gravity and Kaluza--Kleinreduction}, we briefly review four-dimensional charged black holes in Euclidean signature and their thermodynamics with a negative cosmological constant.
In Sec.~\ref{Sec:Two-dimensional gravity via dimensional reduction}, we dimensionally reduce the four-dimensional theory of charged black holes and derive the corresponding two-dimensional effective dilaton gravity, neglecting all KK modes.
In Sec.~\ref{Sec:two-dimensional dilaton gravity and phase transition}, we analyze the phase structure of the black holes from a two-dimensional perspective and demonstrate its agreement with the four-dimensional results.
In Sec.~\ref{Sec:Partition function and effective action}, we construct an effective theory for the KK modes of the electromagnetic field, integrate them out using the heat-kernel method, and compute the retained local one-loop contributions to the effective action.
Finally, Sec.~\ref{Sec:Conclusion} is devoted to the conclusions and a discussion of possible extensions.

\section{Four-dimensional charged black holes}
\label{Sec:Reduced gravity and Kaluza--Kleinreduction}

In this section, we briefly review four-dimensional charged black holes and their thermodynamics at fixed charge, focusing on the Hawking--Page transition.

\subsection{Four-dimensional charged black holes and boundary conditions}
\label{Sec:Reduced gravity from charged black hole}

We begin with charged black holes in four-dimensional Einstein–Maxwell theory.
The Euclidean action\footnote{
	We adopt units where $c = \hbar = k_B = 1$.
} is given by 
\begin{align}
	I[g, A] = & I_\EH[g] + I_\GHY[g] + I_\Maxwell[A^{(1)},g],
	\label{Eq:Einstein-Maxwell-action}
	\\
	I_\EH[g] &= - \frac{1}{16 \pi G_N} \int_{\mc{M}} d^4x \sqrt{\det g_{\mu\nu}} \bigl[
		R_{(4)} - 2 \Lambda
	\bigr],
	\label{Eq:Einstein-Hilbert-action}
	\\
	I_\GHY[g] &=- \frac{1}{8 \pi G_N} \int_{\del \mc{M}} d^3x \sqrt{\det h_{\mu\nu}} K_{(3)},
	\label{Eq:Gibbons-Hawking-York-term}
	\\
	I_\Maxwell[A^{(1)}, g] &= \frac{1}{4e^2} \int_{\mc{M}} d^4x \sqrt{\det g_{\mu\nu}} F_{\mu\nu} F^{\mu\nu} - \frac{1}{e^2} \int_{\del \mc{M}} d^3x \sqrt{\det h_{\mu\nu}} n_{(4)\mu} A_\nu F^{\mu\nu},
	\label{Eq:Maxwell-action}
\end{align}
where $\mc{M}$ and $\del \mc{M}$ are the four-dimensional Euclidean spacetime and its boundary, which is defined by $r = \mr{const.}$ and located outside the event horizon.
$G_N$ denotes the four-dimensional Newton constant.
The Greek indices $\mu,\nu$ run over $0,1,2,3$.
The induced metric on $\del \mc{M}$ is given by $h_{\mu\nu} = g_{\mu\nu} - n_{(4)\mu} n_{(4)\nu}$, where $n_{(4)\mu}$ is the unit normal vector of the boundary.
$R_{(4)}$ is the scalar curvature of $\mc{M}$.
$K_{(3)}$ is the extrinsic curvature of $\del \mc{M}$, which is defined by $K_{(3)} \coloneqq \nabla_\mu n^\mu_{(4)}$.
$\Lambda$ is a four-dimensional cosmological constant.
$A_\mu$ is an abelian gauge field taken to be purely imaginary; $F_{\mu\nu}$ is its field strength, and $e$ is the gauge coupling.
Using differential forms, this is expressed as $F^{(2)} = d A^{(1)}$.
We sometimes use differential forms for the calculations of the gauge field sectors in this paper.
For our notation of the differential forms, see App.~\ref{sec:notation}.

The spherically symmetric electrically charged black hole solution is parameterized by two quantities: the black hole mass $M$ and the electric charge $Q$.
The metric and gauge field configurations are given by 
\begin{align}
	ds^2 &= f(r) dt^2 + \frac{dr^2}{f(r)} + r^2 d \Omega_{(2)}^2,
	\label{Eq:4-dim Riesnner-Nordstrom action}
	\\
	f(r) &\coloneqq 1 - \frac{2 G_NM}{r} + \frac{e^2}{4\pi}\frac{G_NQ^2}{r^2} - \frac{\Lambda}{3} r^2,
	\\
	A^{(1)} &= i\frac{e^2}{4 \pi}\frac{Q}{r} dt,
	\quad 
	F^{(2)} = i\frac{e^2}{4\pi} \frac{Q}{r^2} dt \wedge dr,
	\label{Eq:4-dim gauge field}
\end{align}
where $d \Omega_{(2)}^2 \coloneqq d \theta^2 + \sin \theta^2 d\phi^2$ is the metric on $S^2$.
The coefficient $i$ in Eq.~\eqref{Eq:4-dim gauge field} is introduced so that this solution matches the Lorentzian signature one after Wick rotation.
This solution is called the Reissner--Nordstr\"{o}m (RN)--(A)dS metric.
The black hole horizon is characterized by the equation $f(r_H)=0$, and we write the horizon radius as $r_H$.
The black hole temperature is determined by
the smoothness of the metric at the horizon, which requires periodicity in imaginary time
\begin{align}
    t \sim t+ \beta_{(4)},
     \qquad \beta_{(4)} = 4\pi \left( \frac{df}{dr}(r_H)\right)^{-1}.
     \label{Eq:BH-inverse-temperature-4D}
\end{align}
The metric and gauge field also obey the periodic boundary conditions.
This period $\beta_{(4)}$ is regarded as the four-dimensional inverse temperature of the black hole, which means that the four-dimensional black hole temperature at the horizon becomes
\begin{align}
	T_{(4)} 
    = \frac{1}{4\pi} \frac{df}{dr}(r_H) .
    \label{Eq:BH-temperature-4D}
\end{align}
As an example, we consider the case $\Lambda = 0$. Then the horizon radius $r_H$ is given by
\begin{align}
    r_H =r_\pm \coloneqq  G_NM \pm \sqrt{ G_N^2M^2- \frac{G_N e^2Q^2}{4\pi}},
\end{align}
where $r_+$ ($r_-$) corresponds to the outer (inner) horizon radius.
The temperature at the outer horizon reads 
\begin{align}
	T_{(4)} 
    = \frac{1}{4\pi} \frac{df}{dr}(r_+)
    = \frac{r_+-r_-}{4\pi r_+^2} 
    .
    \label{Eq:BH-temperature-4D for Lambda=0} 
\end{align}
For $M^2=e^2Q^2/(4\pi G_N)$,
the outer and inner horizons coincide, $r_+=r_- = G_NM= eQ\sqrt{G_N/(4\pi)}$. 
In this situation, the black hole temperature vanishes
\begin{align}
T_{(4)}=0.
\end{align}
Black holes at zero temperature are generally called extremal black holes.

\paragraph{Boundary condition for the gauge field and gauge transformations.}

We now comment on the boundary conditions and gauge transformations.
In this paper, we work in the canonical ensemble at a fixed charge~\cite{Hawking:1995ap,Louko:1996dw,Lundgren:2006kt}, following previous studies of the phase transition in AdS black holes~\cite{Hawking:1982dh,Chamblin:1999hg,Chamblin:1999tk}.
The fixed charge requires the following condition on the component of $F_{\mu\nu}$ at the boundary, in addition to the periodic boundary condition in the imaginary time direction:
\begin{align}
	0 = n_{(4)\mu} \delta F^{\mu\nu} \big|_{\del \mc{M}} \propto \delta F^{r \nu} \big|_{\del \mc{M}}.
	\label{Eq:D-dim Neumann (absolute) boundary condition I}
\end{align}
However, the tangential component is not fixed at the boundary, so we introduce the boundary term in the Maxwell action \eqref{Eq:Maxwell-action}.\footnote{
If, instead, the electric potential is fixed, which corresponds to the grand canonical ensemble, it is necessary to fix the tangential components of $A^{(1)}$ on the boundary and omit the corresponding boundary term in the action (see Refs.~\cite{Chamblin:1999hg,Chamblin:1999tk}).
}
In order to evaluate the partition function of the quantum fluctuations in a later section, we use the heat-kernel method and need to impose appropriate boundary conditions on all components of the gauge field.
Then, we also fix the normal component as 
\begin{align}
	0 = n_{(4)}^\mu \delta A_\mu\big|_{\del \mc{M}} \propto \delta A_r \big|_{\del \mc{M}}.
	\label{Eq:D-dim Neumann (absolute) boundary condition II}
\end{align}
The combination of the boundary conditions \eqref{Eq:D-dim Neumann (absolute) boundary condition I} and \eqref{Eq:D-dim Neumann (absolute) boundary condition II}, which imposes a Neumann boundary condition on the tangential component and a Dirichlet boundary condition on the normal component, is called the absolute boundary condition.

For the classical solution~\eqref{Eq:4-dim gauge field}, the on-shell action is gauge invariant.
To see this explicitly, let us consider the gauge transformation of \eqref{Eq:Maxwell-action} with a gauge parameter $\Theta(x)$
\begin{align}
    \delta_\Theta I_\Maxwell &= - \frac{1}{e^2} \int_{\del \mc{M}} d^3x \sqrt{\det h_{\mu\nu}} \Bigl[
        n_{(4)\mu} (\delta_\Theta A_\nu) F^{\mu\nu} + n_{(4)\mu} A_\nu (\delta_\Theta F^{\mu\nu})
    \Bigr]
    \nn \\
    &= - \frac{1}{e^2} \int_{\del \mc{M}} d^3x \sqrt{\det h_{\mu\nu}} n_{(4)\mu} (\del_\nu \Theta) F^{\mu\nu}
    \label{Eq:delta_theta_boundary_term}
     \\
    &= \frac{1}{e^2} \int_{\del \mc{M}} d^3x \sqrt{\det h_{\mu\nu}} \Theta \nabla_\nu ( n_{(4)\mu} F^{\mu\nu} ),
    \label{Eq:delta_theta_boundary_term-2}
\end{align}
where we use the gauge invariance of the field strength $F^{\mu\nu}$ and the fact that $\del \mc{M}$ does not have a boundary.
The integrand in \eqref{Eq:delta_theta_boundary_term-2} is calculated as 
\begin{align}
    \nabla_\nu ( n_{(4)\mu} F^{\mu\nu}) = (\nabla_\nu n_{(4)\mu} )F^{\mu\nu} + n_{(4)\mu} \nabla_\nu F^{\mu\nu} = n_{(4)\mu} \nabla_\nu F^{\mu\nu},
\end{align}
and this vanishes 
when $F_{\mu\nu}$ is the classical solution.
Here, $\nabla_\mu n_{(4)\nu}$ is the extrinsic curvature tensor 
and its indices being symmetric is used in this calculation.
For quantum fluctuations, Eq.~\eqref{Eq:delta_theta_boundary_term} is understood with $F_{\mu\nu}$ replaced by the fluctuating field strength. 
The resulting gauge variation vanishes upon imposing the Neumann boundary condition \eqref{Eq:D-dim Neumann (absolute) boundary condition I}. Thus, the boundary term is invariant within the space of field configurations compatible with the boundary condition.
In addition, the four-dimensional gauge transformation parameter $\Theta(x)$ should satisfy the Neumann boundary condition
\begin{align}
    n_{(4)}^\mu \partial_\mu \Theta|_{\partial \mc{M}} = 0,
    \label{Eq:Neumann boundary condition for the four-dim gauge transformation}
\end{align}
so that the boundary condition \eqref{Eq:D-dim Neumann (absolute) boundary condition II} is also gauge invariant.

\subsection{Hawking--Page phase transition in four dimensions}
\label{Sec:HP-transition}

We briefly review the Hawking--Page transition for $\Lambda <0$, namely the phase transition between thermal AdS space and an AdS black hole \cite{Hawking:1982dh,Chamblin:1999hg,Chamblin:1999tk}.
To this end, we consider the free energy at the semiclassical level. At the tree level, the free energy is given by the on-shell Euclidean action as
\begin{align}
	F_{(4)} = T_{(4)} I[g,A^{(1)}]\big|_{\text{on-shell}},
    \label{Eq:def of 4D free energy}
\end{align}
where $g$ and $A^{(1)}$ in the brackets are solutions of the classical equations of motion (EOMs).
By comparing the free energies of different states, we can analyze their stability and find that the phase transition arises in the asymptotically AdS case.
It is noted that an appropriate counterterm is necessary in addition to the action in~\eqref{Eq:Einstein-Maxwell-action} so that we remove the infrared (IR) divergence and obtain a finite free energy.
The details of the counterterm and the computation of the free energy are discussed in App.~\ref{Sec:Counterterms} and App.~\ref{Sec:free energy}, respectively.

\begin{figure}[t]
    \begin{minipage}[b]{0.52\linewidth}
        \centering
        \includegraphics[scale=0.5]{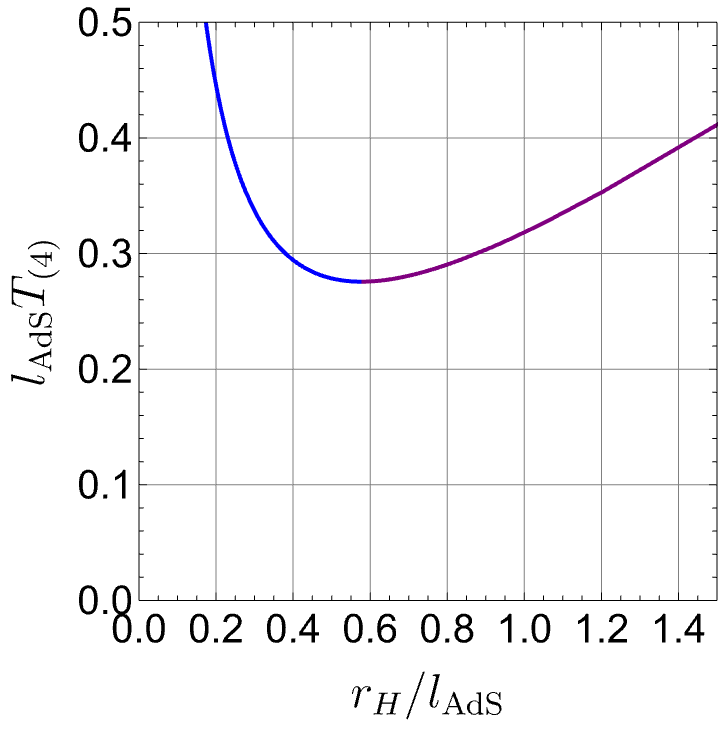}
        \subcaption{$T_{(4)}(r_H)$.}
        \label{Fig:4D_temperature_S_AdS}
    \end{minipage}
    \begin{minipage}[b]{0.52\linewidth}
        \centering
        \includegraphics[scale=0.5]{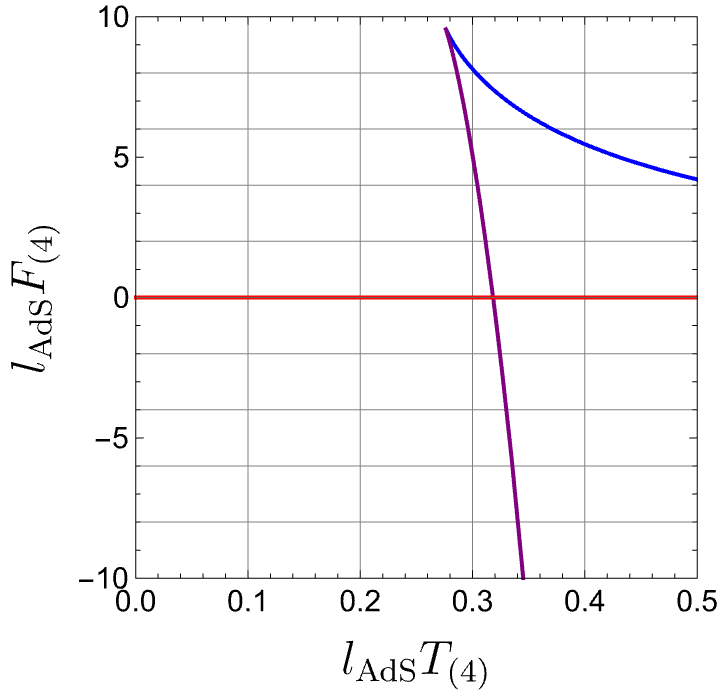}
        \subcaption{$T_{(4)}$ vs. $F_{(4)}$}
        \label{Fig:T_F_plane_S_AdS}
    \end{minipage}
        \caption{
        Plots of (a) the temperature as a function of the horizon radius and (b) the free energy as a function of the temperature for $l_{\rm AdS}/l_{\rm pl}= 10$, $Q=0$, and $e^2=4\pi$.
        The red, blue, and purple curves correspond to the pure AdS, small, and large black hole branches, respectively. 
        At the minimum temperature $T_0$, the small and large black hole branches merge.
        A phase transition between pure AdS and the large black hole occurs at the point where the red and purple curves intersect.
        }
        \label{fig:4D_HP_S_AdS}
\end{figure}

Let us first consider the Schwarzschild--AdS case, $Q=0$.
Since pure (thermal) AdS with $M=0$ has no horizon, the Euclidean geometry is smooth for arbitrary inverse temperature. Its free energy is conventionally taken to be zero (see Eq.~\eqref{eq:Pure-AdS-free-energy-in-4d}).
For Schwarzschild--AdS black holes (with $M\neq0$), Eqs.~\eqref{Eq:BH-temperature-4D} and \eqref{eq:free-energy-in-4d} with $Q=0$ give
\begin{align}
    T_{(4)}(r_H) = \frac{1}{4\pi} \left[ \frac{1}{r_H} + \frac{3r_H}{l_{\rm AdS}^2} \right],
    \quad
    F_{(4)}(r_H) = \frac{1}{4G_N} \left[ r_H - \frac{r_H^3}{l_{\rm AdS}^2}  \right],
    \label{Eq:4D temperature and free energy for S--AdS}
\end{align}
where $r_H$ is the horizon radius defined by $f(r_H)=0$, and $l_{\rm AdS}:=\sqrt{3/|\Lambda|}$ is the four-dimensional AdS length.
Fig.~\ref{fig:4D_HP_S_AdS} shows the plots of the temperature and the free energy for $l_{\rm AdS}/l_{\rm pl} = 10,~Q=0$ and $e^2 =4\pi$.
Here, we introduce $l_{\rm{pl}} \coloneqq \sqrt{G_N}$.
As shown in the figure, the temperature $T_{(4)}(r_H)$ has a minimum\footnote{
The phase transitions associated with extrema of the temperature are briefly discussed at the end of Sec.~\ref{Sec:Thermodynamics of the dilaton gravity with general potential}.
}
at $r_H=l_{\rm AdS}/\sqrt{3}$,
\begin{align}
    T_0 := T_{(4)}\bigg(\frac{l_{\rm AdS}}{\sqrt{3}}\bigg) = \frac{\sqrt{3}}{2\pi l_{\rm AdS}} .
\end{align}
Now let us consider a black hole in a thermal bath with the temperature $T^{\mr{bath}}_{(4)}$.
When the black hole is in thermal equilibrium with the bath, the black hole temperature equals $T_{(4)}^{\rm bath}$.
For $T_{(4)}^{\rm bath}<T_0$, no Schwarzschild--AdS black hole saddle exists because the line $T_{(4)}^{\rm bath}=\mathrm{const.}$ does not intersect the curve $T_{(4)}(r_H)$, and pure AdS is the only relevant saddle.
For $T_{(4)}^{\rm bath}>T_0$, the equation $T_{(4)}(r_H)=T_{(4)}^{\rm bath}=\mathrm{const.}$ has two solutions, corresponding to the small and large black hole branches shown by the blue and purple curves in Fig.~\ref{Fig:4D_temperature_S_AdS}, respectively.
It is noted that the black hole free energy vanishes at $r_H=l_{\rm AdS}$.
Therefore, the free energy of pure AdS coincides with that of the black hole at $r_H=l_{\rm AdS}$, as shown in Fig.~\ref{Fig:T_F_plane_S_AdS}, where the red and purple lines intersect.
This determines the Hawking--Page temperature:
\begin{align}
    T_{(4)\mathrm{HP}} := T_{(4)}(l_{\rm AdS})
    = \frac{1}{\pi l_{\rm AdS}} .
    \label{Eq:4D-HP-Temperature}
\end{align}
In Fig.~\ref{fig:4D_HP_S_AdS}, we find $l_{\rm AdS} T_{(4)\mathrm{HP}} = 1/\pi$ for $l_{\rm AdS}/l_{\rm pl} = 10$.
For $T_0<T_{(4)}^{\rm bath}<T_{(4)\mathrm{HP}}$, although two black hole saddles exist, their free energies are higher than those of pure AdS, as shown in Fig.~\ref{Fig:T_F_plane_S_AdS}.
Thus, pure AdS remains thermodynamically dominant.
For $T_{(4)}^{\rm bath}>T_{(4)\mathrm{HP}}$, the large black hole branch has the lowest free energy and becomes the dominant saddle.
Therefore, at $T_{(4)}^{\rm bath}=T_{(4)\mathrm{HP}}$, the coincidence of the free energies of pure AdS and the large black hole naturally indicates a phase transition between them.
This is known as the Hawking--Page transition.
It should be noted that the transition is first order, as seen in Fig.~\ref{Fig:T_F_plane_S_AdS}.

\begin{figure}[t]
    \begin{minipage}[b]{0.52\linewidth}
        \centering
        \includegraphics[scale=0.5]{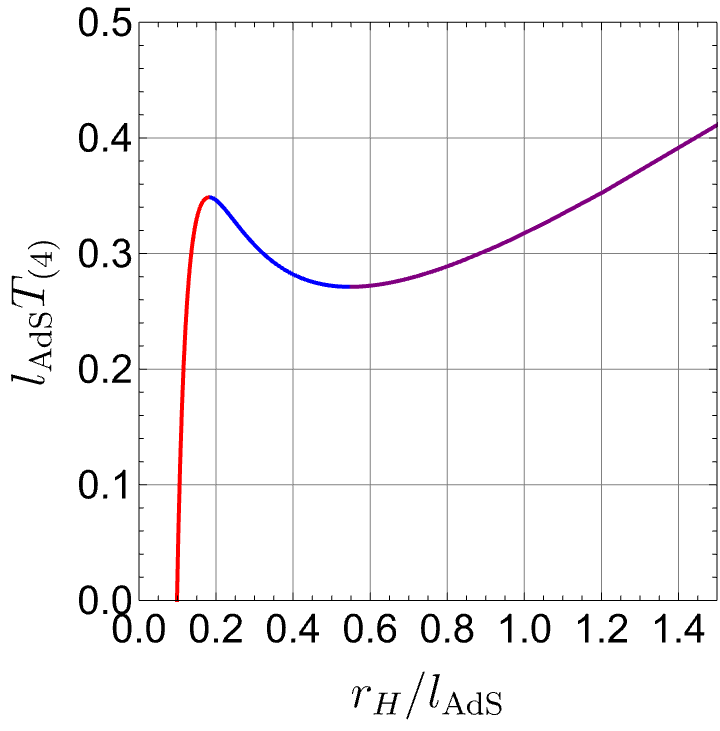}
        \subcaption{$T_{(4)}(r_H)$.}
        \label{Fig:4D_temperature}
    \end{minipage}
    \begin{minipage}[b]{0.52\linewidth}
        \centering
        \includegraphics[scale=0.5]{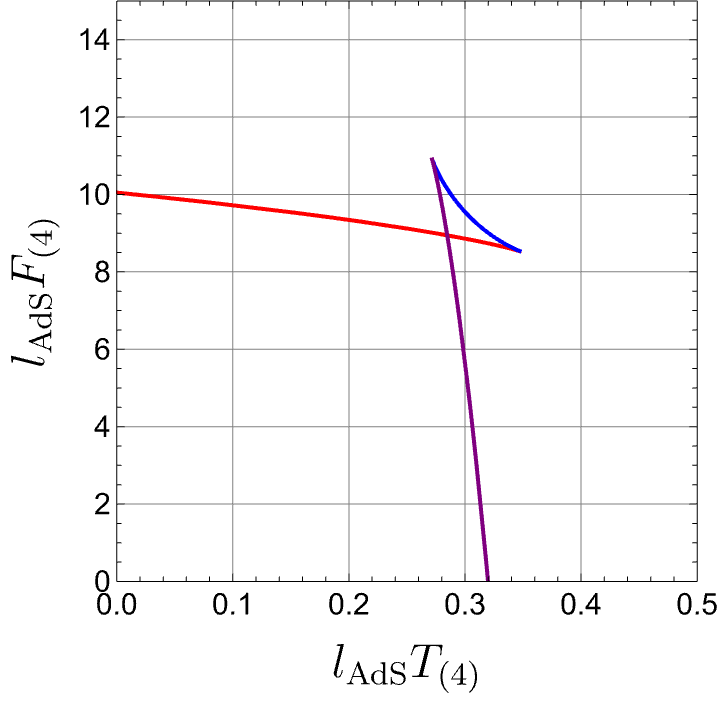}
        \subcaption{$T_{(4)}$ vs. $F_{(4)}$}
        \label{Fig:T_F_plane}
    \end{minipage}
        \caption{
        Plots of (a) the temperature as a function of the horizon radius and (b) the free energy as a function of the temperature for $l_{\rm AdS}/l_{\rm pl}= 10$, $Q=1$ and $e^2=4\pi$. 
        The red, blue, and purple curves correspond to the small, intermediate, and large black hole branches, respectively. 
        A phase transition between the small black hole and the large black hole occurs at the point where the red and purple curves intersect.
        }
        \label{fig:4D_HP}
\end{figure}

Let us next consider the RN--AdS case, $Q\neq0$.
The temperature and the free energy of RN--AdS black holes are given by
\begin{align}
    T_{(4)}(r_H) = \frac{1}{4\pi} \left[ \frac{1}{r_H} + \frac{3r_H}{l_{\rm AdS}^2} - \frac{G_N e^2 Q^2}{4\pi r_H^3}\right],
     \quad
    F_{(4)}(r_H) = \frac{1}{4G_N} \left[ r_H - \frac{r_H^3}{l_{\rm AdS}^2} + \frac{3G_N e^2 Q^2}{4\pi r_H}\right].
    \label{Eq:4D temperature and free energy for RN--AdS}
\end{align}
Apparently, the free energy diverges for the naked singularity solution with $M=0$, corresponding to the limit $r_H \to 0$. Therefore, black hole solutions with a finite horizon radius are always thermodynamically preferred.
For $Q^2<Q_c^2$, $T_{(4)}(r_H)$ has two extrema at
\begin{align}
    r_H = l_{\rm AdS}\sqrt{\frac{1}{6}\left( 1\pm \sqrt{1-\frac{Q^2}{Q_c^2}} \right)},
\end{align}
where the critical charge is defined by
\begin{align}
   Q_c := \sqrt{\frac{\pi l_{\rm AdS}^2}{9G_Ne^2 }}.
   \label{Eq:4D-critical-charge}
\end{align}
In this regime, the solutions split into three branches: small, intermediate, and large black holes, as shown in Fig.~\ref{Fig:4D_temperature}.
As seen from Fig.~\ref{Fig:T_F_plane}, the thermodynamically preferred branch changes from the small to the large black hole branch at a certain temperature, where the red and purple lines intersect.
This signals a phase transition in RN--AdS black holes~\cite{Chamblin:1999hg,Chamblin:1999tk}.
For $Q^2>Q_c^2$, the small and intermediate branches merge and disappear, leaving only a single branch.
The endpoint of the transition curve in the $(T_{(4)},Q)$-plane is located at $(T_{(4)c},Q_c)$, with
\begin{align}
    T_{(4)c} \coloneqq T_{(4)}\bigg(\frac{l_{\rm AdS}}{\sqrt{6}}\bigg)\bigg|_{Q=Q_c}
    =\sqrt{\frac{2}{3}}T_{(4)\text{HP}}.
    \label{Eq:4D-critical-temp}
\end{align}
As illustrated in Fig.~\ref{Fig:AdS_phase_diagram}, this phase structure is analogous to that of the van der Waals--Maxwell liquid-gas system.

\section{Two-dimensional effective dilaton gravity via dimensional reduction}
\label{Sec:Two-dimensional gravity via dimensional reduction}

In this section, we dimensionally reduce the four-dimensional gravitational theory describing charged black holes and derive the two-dimensional effective dilaton gravity, neglecting the massive KK modes. We also discuss its relation to JT gravity from this perspective.

In general, a theory can be explicitly reduced to a lower-dimensional one when spacetime enjoys symmetries.
In particular, four-dimensional spherically symmetric black holes can be studied using a two-dimensional theory obtained via dimensional reduction, in which the $S^2$ directions are integrated out~\cite{Grumiller:2002nm,Iliesiu:2020qvm,Banerjee:2021vjy,Gukov:2022oed}.
To obtain such an effective two-dimensional gravity model describing the black hole, we make the following assumptions:
\begin{itemize}
	\item The four-dimensional spacetime $\mc{M}$ can be decomposed into a two-dimensional spacetime $\Sigma$ and a two-dimensional sphere $S^2$:
	\begin{align}
		\mc{M} = \Sigma \times S^2,
		\qquad 
		\del \mc{M} = \del \Sigma \times S^2.
		\label{Eq:D=4 geometric assumption}
	\end{align}

	\item The four-dimensional metric is given by 
	\begin{align}
		ds^2 = \frac{\Phi(x)}{r_0^2} \Bigl[
			g_{ab}(x) dx^a dx^b + r_0^2 d\Omega_{(2)}^2
		\Bigr],
		\label{Eq:4-dim metric anzatz}
	\end{align}
	where $x^a$ $(a=0,1)$ are coordinates on $\Sigma$, and $r_0$ is a constant with the dimension of length corresponding to the compactification radius.
	Note that the dilaton $\Phi$ is a scalar field corresponding to the size of $S^2$ and has the dimension of length squared.
    Comparing the ansatz \eqref{Eq:4-dim metric anzatz} with the classical solution \eqref{Eq:4-dim Riesnner-Nordstrom action}, we find that $\Phi = r^2$ at the classical level.
	
	\item The four-dimensional gauge field $A^{(1)}$ depends only on $x$, and the flux lies along $\Sigma$
	\begin{align}
		A^{(1)}(x,y) = \frac{1}{\sqrt{4\pi}} a_a(x) dx^a.
		\label{Eq:4-dim Elemag anzatz}
	\end{align}
	Here, $y$ denotes coordinates on $S^2$.
\end{itemize}
The first assumption implies that the four-dimensional spacetime has spherical symmetry, and the second and third assumptions imply the truncation of non-zero (massive) KK modes.
This truncation is often justified in the near-extremal low-temperature regime, in which excitations of massive KK modes are expected to be suppressed~\cite{Iliesiu:2020qvm}.
The normalization factor of the gauge field $1/\sqrt{4\pi}$, which is the inverse square root of the $S^2$ volume, is included to ensure consistency with later results.

The two-dimensional effective action is obtained by substituting the ansatz \eqref{Eq:4-dim metric anzatz} and \eqref{Eq:4-dim Elemag anzatz} into the four-dimensional action \eqref{Eq:Einstein-Maxwell-action} and integrating over $S^2$.
From the metric ansatz \eqref{Eq:4-dim metric anzatz}, the four-dimensional scalar curvature is evaluated as 
\begin{align}
	R_{(4)} = \frac{r_0^2}{\Phi} \biggl[
		R_{(2)} + \frac{2}{r_0^2} + \frac{3}{2} \biggl(
			\frac{(\nabla_a \Phi)(\nabla^a \Phi)}{\Phi^2} - 2 \frac{\nabla_a \nabla^a \Phi}{\Phi}
		\biggr)
	\biggr],
\end{align}
where $2/r_0^2$ in the bracket is the scalar curvature of $S^2$. Hence, the Einstein--Hilbert action \eqref{Eq:Einstein-Hilbert-action} is expressed as 
\begin{align}
	I_\EH[g,\Phi] = - \frac{1}{4 G_N} \int_\Sigma d^2x \sqrt{\det g_{ab}} \biggl[
		\Phi \Bigl(
			R_{(2)} + \frac{2}{r_0^2}
		\Bigr) - \frac{2\Lambda}{r_0^2} \Phi^2 + \frac{3}{2\Phi} (\nabla_a \Phi)(\nabla^a \Phi) - 3 \nabla^2 \Phi
	\biggr].
	\label{Eq:2-dim-Eintein-Hilbert-action-1}
\end{align}

Let $n_{(4)} = n^\mu_{(4)} \del_\mu$
and $n_{(2)} = n^a_{(2)} \del_a$ be four- and two-dimensional unit normal vectors satisfying $g_{\mu\nu} n^\mu_{(4)} n^\nu_{(4)} =1$ and $g_{ab} n^a_{(2)} n^b_{(2)} = 1$, respectively.
Using the assumptions \eqref{Eq:D=4 geometric assumption} and \eqref{Eq:4-dim metric anzatz} and the normalization conditions for $n_{(4)}$ and $n_{(2)}$, we obtain the relation
\begin{align}
	n_{(4)} = \frac{r_0}{\Phi^{1/2}} n_{(2)}.
\end{align}
By the definition of $K_{(3)}$, we find that 
\begin{align}
	K_{(3)} = \frac{r_0}{\Phi^{1/2}} \biggl[
		\frac{3}{2\Phi} n^a_{(2)} \nabla_a \Phi + K_{(1)}
	\biggr],
\end{align}
where $K_{(1)} \coloneqq \nabla_a n^a_{(2)}$ is the extrinsic curvature of $\del \Sigma$.
Because the three-dimensional volume factor of $\del \mc{M}$ is proportional to $4 \pi \Phi \times \Phi^{1/2}$, the Gibbons--Hawking--York term \eqref{Eq:Gibbons-Hawking-York-term} becomes
\begin{align}
	I_\GHY[g,\Phi] = - \frac{1}{2 G_N} \int_{\del \Sigma} d\tau \sqrt{\det h_{ab}} \Bigl(
		\Phi K_{(1)} + \frac{3}{2} n^a_{(2)} \del_a \Phi
	\Bigr),
	\label{Eq:2-dim-Gibbons-Hawking-York-term-1}
\end{align}
where we write the coordinate on the boundary $\del \Sigma$ as $\tau$, and define the induced metric on $\del \Sigma$ by $h_{ab} = g_{ab} - n_{(2)a} n_{(2)b}$.
Note that the second term of Eq.~\eqref{Eq:2-dim-Gibbons-Hawking-York-term-1} cancels the last term of Eq.~\eqref{Eq:2-dim-Eintein-Hilbert-action-1}.
Next, the four-dimensional Maxwell action is simply reduced to the two-dimensional Maxwell action:
\begin{align}
	I_\Maxwell[a^{(1)}, g] = \frac{r_0^2}{4 e^2} \int_\Sigma d^2x \sqrt{\det g_{ab}} f_{ab} f^{ab} - \frac{r_0^2}{e^2} \int_{\del \Sigma} d\tau \sqrt{\det h_{ab}} n_{(2)a} a_b f^{ab}, 
    \quad
	f^{(2)} = d a^{(1)}.
	\label{Eq:2-dim Maxwell action}
\end{align}
The factor $e/(\sqrt{4 \pi} r_0)$ can be interpreted as a two-dimensional effective coupling constant.
Then, the total action consists of a two-dimensional dilaton gravity term and a two-dimensional Maxwell term:
\begin{align}
	I[g,\Phi,a] &= - \frac{1}{4 G_N} \int_{\Sigma} d^2x \sqrt{\det g_{ab}} \biggl[
        \Phi \Bigl(
			R_{(2)} + \frac{2}{r_0^2}
		\Bigr) - \frac{2\Lambda}{r_0^2} \Phi^2
        + \frac{3}{2\Phi} \del_a \Phi \del^a \Phi + \frac{2 G_Nr_0^2}{e^2} f_{ab} f^{ab}
	\biggr]
	\nn \\
	& \qquad 
	- \frac{1}{2 G_N} \int_{\del \Sigma} d\tau \sqrt{\det h_{ab}} \biggl[
		\Phi K_{(1)} + \frac{2 G_N r_0^2}{e^2} n_{(2)a} a_b f^{ab}
	\biggr].
	\label{Eq:action-before-weyl}
\end{align}
We integrate out the gauge field in the above action to obtain an effective dilaton gravity theory.
To this end, first, we check the two-dimensional classical solution obtained from the four-dimensional classical one \eqref{Eq:4-dim gauge field}.
The classical solution of the two-dimensional gauge field is 
\begin{align}
	\nabla_a f^{ab} = 0.
    \label{Eq:EOM of 2D gauge field}
\end{align}
The solution is parametrized by a constant $C$ as
\begin{align}
	f^{(2)} = da^{(1)} = \frac{\sqrt{\det g_{ab}}}{2!} C \varepsilon_{ab} dx^a \wedge dx^b
	\label{Eq:flux-2d}
\end{align}
where $\varepsilon_{ab}$ is the two-dimensional antisymmetric symbol.
This $C$ is fixed so that the two-dimensional classical solution matches the four-dimensional one.
The four-dimensional gauge field \eqref{Eq:4-dim Elemag anzatz} is given by
\begin{align}
	d A^{(1)}= \frac{1}{\sqrt{4 \pi}}\frac{1}{2} \sqrt{\det \Bigl( \frac{\Phi}{r_0^2} g_{ab} \Bigr)}~\frac{r_0^2}{\Phi} C\varepsilon_{ab} dx^a \wedge dx^b
	= \frac{r_0^2}{\sqrt{4\pi}} \frac{C}{r^2} dt \wedge dr,
	\label{Eq:flux-4d}
\end{align}
where we used \eqref{Eq:4-dim metric anzatz} and the condition $g_{tt} g_{rr} = 1$ for the four-dimensional metric \eqref{Eq:4-dim Riesnner-Nordstrom action}, together with $\Phi = r^2$.
By comparing the two-dimensional classical solution \eqref{Eq:flux-4d} with the four-dimensional classical solution \eqref{Eq:4-dim gauge field}, the constant $C$ is expressed in terms of the black hole charge $Q$ as
\begin{align}
	C = \frac{i}{\sqrt{4\pi}} \frac{e^2}{r_0^2} Q.
\end{align}
Hence, the on-shell Maxwell action is evaluated as 
\begin{align}
    I_\Maxwell\big|_{\text{on-shell}} &=  \frac{r_0^2}{4 e^2} \int_{\Sigma} d^2 x \sqrt{\det g_{ab}} f_{ab} f^{ab} - \frac{r_0^2}{e^2} \int_{\Sigma} d^2x \sqrt{\det g_{ab}} \nabla_a ( a_b f^{ab})
    \nn\\
    &= - \frac{r_0^2}{4e^2} \int_\Sigma d^2x \sqrt{\det g_{ab}} f_{ab} f^{ab} - \frac{r_0^2}{e^2} \int_\Sigma d^2x \sqrt{\det g_{ab}} a_b \nabla_a f^{ab}
    \nn \\
    &= \frac{1}{2} \int_\Sigma d^2x \sqrt{\det g_{ab}} \frac{e^2}{4 \pi r_0^2} Q^2,
    \label{Eq:2-dim Maxwell action on-shell}
\end{align}
where in the first line we rewrite the boundary term using Gauss's theorem and then use the relation $(\nabla_a a_b) f^{ab} = \frac{1}{2} f_{ab} f^{ab}$ together with the EOM~\eqref{Eq:EOM of 2D gauge field}.
Thus, in the effective dilaton gravity action, the Maxwell term appears as a cosmological constant term proportional to $Q^2$:
\begin{align}
    I_{\text{eff}}[g,\Phi]
    &= - \frac{1}{4 G_N} \int_{\Sigma} d^2x \sqrt{\det g_{ab}}
    \left[ 
        \Phi\left(R_{(2)}+\frac{2}{r_0^2}\right) -\frac{2\Lambda}{r_0^2} \Phi^{2}
        +\frac{3}{2\Phi} (\nabla \Phi)^2
        -\frac{G_N e^2}{2\pi r_0^2}Q^2
    \right]
    \notag\\
    &\qquad-\frac{1}{2G_N}\int_{\partial \Sigma} dx \sqrt{\det h_{ab}}~\Phi K_{(1)}.
    \label{eq:total-tree-action}
\end{align}
We now eliminate the dilaton kinetic term by performing the following Weyl transformation to simplify the analysis:
\begin{align}
    g_{ab} \mapsto  \frac{\Phi^{\frac{3}{2}}}{r_0^3} g_{ab}, \quad h_{ab} \mapsto  \frac{\Phi^{\frac{3}{2}}}{r_0^3} h_{ab}.
	\label{Eq:Weyl transformation}
\end{align}
Finally, the effective dilaton gravity action is given by
\begin{align}
	I_\eff[g,\Phi] &= - \frac{1}{4G_N} \int_\Sigma d^2x \sqrt{\det {g}_{ab}} \biggl[
		\Phi {R}_{(2)} + V_\eff(\Phi)
	\biggr] - \frac{1}{2 G_N} \int_{\del \Sigma} dx \sqrt{\det {h}_{ab}} \Phi {K}_{(1)},
    \label{Eq: effective dilaton action w/o massive KK mode}
	\\
	V_\eff(\Phi)&= 2 r_0 
    \left[\Phi^{-\frac{1}{2}} -  \Lambda \Phi^{\frac{1}{2}} - \frac{G_N e^2}{4\pi} Q^2 \Phi^{-\frac{3}{2}}\right].
    \label{Eq: effective dilaton potential w/o massive KK mode}
\end{align}
It should be noted that $V_{\mathrm{eff}}(\Phi)$ consists of the curvature of $S^2$ (coming from the four-dimensional curvature), the four-dimensional cosmological constant, and the four-dimensional energy density of the electric field, together with the contribution from the Weyl transformation \eqref{Eq:Weyl transformation}.\footnote{Note that the sign of $V_{\mathrm{eff}}$ is opposite to the conventional choice.}
So far, we have neglected quantum fluctuations of the two-dimensional gauge field, whose contributions will be discussed in Sec.~\ref{Sec:Zero modes}.

\subsubsection*{Connection to JT gravity}

Before turning to the thermodynamic analysis with the dilaton potential \eqref{Eq: effective dilaton potential w/o massive KK mode}, we briefly comment on a relation to JT gravity.
As discussed above, the effective theory of four-dimensional spherically symmetric charged black holes is a dilaton gravity theory whose potential is given by Eq.~\eqref{Eq: effective dilaton potential w/o massive KK mode}. 
In a certain limit, the effective theory reduces to JT gravity\cite{Iliesiu:2020qvm,Banerjee:2021vjy,Gukov:2022oed}.
We now focus on the near-horizon region $r\sim r_H$ and parameterize the dilaton as $\Phi(r) = r_H^2 + \tilde{\phi}(r)$.
Then, in the near-horizon limit where $\tilde{\phi}$ is sufficiently small, the potential is approximately given by
\begin{align}
    V_\eff (\Phi) \approx \tilde{V}_0 + 2 \lambda \tilde{\phi},
	\label{effective dilaton potential in flat case}
\end{align}
where $\tilde{V}_0 \coloneqq V_\eff(r_H^2)$ and $\lambda$ are constants.
Here, we take the near-extremal limit in which $\tilde{V}_0$ vanishes because the black hole temperature is proportional to this constant, as shown in the next section.
In this limit, the effective action becomes
\begin{align}
	I_{\mr{JT}}[\tilde{g}, \Phi] 
    = 
    &-\frac{\pi r_H^2}{G_N} \underbrace{\left[\frac{1}{4\pi} \int_{\Sigma} d^2 x \sqrt{\det \tilde{g}_{ab}} R_{(2)} + \frac{1}{2\pi}\int_{\del \Sigma} d\tau \sqrt{\det \tilde{h}_{ab}} K_{(1)}\right]}_{\displaystyle =\chi(\Sigma)}
    \notag\\
    &- \frac{1}{4G_N} \int_{\Sigma} d^2 x \sqrt{\det \tilde{g}_{ab}} ~\tilde{\phi} \Bigl( \tilde{R}_{(2)} + 2 \lambda \Bigr) - \frac{1}{2G_N}\int_{\del \Sigma} d\tau \sqrt{\det \tilde{h}_{ab}} ~\tilde{\phi} \tilde{K}_{(1)},
	\label{Eq:JT-gravity}
\end{align}
where $\chi(\Sigma)$ is the Euler characteristic of two-dimensional space $\Sigma$.
This is nothing but JT gravity~\cite{Teitelboim:1983ux,Jackiw:1984je} in the Euclidean signature.
Therefore, in the near-horizon and near-extremal limit, the two-dimensional effective theory of four-dimensional spherically symmetric charged black holes is approximately described by JT gravity~\cite{Almheiri:2014cka,Maldacena:2016upp,Nayak:2018qej,Gukov:2022oed, Iliesiu:2020qvm,Banerjee:2021vjy}.
This discussion is highly simplified, and a more detailed analysis is necessary \cite{Gukov:2022oed, Iliesiu:2020qvm,Banerjee:2021vjy}.

The advantage of JT gravity is that its path integral can be  calculated exactly~\cite{Saad:2019lba,Iliesiu:2019xuh}.
Since the dilaton potential is linear, the path integral over the dilaton imposes the constraint $R=-2\lambda$.
The functional integral over metrics can be evaluated using Mirzakhani’s recursion relations associated with hyperbolic geometry~\cite{Mirzakhani:2006eta}.
Furthermore, the boundary action with the extrinsic curvature becomes the Schwarzian action, which is a one-dimensional system.
The partition function can be calculated exactly \cite{Stanford:2017thb} using the Duistermaat--Heckman formula \cite{Duistermaat:1982vw}.
Thus, the full partition function of JT gravity can be evaluated exactly.
This fact enables us to estimate quantum gravitational effects in the near-extremal black hole, thereby providing a possible resolution of the mass gap problem~\cite{Iliesiu:2020qvm}.

We can see that JT gravity serves as a good model to study the quantum effects of spherically symmetric charged black holes in the near-extremal and near-horizon limits.
However, JT gravity has the following issues:
\begin{enumerate}[label={Issue~\arabic*}]
    
    \item JT gravity cannot probe non-extremal black holes or the geometric information outside the near-horizon region. 
    For example, the Hawking--Page transition does not occur in JT gravity.
    
    \item JT gravity does not include contributions to the potential from massive KK modes of the gravitational and electromagnetic fields owing to the low temperature of the near-extremal black hole.
    In other words, quantum fluctuations around the spherically symmetric classical solution \eqref{Eq:4-dim Riesnner-Nordstrom action} and \eqref{Eq:4-dim gauge field} are not fully taken into account.
    
\end{enumerate}
Issue 1 can be addressed by considering dilaton gravity with a nonlinear dilaton potential.
Although deformed JT gravity and its phase structure have been discussed by Witten in Ref.~\cite{Witten:2020ert}, the dilaton gravity obtained in this paper by dimensional reduction from higher-dimensional gravity cannot be regarded simply as a deformation of JT gravity.
This is because the dilaton potential contains fractional powers of the dilaton field, as shown in Eq.~\eqref{Eq: effective dilaton potential w/o massive KK mode}.
This suggests that Witten's discussion requires extension to a more general setup.
We discuss these details in Sec.~\ref{Sec:two-dimensional dilaton gravity and phase transition}.
Issue 2 can be addressed by performing KK reduction and integrating out massive KK modes.
This requires a generalization of the ansatz in Eqs.~\eqref{Eq:4-dim metric anzatz} and \eqref{Eq:4-dim Elemag anzatz}.
We discuss the details of the KK reduction of the photon field in Sec.~\ref{Sec:Partition function and effective action}, following earlier studies of $S^2$ compactification~\cite{Michelson:1999kn,Iliesiu:2020qvm}.

\section{Phase transitions in two-dimensional dilaton gravity}
\label{Sec:two-dimensional dilaton gravity and phase transition}

In this section, we study the thermodynamics of two-dimensional dilaton gravity.
In the previous section, we derived a two-dimensional effective dilaton gravity with the potential \eqref{Eq: effective dilaton potential w/o massive KK mode} from the four-dimensional RN--(A)dS black hole in Einstein--Maxwell theory.
Following the approach of Ref.~\cite{Witten:2020ert}, the semiclassical thermodynamics can be analyzed in a way that is largely independent of the detailed form of the dilaton potential.
To this end, we consider the general action\footnote{
    We keep the four-dimensional Newton constant $G_N$ so that the action matches the one obtained by dimensional reduction.
    In the conventional normalization of two-dimensional dilaton gravity, the overall coefficient is often written as $1/2$ or $1/(16\pi G_{N(2)})$, where $G_{N(2)}$ denotes the two-dimensional Newton constant, and the dilaton is typically taken to be dimensionless.}
\begin{align}
	I[g,\Phi] &= - \frac{1}{4G_N} \int_\Sigma d^2x \sqrt{\det g_{ab}} \bigl[
		\Phi R_{(2)} + V(\Phi)
	\bigr] - \frac{1}{2G_N}\int_{\del \Sigma} d\tau \sqrt{\det h_{ab}} \Phi K_{(1)}
    \\
	&\eqqcolon I_\bulk [g,\Phi] + I_{\mathrm{boundary}}[g,\Phi].
    \label{Eq:2-dim. dilaton gravity action}
\end{align}
As in the case of deformations of JT gravity~\cite{Witten:2020ert}, the phase structure is determined from the free energy obtained from the semiclassical partition function.
In particular, we show that the semiclassical thermodynamics is governed by the form of the dilaton potential.
We then apply this framework to the effective potential \eqref{Eq: effective dilaton potential w/o massive KK mode} and demonstrate that the resulting phase structure agrees with that of four-dimensional charged black holes.

\subsection{Thermodynamics of dilaton gravity with a general potential}
\label{Sec:Thermodynamics of the dilaton gravity with general potential}

First, let us find a classical black hole solution with a general dilaton potential $V(\Phi)$.
By a suitable choice of coordinates, the metric and dilaton are taken to be of the form
\begin{align}
    ds^2 = H(x)\, d\tau^2 + \frac{dx^2}{G(x)}, 
    \qquad
    \Phi = \Phi(x).
\end{align}
Note that the coordinates $(\tau,x )$ are not the same as $(t,r)$ in four dimensions, and the relation among them is discussed later.
Substituting this ansatz into the action gives
\begin{align}
   I_{\mathrm{bulk}}[g,\Phi] 
    &= -\frac{\beta_{(2)}}{4G_N} \int dx\left[ -\Phi(x)\, \frac{d}{dx}\left( \sqrt{\frac{G(x)}{H(x)}} \frac{dH}{dx}(x) \right) + 
    \sqrt{\frac{H(x)}{G(x)}}
    V(\Phi)\right], 
    \label{Eq:bulk-action}
    \\
    I_{\mathrm{boundary}}[g,\Phi]  &= - \frac{\beta_{(2)}}{4G_N} \Phi \sqrt{\frac{G}{H}}\frac{dH}{dx}\bigg|_{x={\rm boundary}} .
    \label{Eq:boundary-action}
\end{align}
Here, we assumed that the two-dimensional Euclidean time $\tau$ has a period $\beta_{(2)}$. 
The EOMs for $G$, $H$, and $\Phi$ read
\begin{align}
    \left(\frac{G}{H}\right)
    \frac{d\Phi}{dx} \frac{dH}{dx} - 
    V(\Phi) &=0,
    \label{Eq:EOM1}
    \\
    \frac{d}{dx} \left( \sqrt{\frac{G}{H}} \frac{d\Phi}{dx} \right)
    &=0,
    \label{Eq:EOM2}
    \\
    \sqrt{\frac{G}{H}}
    \frac{d}{dx}\left( \sqrt{\frac{G}{H}} \frac{dH}{dx} \right) - 
    \frac{dV}{d\Phi}(\Phi) &=0.
    \label{Eq:EOM3}
\end{align}
Here, we used the first equation \eqref{Eq:EOM1} to simplify Eq.~\eqref{Eq:EOM2}.
It is noted that, upon varying $H$, the surface term arising from the bulk action that involves derivatives of $\delta H$ is canceled by the boundary contribution:
\begin{align}
    \delta_H I_{\rm bulk} \ni \frac{\beta_{(2)}}{4G_N}\Phi \sqrt{\frac{G}{H}}\Bigl(\frac{d}{dx}\delta H\Bigr)\bigg|_{x={\rm boundary}},
    \quad 
    \delta_H I_{\mathrm{boundary}} \ni - \frac{\beta_{(2)}}{4G_N}\Phi \sqrt{\frac{G}{H}}\Bigl(\frac{d}{dx}\delta H\Bigr)\bigg|_{x={\rm boundary}}.
\end{align}
We can reparameterize the radial coordinate $x$ as $\rho$ to set $H=G$:
\begin{align}
    d\rho := \sqrt{\frac{H}{G}} dx, \quad
    ds^2 = H(\rho) d\tau^2 + \frac{d\rho^2}{H(\rho)}.
    \label{Eq:2-dim. simplified metric}
\end{align}
The EOMs \eqref{Eq:EOM1}--\eqref{Eq:EOM3} are simplified to
\begin{align}
    \frac{d\Phi}{d\rho} \frac{dH}{d\rho} - V(\Phi) =0,
    \qquad
    \frac{d^2\Phi}{d\rho^2}=0,
    \qquad   
    \frac{d^2 H}{d\rho^2} - \frac{dV}{d\Phi}(\Phi) =0.
    \label{Eq:EOMs for Phi and H}
\end{align}
The family of solutions is parameterized by a dilaton value $\Phi_H$ as 
\begin{align}
    \Phi(\rho) = \gamma \rho, 
    \qquad
    H(\rho) = \frac{1}{\gamma^2}\int_{\Phi_H}^{\gamma\rho} V(\Phi)\, d\Phi ,
    \label{Eq:classical-solution}
\end{align}
where $\gamma$ is introduced and has dimensions of length.\footnote{
    This constant can be eliminated by a coordinate transformation and a redefinition of $H$, since the metric \eqref{Eq:2-dim. simplified metric} is invariant under $(\tau,\rho,H) \mapsto (\tau/\gamma', \gamma' \rho, \gamma'^2 H)$. 
    Here $\gamma'$ is a constant. 
    However, we retain $\gamma$ for later convenience.
}
This metric describes a black hole geometry, and $\rho_H \coloneqq \Phi_H/\gamma$ gives the horizon radius defined by $H(\rho_H)=0$.
In the following discussion, we focus on the region outside the horizon, $\rho_H \le \rho < \infty$, assuming that the boundary is located at $\rho= \infty$ and that $V(\Phi) \geq 0$ throughout the exterior region of the event horizon.
Let us consider the temperature of this black hole solution.
As in Eq.~(\ref{Eq:BH-inverse-temperature-4D}), the smoothness condition at the horizon requires the Euclidean time periodicity, $\tau \sim \tau + \beta_{(2)}$ with
\begin{align}
    \beta_{(2)} 
    = 4\pi\bigg(\frac{dH}{d\rho}(\rho_H)\bigg)^{-1}
    = \frac{4\pi\gamma}{V(\Phi_H)}.
    \label{dilaton gravity inverse temperature}
\end{align}
$T_{(2)}=\beta^{-1}_{(2)} = V(\Phi_H)/(4\pi\gamma)$ is regarded as 
the Hawking temperature.
The relation between $\beta_{(2)}$ and $\beta_{(4)}$ in Eq.~\eqref{Eq:BH-inverse-temperature-4D} is also discussed later.

Next, we compute the semiclassical partition function
\begin{align}
    Z(\beta_{(2)}) \approx \exp\left[-I[g,\Phi]\big|_{\text{on-shell}}\right],
\end{align}
where the on-shell action is obtained by substituting the classical solutions of \eqref{Eq:classical-solution}:
\begin{align}
    I[g,\Phi]\big|_{\text{on-shell}}
    &= \frac{\beta_{(2)}}{4G_N\gamma}\int_{\Phi_H}^{\infty} \left(
    \Phi \frac{dV}{d\Phi} - V
    \right) d\Phi
    -  \frac{\beta_{(2)}}{4G_N\gamma}\Phi V|_{\Phi = \infty} 
    \nn \\
    &= -\frac{\pi\Phi_H}{G_N} - \frac{\beta_{(2)}}{2G_N \gamma} \int_{\Phi_H}^{\infty} V(\Phi)\, d\Phi.
\end{align}
Here, the relation $\Phi =\gamma \rho$ and the temperature given in \eqref{dilaton gravity inverse temperature} are used.
The last term in the second line is divergent if $V(\Phi)$ contains terms behaving, for instance, as $\Phi^{\epsilon}$ with $-1 \le \epsilon$.
Therefore, we introduce the following counterterm to render the on-shell action finite \cite{Grumiller:2007ju}:
\begin{align}
    I_c[g,\Phi] 
    &= \frac{1}{2G_N}\int_{\partial \Sigma} 
    d\tau \sqrt{\det h_{ab}}
    \sqrt{W(\Phi)} 
    = \frac{1}{2G_N}\left.\int_0^{\beta_{(2)}} d\tau\, \sqrt{H(\rho) W(\Phi)}\right|_{\rho, \Phi \to\infty},
    \label{Eq:2-dim counterterm}
\end{align}
where $W(\Phi)$ is a prepotential defined by
\begin{align}
    V(\Phi) \eqqcolon \frac{dW}{d\Phi}(\Phi),
\end{align}
and hence $H(\rho)$ reads 
\begin{align}
    H(\rho) = \frac{[W(\Phi)-W(\Phi_H)]}{\gamma^2}
\end{align}
for the on-shell $H$.
The prepotential has an ambiguity up to an additive constant.
Hereafter, we set the constant term equal to zero for simplicity without loss of generality. 
The choice of the constant reflects the zero point of the (free) energy, as seen later.
We rewrite the on-shell action and counterterm as
\begin{align}
    I[g,\Phi]\big|_{\text{on-shell}}
    &= -\frac{\pi\Phi_H}{G_N} - \frac{\beta_{(2)}}{2G_N\gamma} \lim_{\Phi\to\infty} W(\Phi) + \frac{\beta_{(2)}}{2G_N\gamma} W(\Phi_H), \\
    I_c[g,\Phi] 
    &= \frac{\beta_{(2)}}{2G_N\gamma} \lim_{\Phi\to\infty} \sqrt{(W(\Phi)-W(\Phi_H))W(\Phi)}
    \nn \\
    &= \frac{\beta_{(2)}}{2G_N\gamma} \biggl[\lim_{\Phi\to\infty} W(\Phi) - \frac{1}{2} W(\Phi_H)
    \biggr],
\end{align}
where we assumed $\lim_{\Phi\to\infty}[{W(\Phi_H)}/{W(\Phi)}] \to 0$.
The divergent terms turn out to cancel between the two actions, and the renormalized on-shell action reads
\begin{align}
    I_{\mathrm{ren}}[g,\Phi]\big|_{\text{on-shell}}
    = I[g,\Phi]\big|_{\text{on-shell}} + I_c[g,\Phi]
    = -\frac{\pi\Phi_H}{G_N} + \frac{\beta_{(2)}}{4G_N\gamma} W(\Phi_H).
\end{align}
Using the definition of the free energy,
\begin{align}
    \beta_{(2)} F_{(2)}=-\log Z =  - S + \beta_{(2)} E_{(2)} \approx  I_{\mathrm{ren}}[g,\Phi]\big|_{\text{on-shell}}, 
\end{align}
where $E$ and $S$ denote the energy and entropy of the black hole, respectively, and we obtain
\begin{align}
    F_{(2)}=\frac{1}{4G_N \gamma}(W(\Phi_H)-\Phi_H V(\Phi_H)),
    \qquad
    E_{(2)} = \frac{1}{4G_N\gamma} W(\Phi_H),
    \qquad 
    S = \frac{\pi\Phi_H}{G_N}. 
    \label{Eq:dilaton-gravity-entropy-energy}
\end{align}
Note that an additive constant in $W(\Phi)$ shifts the (free) energy if present.
The above entropy coincides with the Bekenstein--Hawking formula $S=\frac{\pi r_H^2}{G_N}$ in four dimensions upon identifying $r_H^2 = \Phi_H$ as in Sec.~\ref{Sec:Reduced gravity from charged black hole}.
Although we have introduced an appropriate two-dimensional term to render the on-shell action finite, this is also naturally obtained by dimensional reduction of the corresponding four-dimensional boundary term. 
Details are presented in App.~\ref{Sec:Counterterms}.

\begin{figure}[tbp]
    \begin{minipage}[t]{0.3\columnwidth}
        \centering
        \includegraphics[scale=0.3]{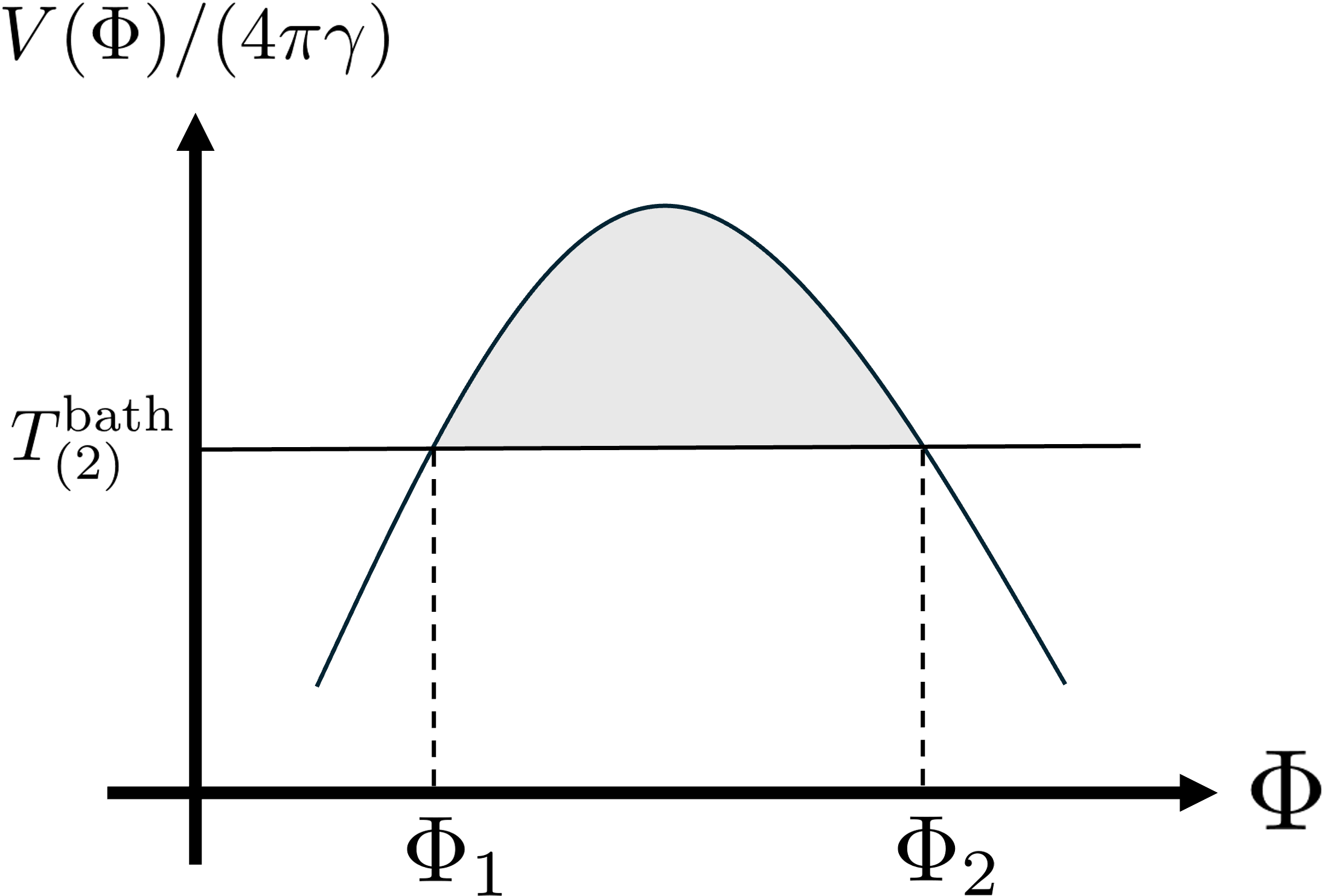}
        \subcaption{$\Delta F_{(2)}>0$: the smaller black hole is more stable than the larger one.}
        \label{Fig:Delta_F_positive}
    \end{minipage}
    \quad
    \begin{minipage}[t]{0.3\columnwidth}
        \centering
        \includegraphics[scale=0.3]{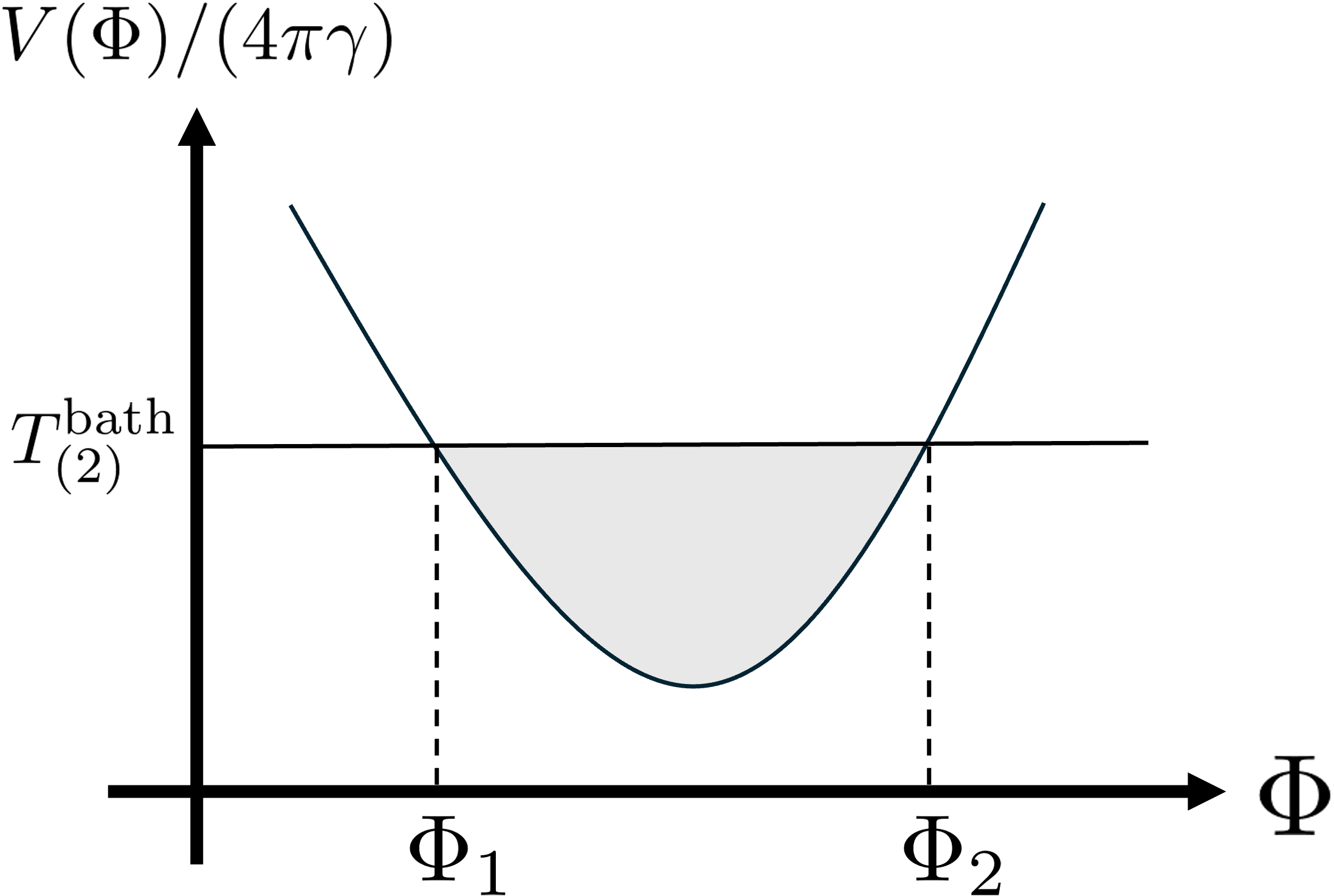}
        \subcaption{$\Delta F_{(2)}<0$: the larger black hole is more stable than the smaller one.}
        \label{Fig:Delta_F_negative}
    \end{minipage}
    \quad
    \begin{minipage}[t]{0.3\columnwidth}
         \centering
        \includegraphics[scale=0.3]{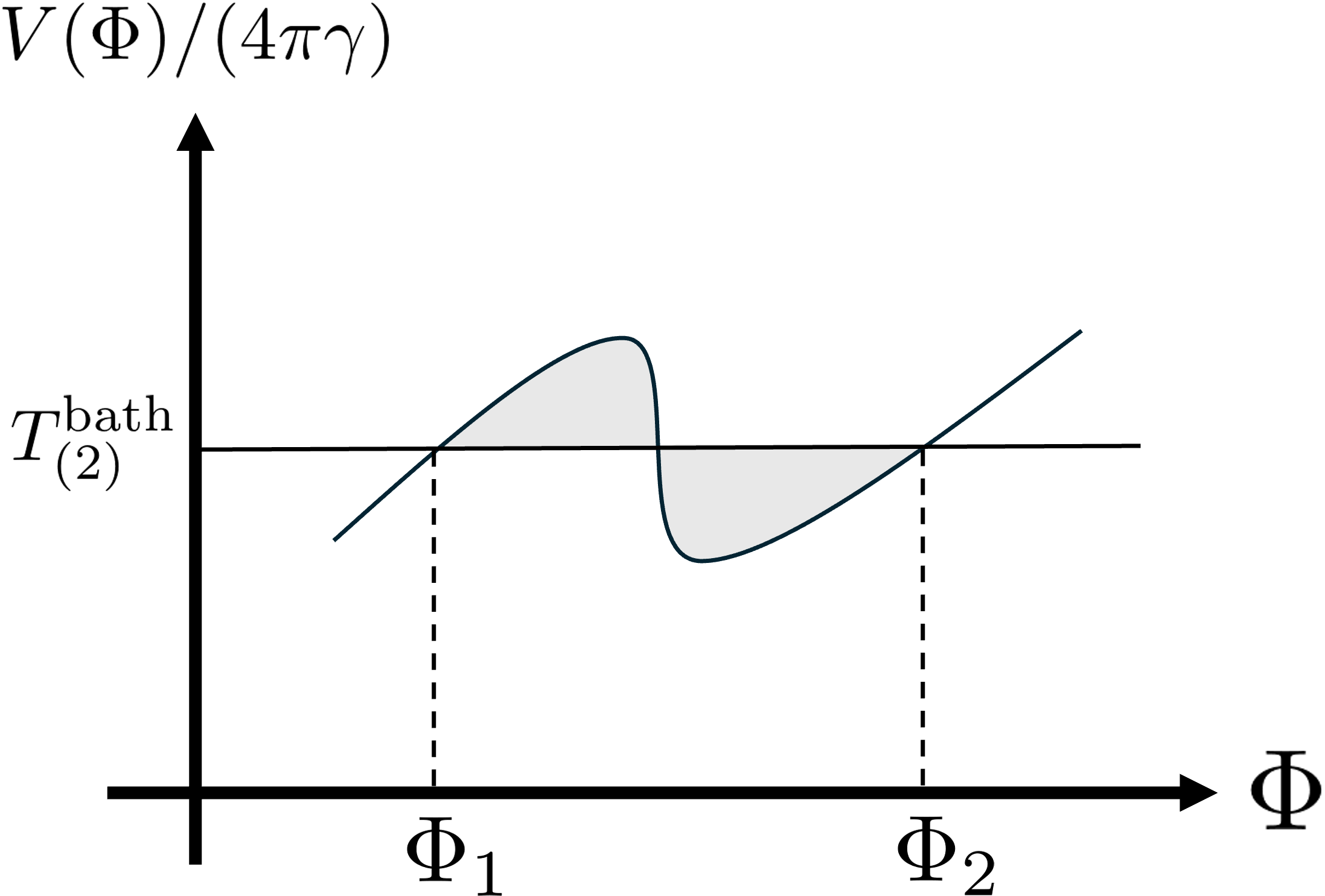}
        \subcaption{$\Delta F_{(2)}=0$: a 
         phase transition can occur between the smaller black hole and the larger one.}
         \label{Fig:Delta_F_zero}
     \end{minipage}
    \caption{
    Three examples of the relations between the black hole temperature and dilaton potential in the $(\Phi, V(\Phi)/(4\pi\gamma))$-plane.
    The area of the shaded region corresponds to $\Delta F$.
    Recalling that $\Phi=\Phi_H$ is proportional to the entropy as in Eq.~\eqref{Eq:dilaton-gravity-entropy-energy}, this is analogous to the temperature-entropy diagram. 
    $\Phi_H$ denotes the dilaton value at an intersection point between the line of $T_{(2)}^{\rm bath}={\rm const.}$ and the curve $V(\Phi)/(4\pi\gamma)$ in the figures, corresponding to a horizon radius of a black hole state.
    }
    \label{fig:T_diagram}
\end{figure}

Now we consider the phase structure of dilaton gravity using the temperature \eqref{dilaton gravity inverse temperature} and the free energy \eqref{Eq:dilaton-gravity-entropy-energy}.
Suppose that there exist two horizons, $\Phi_H=\Phi_1$ and $\Phi_H=\Phi_2$, such that 
$V(\Phi_1) = V(\Phi_2)$ with $\Phi_1 < \Phi_2$. 
In this case, the ``smaller'' black hole has its event horizon at $\rho=\Phi_1/\gamma$, while the ``larger'' one has its event horizon at $\rho=\Phi_2/\gamma$.
Then, we consider black holes in thermal equilibrium with a heat bath at temperature $T_{(2)}^{\rm bath}$.
These two black holes have the same temperature
\begin{align}
    T_{(2)}= \frac{V(\Phi_1)}{4\pi\gamma} =\frac{V(\Phi_2)}{4\pi\gamma}
    =T_{(2)}^{\rm bath},
    \label{Eq:temperature-2}
\end{align}
but they may have different entropy, energy, and free energy.
A comparison of the free energies of the two black holes allows us to assess their relative thermodynamic stability.
The differences in entropy, energy, and free energy are given by
\begin{align}
    \Delta S &= \frac{\pi}{G_N} (\Phi_2 - \Phi_1),
    \qquad 
    \Delta E = \frac{1}{4G_N\gamma} \int_{\Phi_1}^{\Phi_2} V(\Phi) d \Phi,
    \\
    \Delta F_{(2)} &= \Delta E_{(2)} - T_{(2)}^{\rm bath} \Delta S = \frac{\pi}{G_N} \int_{\Phi_1}^{\Phi_2}\bigg[ \frac{V(\Phi)}{4\pi \gamma} - T_{(2)}^{\rm bath} \bigg] d\Phi,
    \label{Eq:Delta F}
\end{align}
because these black holes share the same temperature $T_{(2)}^{\rm bath}$.
Eq.~\eqref{Eq:Delta F} indicates that $\Delta F_{(2)}$ is given by the area bounded by the curve $V(\Phi)/(4\pi \gamma)$ and the line of the constant $T_{(2)}^{\rm bath}$, as shown by the shaded region in Fig.~\ref{fig:T_diagram}. 
Here, an intersection point at a given $\Phi = \Phi_H$ between the line of $T_{(2)}^{\rm bath}={\rm const.}$ and the curve $V(\Phi)/(4\pi \gamma)$ represents a black hole state with a temperature $V(\Phi_H)/(4\pi \gamma)$ and horizon located at $\rho=\Phi_H/\gamma$.
The relative phase stability is determined by the shape of the dilaton potential.
As shown in Fig.~\ref{Fig:Delta_F_positive}, when the line $T_{(2)}^{\rm bath}=\mathrm{const.}$ intersects a locally convex region of the dilaton potential at two points, $\Phi_1$ and $\Phi_2$, the free-energy difference $\Delta F_{(2)}$ between them is positive.  
In this case, the smaller black hole is thermodynamically favored over the larger one, even at the same temperature $T_{(2)}$.
Conversely, when the line $T_{(2)}^{\rm bath}=\mathrm{const.}$ intersects a locally concave region of the dilaton potential at two points, the free-energy difference $\Delta F_{(2)}$ becomes negative, as shown in Fig.~\ref{Fig:Delta_F_negative}.  
In this case, the larger black hole is thermodynamically favored over the smaller one at the same temperature.
When the dilaton potential has two extrema and $\Delta F_{(2)}$ vanishes, as shown in Fig.~\ref{Fig:Delta_F_zero}, a phase transition between the small and large black holes can occur. 
This provides a two-dimensional interpretation of the Hawking--Page transition between asymptotically AdS geometries discussed in Sec.~\ref{Sec:HP-transition}.

\begin{figure}[t]
    \centering
    \includegraphics[scale=0.4]{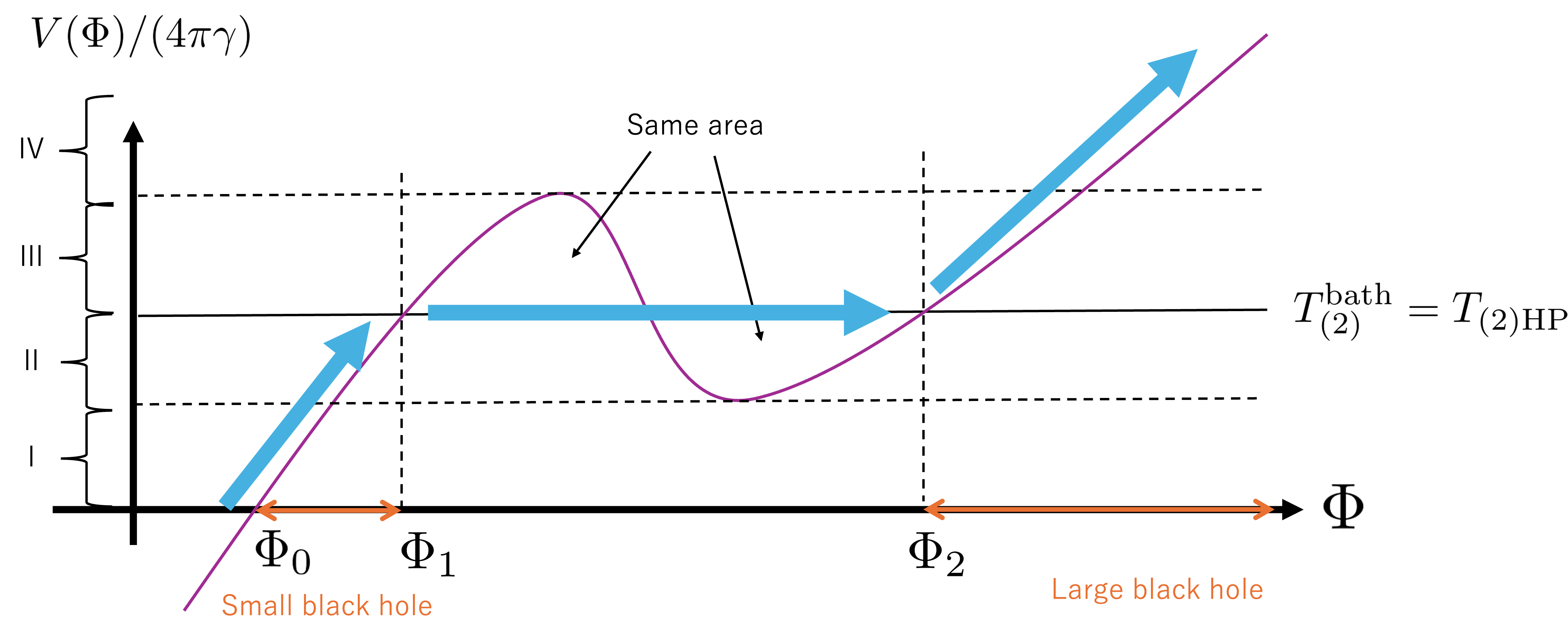}
    \caption{
        Schematic behavior of the phase transition in the $(\Phi, V(\Phi)/(4\pi \gamma))$-plane.
        If the temperature lies in region I, one stable state exists.
        In region II, where three states coexist, the smallest black hole is the most stable.
        At $T_{(2)}^{\rm bath} = T_{(2){\rm HP}}$, a first-order phase transition occurs and $\Phi_H$ jumps from $\Phi_1$ to $\Phi_2$.
        In region III, where three states coexist as in region II, the largest black hole is more stable than the smaller ones.
        In region IV, only one state exists.
    }
    \label{fig:Schematic_potential_graph.}
\end{figure}

Let us consider in more detail the case shown in Fig.~\ref{Fig:Delta_F_zero} and the associated black hole phase transition.
The dilaton potential is shown schematically in Fig.~\ref{fig:Schematic_potential_graph.}. 
There exist intersection points at $\Phi=\Phi_H$ between the constant line $T_{(2)}^{\rm bath}$ and the curve $V(\Phi)/(4\pi\gamma)$.
If the bath temperature is sufficiently low (high), the smallest (largest) black hole is realized, depending on the sign of the free-energy difference.
For $T_{(2)}^{\rm bath}=T_{(2)\rm HP}$, where $\Delta F_{(2)}=0$, the system can undergo a phase transition from the smallest black hole with the horizon at $\rho=\Phi_1/\gamma$ to the largest black hole with the horizon at $\rho=\Phi_2/\gamma$, as indicated by the blue arrows in Fig.~\ref{fig:Schematic_potential_graph.}, as the bath temperature increases.
It should be noted that the transition is first order, since the entropy $\pi \Phi_H/G_N$ (and the horizon size $\Phi_H/\gamma$), obtained from the derivative of the free energy, exhibit discontinuous jumps between $\Phi_1$ and $\Phi_2$.\footnote{
    In this paper, we do not evaluate the transition rate.
    Such dynamical analyses have been performed in Refs.~\cite{Li:2020nsy,Li:2021zep,Liu:2021lmr,Li:2022ayz,Afshar:2024mwy}.
}

Thermodynamic stability is also discussed in terms of specific heat, which is given by 
\begin{align}
    \frac{dE_{(2)}}{dT_{(2)}} = \frac{dE_{(2)}}{d\Phi} \frac{d\Phi}{dT_{(2)}}\bigg|_{\Phi=\Phi_H} 
    = \frac{4\pi^2 \gamma T_{(2)}}{G_N} \left( \frac{dV}{d\Phi} \right)^{-1}\bigg|_{\Phi=\Phi_H},
    \label{Eq:specific heat}
\end{align}
where Eqs.~\eqref{dilaton gravity inverse temperature} and \eqref{Eq:dilaton-gravity-entropy-energy} are used.
This equation means that $\frac{d V}{d \Phi}$ provides information about thermodynamic stability.
If $\frac{dV}{d\Phi}(\Phi_H)<0$, the specific heat is negative, and the corresponding black hole is thermodynamically unstable against temperature fluctuations due to evaporation.
For instance, in Fig.~\ref{Fig:Delta_F_positive} (Fig.~\ref{Fig:Delta_F_negative}), the smaller (larger) black hole with its horizon located at $\rho=\Phi_1/\gamma$ ($\Phi_2/\gamma$) is stable owing to the positive gradient of $V(\Phi)$, while the larger (smaller) one with its horizon located at $\rho=\Phi_2/\gamma$ ($\Phi_1/\gamma$) is unstable. 
In Fig.~\ref{Fig:Delta_F_zero}, two black holes with horizons located at $\rho=\Phi_1/\gamma$ and $\rho=\Phi_2/\gamma$ are thermodynamically stable.
This can also be seen from the viewpoint of the free energy,
\begin{align}
    \frac{dE_{(2)}}{dT_{(2)}} = - \frac{\pi^2\Phi_H T_{(2)}}{G_N^2} \left(\frac{dF_{(2)}}{d\Phi}\right)^{-1}\bigg|_{\Phi=\Phi_H}.
    \label{Eq:dE&dF}
\end{align}
If $\frac{dF_{(2)}}{d\Phi}(\Phi_H)$ is negative (positive), the specific heat is positive (negative), and the corresponding black hole branch is thermodynamically stable (unstable).
Since the specific heat of the black hole is proportional to $\left(dV/d\Phi\right)^{-1}$ or $\left(dF/d\Phi\right)^{-1}$, it diverges at the extrema of $V$, or equivalently of $F$. 
Moreover, its sign changes across such points.  
Such a point is known as the Davies point~\cite{Davies:1978zz,Davies:1977bgr}. 
Near this point, a transition from an unstable black hole to a stable one is expected to occur. 
The condition $dV/d\Phi=0$ can also be regarded as defining a critical point where the phase transition disappears.  
It should be noted that this transition is driven by instability, while the Hawking--Page transition occurs between stable states.

Finally, we comment on the treatment of solutions with $\Phi_H=0$. 
Such solutions can be smooth and horizonless, and hence the temperature $T_{(2)}$ is arbitrary. 
To determine the phase diagram, one needs to compare the thermodynamic stability of the horizonless state and the black hole state at a given temperature. 
This is analogous to the original Hawking--Page transition between pure AdS and the Schwarzschild--AdS black hole discussed in Sec.~\ref{Sec:HP-transition}.

\subsection{Examples of thermodynamics in effective dilaton gravity}
\label{sec:thermodynamics-of-effectivel-dilaton-gravity}

Applying the above discussion to the effective dilaton gravity with the potential \eqref{Eq: effective dilaton potential w/o massive KK mode},
\begin{align*}
    V_\eff(\Phi)&= 2 r_0 
    \left[\Phi^{-\frac{1}{2}} -  \Lambda \Phi^{\frac{1}{2}} - \frac{G_N e^2}{4\pi} Q^2 \Phi^{-\frac{3}{2}}\right],
\end{align*}
we solve the EOM and show that the resulting phase structure agrees with the four-dimensional one.
Using Eq.~\eqref{Eq:classical-solution}, the classical solutions for the above effective potential are given by
\begin{align}
    \Phi &= r_0\rho, \label{Eq:2-dim. dilaton solution}\\
    ds^2 &= H(\rho) d\tau^2 + \frac{d\rho^2}{H(\rho)}, \quad
    H(\rho) = \frac{1}{r_0^2} \left[W(\rho) - W(\rho_H)\right],
    \label{Eq: 2-dim. metric solution}
    \\
    W(\rho) &= 
    4 r_0
    \left[
        (r_0 \rho)^{\frac{1}{2}} 
        +\frac{e^2}{4\pi}G_N Q^2 (r_0 \rho)^{-\frac{1}{2}} 
        -\frac{\Lambda}{3} (r_0 \rho)^{\frac{3}{2}}
    \right],
    \label{Eq: 2-dim. effective prepotential}
\end{align}
where we set $\gamma=r_0$ for simplicity.
For $\Lambda \leq 0$, the equation $H(\rho) = 0$ may admit multiple solutions, such as $\rho_{H_1}$ and $\rho_{H_2}$ ($\rho_{H_1} > \rho_{H_2}$), as in the four-dimensional case. 
The solutions $\rho_{H_1}$ and $\rho_{H_2}$ correspond to the outer and inner horizon radii, respectively. 
For $\Lambda > 0$, there is an additional solution corresponding to a cosmological horizon at finite $\rho$. 
Hereafter, we restrict our attention to the region outside the outer horizon for $\Lambda \leq 0$, since the integration over $\Phi$ in the action becomes subtle otherwise.
Performing the inverse transformation of \eqref{Eq:Weyl transformation} and using the four-dimensional metric ansatz \eqref{Eq:4-dim metric anzatz}, we obtain 
the four-dimensional RN metric with mass $M$ and charge $Q$:
\begin{align}
    ds^2 &= f(r) dt^2 + \frac{dr^2}{f(r)} + r^2 d\Omega^2_{(2)},
    \label{Eq: 4-dim. metric solution from 2-dim. effective action}
    \\
    f(r) &:= 1 - \frac{2G_N M}{r} + \frac{e^2}{4\pi}\frac{G_NQ^2}{r^2} -\frac{\Lambda}{3}r^2, 
    \label{Eq:f(r) w/ eff charge}
    \\
    M &:= \frac{1}{2}\cdot\frac{W(\rho_H)}{4 G_N r_0},
     \label{Eq:mass-W}
\end{align}
where we redefine the coordinates as
\begin{align}
    (t,r) := (2\tau,\sqrt{r_0\rho})=(2\tau,\sqrt{\Phi}).
    \label{Eq: cdnt transformation 4d to 2d}
\end{align}
From Eqs.~\eqref{Eq: 2-dim. metric solution} and \eqref{Eq:f(r) w/ eff charge}, it follows that
\begin{align}
    H(\rho) = 4\sqrt{\frac{\rho}{r_0}}\, f(\sqrt{r_0 \rho}).
\end{align}
Thus, the horizon position $\rho_H$ is related to the four-dimensional outer horizon radius $r_H$ via
$r_H \coloneqq \sqrt{r_0 \rho_H} = \sqrt{\Phi_H}$, i.e., $H(\rho_H) = f(r_H) = 0$.

Now let us examine the thermodynamics of the effective dilaton gravity.
Using the semiclassical approximation, the free energy, energy, and entropy are given by Eq.~\eqref{Eq:dilaton-gravity-entropy-energy}
\begin{align}
    F_{(2)}(\Phi_H,Q) &= \frac{1}{2G_N} \left(\Phi^{\frac{1}{2}}_H + \frac{\Lambda}{3} \Phi^{\frac{3}{2}}_H + \frac{3G_N e^2 Q^2}{4\pi} \Phi_H^{-\frac{1}{2}}\right),
    \label{Eq:2-dim free energy w/o KK modes}
    \\
    E_{(2)}(\Phi_H, Q) &= \frac{1}{G_N} \left(
        \Phi_H^{\frac{1}{2}} - \frac{\Lambda}{3} \Phi_H^{\frac{3}{2}} + \frac{G_N e^2 Q^2}{4\pi} \Phi_H^{- \frac{1}{2}} 
    \right),
    \label{Eq:2-dim energy w/o KK modes}
    \\
    S(\Phi_H) &= \frac{\pi \Phi_H}{G_N}.
    \label{Eq:2-dim entropy w/o KK modes}
\end{align}
The two-dimensional Hawking temperature is also given by
\begin{align}
    T_{(2)}(\Phi_H) 
    &= \frac{V_\eff(\Phi_H)}{4\pi r_0}
    = \frac{1}{2\pi} 
    \left(
        \Phi_H^{-\frac{1}{2}} - \Lambda \Phi_H^{\frac{1}{2}} - \frac{G_N e^2}{4\pi}Q^2 \Phi_H^{-\frac{3}{2}}
    \right).
\end{align}
From the relation between the Euclidean times, $t = 2\tau$ in Eq.~\eqref{Eq: cdnt transformation 4d to 2d}, the above two-dimensional temperature (periodicity) is rewritten in terms of the four-dimensional one in Eq.~\eqref{Eq:BH-temperature-4D},
\begin{align}
    T_{(4)} = \frac{1}{2} T_{(2)}.
    \label{Eq:T(4)&T(2)}
\end{align}
Thus, free energies are also related as 
\begin{align}
    F_{(4)}  = \frac{1}{2} F_{(2)}.
        \label{Eq:physical-free-energy}
\end{align}
Similarly, the four-dimensional black hole mass in Eq.~\eqref{Eq:mass-W} is related to the two-dimensional energy as $M = \frac{1}{2} E_{(2)}$.
Thus, as expected, the two-dimensional free energy and temperature are obtained from the four-dimensional expressions in Eq.~\eqref{Eq:4D temperature and free energy for RN--AdS} by replacing $r_H^2$ with $\Phi_H$.

As in Sec.~\ref{Sec:Thermodynamics of the dilaton gravity with general potential}, the phase structure is discussed below.
At fixed charge $Q$, we compare the free energies of the possible states and determine which states are thermodynamically favored.
We mainly focus on the asymptotically AdS case ($\Lambda<0$), where the phase transition of interest occurs.
We also discuss the asymptotically flat case at the end of this section, but the phase transition does not arise in this case.
The asymptotically dS ($\Lambda>0$) case involves some subtleties due to the presence of a cosmological horizon, and we leave the dS case for future work.\footnote{
    See Refs.~\cite{Easson:2025ekn,Easson:2026kng} for black hole physics in dS case.
}

\begin{figure}[t]
    \begin{minipage}[b]{0.52\linewidth}
        \centering
        \includegraphics[scale=0.5]{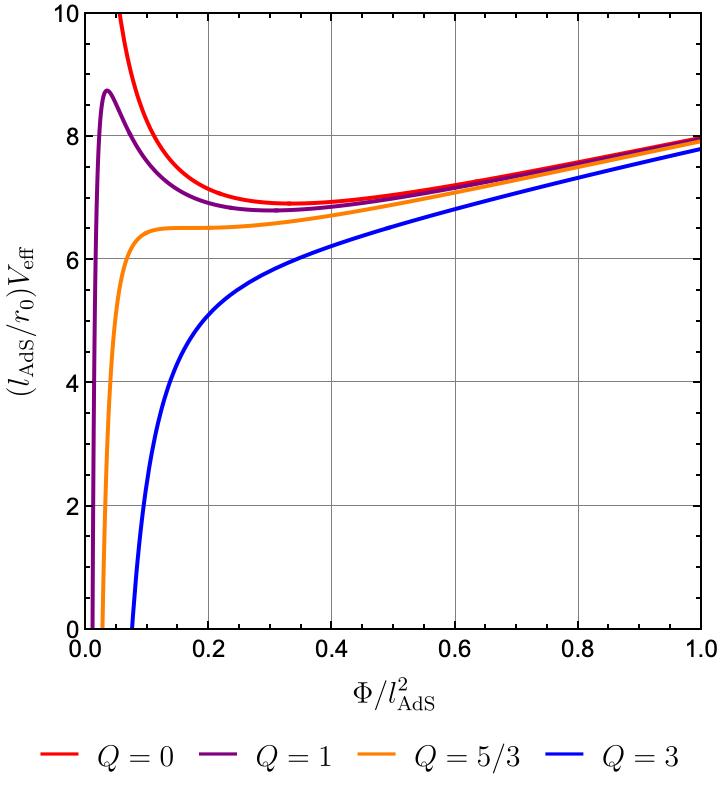}
        \subcaption{$V_{\mathrm{eff}}(\Phi)$.}
        \label{Fig:S2_AdS_potential}
    \end{minipage}
    \begin{minipage}[b]{0.52\linewidth}
        \centering
        \includegraphics[scale=0.5]{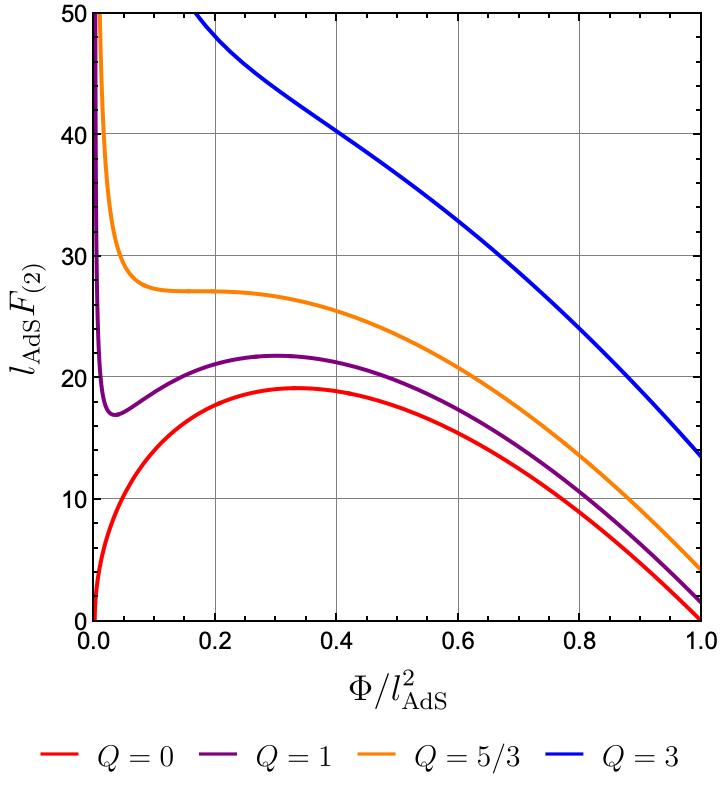}
        \subcaption{$F(\Phi)$.}
        \label{Fig:S2_AdS_free_energy}
    \end{minipage}
        \caption{
            Plots of (a) the dilaton potential $V_{\mathrm{eff}}(\Phi)$ and (b) the free energy $F_{(2)}(\Phi)$ for $l_{\rm AdS}/l_{\mathrm{pl}}=10$ and $e^2=4\pi$.  
            The red curve corresponds to $Q=0$. The orange curve represents the critical charge $Q_c=5/3$. The purple curve corresponds to $Q=1<Q_c$ and has two extrema, for which the Hawking--Page transition can occur. The blue curve corresponds to $Q=3>Q_c$ and has no extrema, implying that the transition disappears.
        }
        \label{fig:S2_AdS_potential_free_energy}
\end{figure}

\subsubsection*{Phase transition of AdS charged black hole via effective dilaton gravity}

For $\Lambda <0$, the dilaton potential and free energy are expressed as follows:
\begin{align}
    V_\eff(\Phi) &= 2 r_0 \left(
        \Phi^{-\frac{1}{2}} + \frac{3}{l_{\rm AdS}^2} \Phi^{\frac{1}{2}} - \frac{G_Ne^2}{4\pi} Q^2 \Phi^{- \frac{3}{2}}
    \right),
    \\
    F_{(2)}(\Phi_H, Q) &= \frac{1}{2 G_N} \left(
        \Phi^{\frac{1}{2}}_H - \frac{1}{l_{\rm AdS}^2} \Phi^{\frac{3}{2}}_H + \frac{3 G_N e^2 }{4\pi} Q^2 \Phi^{-\frac{1}{2}}
    \right),
\end{align}
and the extrema of this potential are given by
\begin{align}
    \Phi = \frac{l_{\rm AdS}^2}{6} \left(
        1 \pm \sqrt{1 - \frac{Q^2}{Q_c^2}}
    \right).
    \label{Eq:AdS-extremumdilaton}
\end{align}
Here, $l_{\rm AdS} := \sqrt{3/|\Lambda|}$ denotes the four-dimensional AdS radius.
We have introduced the critical charge 
\begin{align}
    Q_c \coloneqq \sqrt{\frac{\pi l_{\rm AdS}^2}{9G_Ne^2 }},
    \label{Eq:critical-temperature}
\end{align}
for which the square root in Eq.~\eqref{Eq:AdS-extremumdilaton} vanishes.
The plots of the dilaton potential and free energy for $\Lambda < 0$ are shown in Fig.~\ref{fig:S2_AdS_potential_free_energy}, where we set $l_{\rm AdS}/l_{\mathrm{pl}}=10$ and $e^2 = 4 \pi$ so that $Q_c = 5/3$.
The plots show that the thermodynamic stability of the states depends on the black hole charge.
Since the two-dimensional free energy and temperature coincide exactly with the four-dimensional ones under $\Phi_H = r_H^2$ together with Eqs.~\eqref{Eq:T(4)&T(2)} and \eqref{Eq:physical-free-energy}, Fig.~\ref{fig:S2_AdS_potential_free_energy} is identical to Figs.~\ref{fig:4D_HP_S_AdS} and \ref{fig:4D_HP}.

Although the discussion of thermodynamic stability is essentially the same as that in Sec.~\ref{Sec:HP-transition}, since the thermodynamic quantities coincide with those in four dimensions upon replacing $r_H^2$ with $\Phi_H$, we present it here from the two-dimensional perspective in terms of the dilaton $\Phi$.
First, we consider the $Q=0$ case, corresponding to the red lines in Fig.~\ref{fig:S2_AdS_potential_free_energy}.
In this case, both the dilaton potential and the free energy exhibit an extremum at $\Phi = l_{\rm AdS}^2/3$.
For $T_{(2)}^{\rm bath} <  T_{(2)}(\Phi_H=l_{\rm AdS}^2/3, Q= 0)$, pure AdS (with $\Phi_H = 0$) is thermally preferred because there is no intersection between the constant line $T_{(2)}^{\rm bath}$ and the curve of the dilaton potential. 
Here, $T_{(2)}^{\rm bath}$ denotes the temperature of the thermal bath.
Note that the free energy vanishes at $\Phi = 0$ and $\Phi = l_{\rm AdS}^2$: 
\begin{align}
F_{(2)}\biggl(\Phi_H = l_{\rm AdS}^2\biggr) - F_{(2)}\biggl(\Phi_H = 0\biggr) = 0. 
\end{align}
Thus, let us define 
\begin{align}
    T_{(2)\mr{HP}} \coloneqq \frac{1}{4 \pi r_0}V_\eff\bigg(\Phi_H=l_{\rm AdS}^2, Q= 0\bigg) = \frac{2}{\pi l_{\rm AdS}},
\end{align}
where $l_{\rm AdS}/l_{\rm{pl}}=10$ in Fig.~\ref{fig:S2_AdS_potential_free_energy}.
For $T_{(2)}(\Phi_H=l_{\rm AdS}^2/3, Q= 0)<T_{(2)}^{\rm bath} < T_{(2)\mathrm{HP}}$, pure AdS is thermally stable since it has a lower free energy than that of the AdS black hole (with $ \Phi_H >0$). 
In contrast, AdS black holes are thermally preferred for $T_{(2)}^{\rm bath} > T_{(2)\mathrm{HP}}$.
At the temperature of $T_{(2)}^{\rm bath} = T_{(2)\mr{HP}}$, a phase transition between the pure AdS and the black hole with $\Phi_H = l_{\rm AdS}^2$ occurs.
This is nothing but the Hawking--Page phase transition between the pure AdS and AdS black holes~\cite{Hawking:1982dh} as discussed in Sec.~\ref{Sec:HP-transition}.
Next, we consider the cases $Q \neq 0$. 
As shown in Fig.~\ref{Fig:S2_AdS_free_energy}, black holes with $\Phi_H > 0$ are thermodynamically favored over the $\Phi_H = 0$ configuration because their free energy is lower.
For $Q < Q_c$, the effective potential has two distinct extrema, given by Eq.~\eqref{Eq:AdS-extremumdilaton}, as shown by the purple curve in Fig.~\ref{Fig:S2_AdS_potential}. In this case, a Hawking--Page phase transition between small and large black holes can occur. 
On the other hand, the case $Q = Q_c$ corresponds to the critical point where the phase transition disappears, since the effective potential has a single extremum at $\Phi = l_{\rm AdS}^2/6$, as shown by the orange curve in Fig.~\ref{Fig:S2_AdS_potential}.
For $Q > Q_c$, the effective potential (free energy) becomes a monotonically increasing (decreasing) function of $\Phi$, as shown by the blue curve in Fig.~\ref{fig:S2_AdS_potential_free_energy}, and only a single black hole state with positive specific heat exists.

\begin{figure}[t]
    \centering
    \includegraphics[scale=0.4]{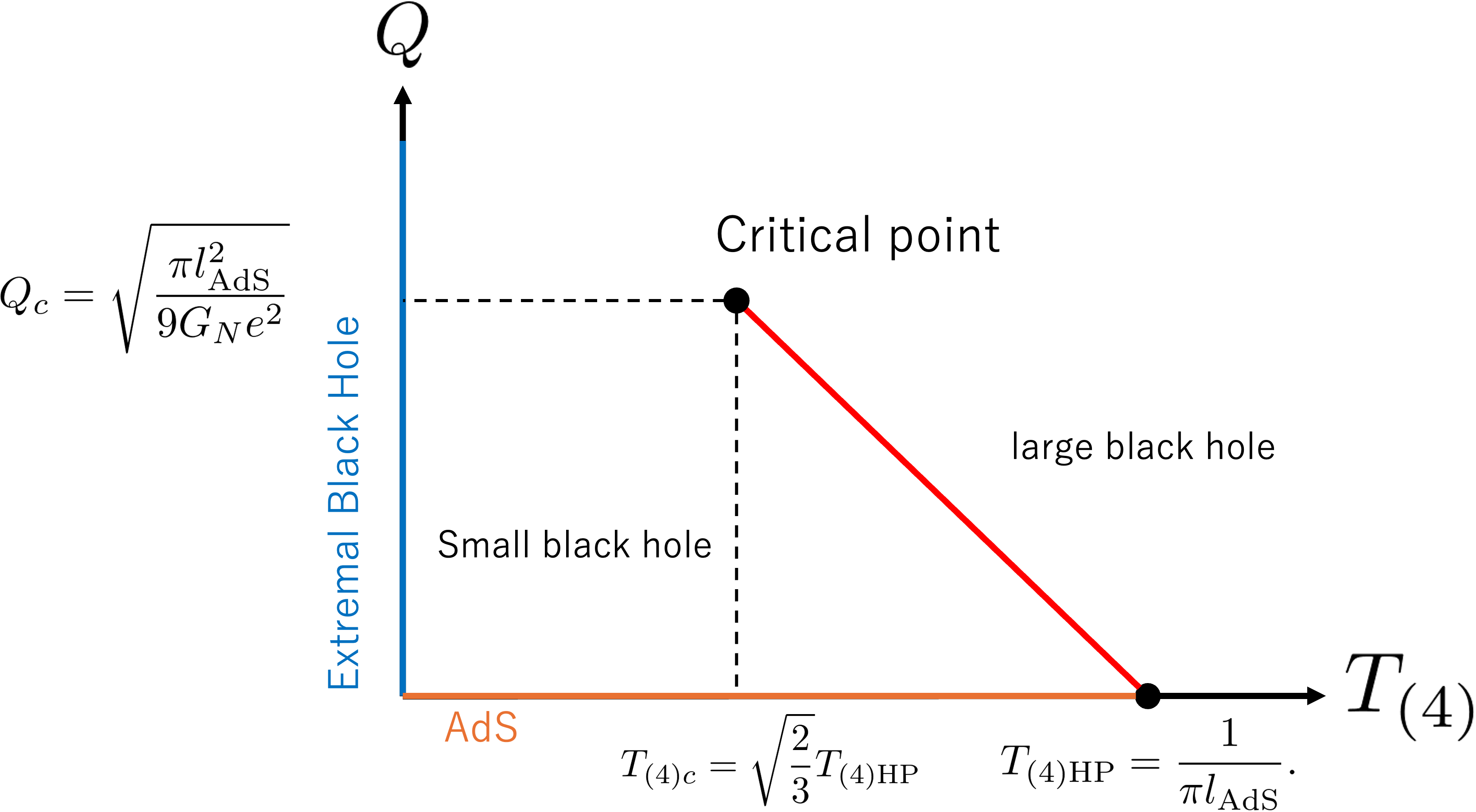}
    \caption{
        A schematic phase diagram in the $(T_{(4)},Q)$ plane for $\Lambda<0$. Here, $T_{(4)\mathrm{HP}}$ denotes the four-dimensional Hawking--Page temperature, $Q_c$ the critical charge, and $T_{(4)c}$ the four-dimensional critical temperature. 
        The red curve represents the phase transition line where the two phases have equal free energy.
        Along this curve, the small-black-hole branch continuously approaches pure AdS in the $Q\to0$ limit, as shown in Fig.~\ref{fig:S2_AdS_potential_free_energy}. Consequently, the phase transition line meets the $Q=0$ axis at $T=T_{(4)\mathrm{HP}}$.
    }
    \label{Fig:AdS_phase_diagram}
\end{figure}

Taking Eqs.~\eqref{Eq:T(4)&T(2)} and \eqref{Eq:physical-free-energy} into account, we estimate the original four-dimensional quantities from a two-dimensional perspective.
Fig.~\ref{Fig:AdS_phase_diagram} shows the four-dimensional phase diagram in the $(T_{(4)},Q)$ plane, where the phase transition occurs when the free energies of the two black-hole branches coincide.  
The phase transition curve has an endpoint at $(T_{(4)},Q)=(T_{(4)c},Q_c)$, where $T_{(4)c}={V_{\mathrm{eff}}\left( {l_{\rm{AdS}}^2}/{6}, Q_c \right)}/{(8\pi r_0)}$.
This result agrees with the four-dimensional results in Eqs.~\eqref{Eq:4D-critical-charge} and~\eqref{Eq:4D-critical-temp}.

\paragraph{Phase structure of asymptotically flat charged black hole.}

\begin{figure}[t]
    \begin{minipage}[b]{0.5\linewidth}
        \centering
        \includegraphics[scale=0.6]{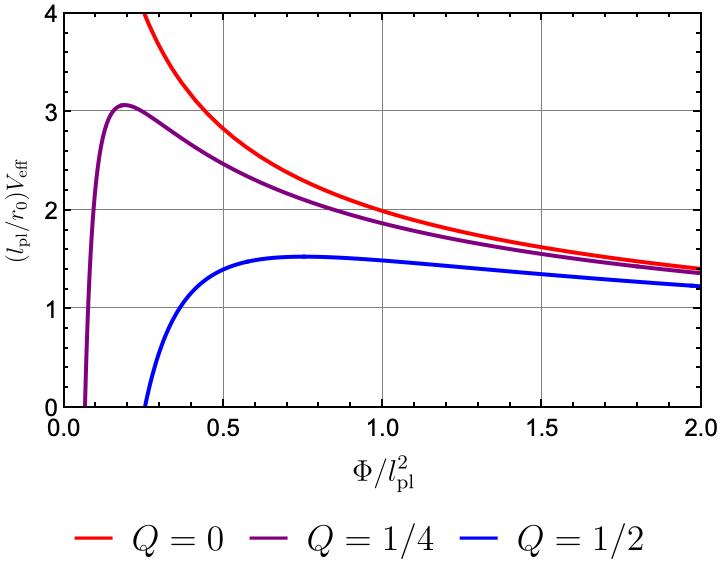}
        \subcaption{$V_{\mathrm{eff}}(\Phi)$ for $\Lambda=0$.}
        \label{Fig:S2_flat_potential}
    \end{minipage}
    \begin{minipage}[b]{0.5\linewidth}
        \centering
        \includegraphics[scale=0.6]{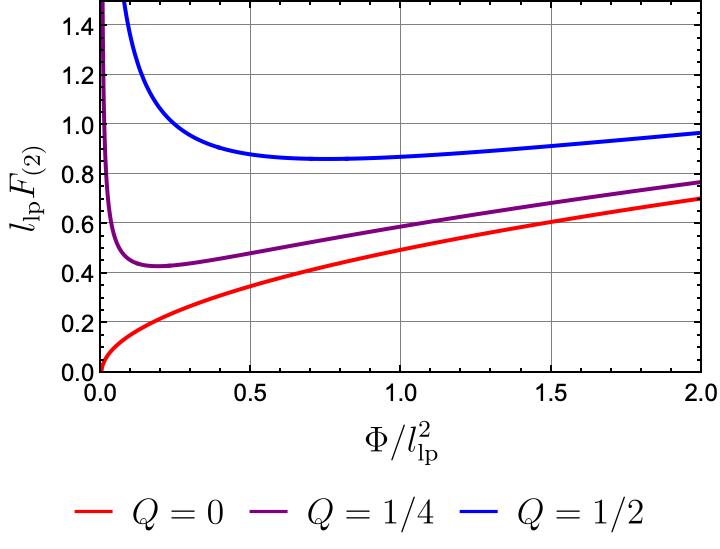}
        \subcaption{$F(\Phi)$ for $\Lambda=0$.}
        \label{Fig:S2_flat_free_energy}
    \end{minipage}
    \caption{
        Plot of (a) the dilaton potential $V_{\text{eff}}(\Phi)$ and (b) free energy $F(\Phi)$ for $\Lambda=0$ and $e^2=4\pi$.
    }
    \label{S2_flat_potential_free_energy}
\end{figure}

We apply a similar analysis to the asymptotically flat case with $\Lambda=0$.
The behavior of the dilaton potential and the free energy is shown in Fig.~\ref{S2_flat_potential_free_energy}.
In this figure, we set $e^2 = 4 \pi$.
For $Q=0$, as shown by the red lines in Fig.~\ref{S2_flat_potential_free_energy}, the geometry with $\Phi=0$
has zero free energy, whereas black hole states have positive free energy.
Thus, the former geometry (no black hole) is thermodynamically preferred.
Conversely, for $Q\neq0$, black-hole states have lower free energy and are always thermodynamically preferred, as shown by the purple and blue lines in Fig.~\ref{S2_flat_potential_free_energy}.

For the charged black hole, the dilaton potential and the free energy have only one extremum point at $\Phi=\Phi_0 := \frac{3e^2 G_N Q^2}{4 \pi}$, as shown in Fig.~\ref{S2_flat_potential_free_energy}.
As in the case shown in Fig.~\ref{Fig:Delta_F_positive}, the small black hole $(\Phi_H<\Phi_0)$ is always more stable than the large black hole $(\Phi_H>\Phi_0)$.
This implies that there is no Hawking--Page-type phase transition for $\Lambda=0$.
Note that no static black-hole configuration can remain in thermal equilibrium with the heat bath for $T_{(2)}^{{\rm bath}} > \frac{V_{\mathrm{eff}}(\Phi_0,Q)}{4\pi r_0}$.
The absence of Hawking--Page-type phase transition in the asymptotically flat case can also be seen from the behavior of the free energy in Fig.~\ref{Fig:S2_flat_free_energy}.
When $Q=0$, the free energy increases monotonically, as shown by the red line, and there are no distinct states with the same free energy.
When $Q\neq0$, there are two states with the same free energy, as shown by the purple and blue lines.
However, the heat capacity in Eq.~\eqref{Eq:dE&dF} is negative in the larger-$\Phi$ region because $dF_{(2)}/d\Phi>0$ there.
This means that the larger black hole is thermodynamically unstable.

\section{Partition function and effective action with KK modes}
\label{Sec:Partition function and effective action}

In this section, we analyze the contributions of the KK modes of the four-dimensional electromagnetic field on $S^2$ to the two-dimensional effective theory and its thermodynamics.
We integrate out the KK modes and derive the regularized one-loop effective action for two-dimensional dilaton gravity, retaining only the leading local terms. 
Higher-derivative and nonlocal terms are neglected.

\subsection{KK reduction of the electromagnetic gauge field}
\label{Sec:Kaluza--Kleinreduction}

We first perform the KK reduction of the four-dimensional gauge field and obtain the two-dimensional action for the KK modes, which are treated as quantum fluctuations around the classical two-dimensional gravity background \cite{Michelson:1999kn,Donnelly:2015hxa,Iliesiu:2020qvm}.

We decompose the four-dimensional gauge field $A^{(1)} = A_\mu(x,y) dx^\mu$ as
\begin{align}
	A_\mu(x,y) dx^\mu = A_a(x,y) dx^a + A_k(x,y)dy^k
\end{align}
on the background $\Sigma \times S^2$. Here, $A_a(x,y)$ and $A_k(x,y) dy^k$ are a zero-form (scalar) and a one-form (vector) on $S^2$, respectively. 
Expanding the scalar field in spherical harmonics $Y^{(l,m)}(y)$, we obtain
\begin{align}
	A_a(x,y) = \sum_{l=0}^\infty \sum_{|m|\leq l} a_{a(l,m)}(x) Y^{(l,m)}(y),
	\label{Eq:4-dim gauge field A_a decomposition}
\end{align}
where $a_{a(l,m)}(x)$ is a two-dimensional vector field on $\Sigma$.
The spherical harmonics satisfy
\begin{align}
\Delta_{S^2} Y^{(l,m)} = l(l+1)Y^{(l,m)},
\qquad 
\int_{S^2} Y^{(l,m)} \ast_{S^2} Y^{(l',m')} = \delta_{l,l'}\delta_{m,m'}.
\label{Eq:conditions-Y^lm}
\end{align}
Throughout, we use the language of differential forms on $S^2$. 
Here, $\ast_{S^2}$ is the Hodge star on the $S^2$ and $\Delta_{S^2} \coloneqq - \ast_{S^2} d \ast_{S^2} d- d \ast_{S^2} d \ast_{S^2}$ is the Laplace operator on $S^2$.
Similarly, a one-form $A_k(x,y)$ on $S^2$ can be expanded in one-form spherical harmonics as
\begin{align}
	A_k(x,y) dy^k &= \sum_{l \geq 1} \sum_{|m|\leq l} \frac{1}{m_l} \Bigl[
		\tilde{\varphi}_{(l,m)}(x) dY^{(l,m)}(y) + \varphi_{(l,m)}(x) \ast_{S^2} dY^{(l,m)}(y)
	\Bigr],
	\label{Eq:4-dim gauge field A_k decomposition}
\end{align}
where we define the KK mass as
\begin{align}
    m_l \coloneqq \frac{\sqrt{l(l+1)}}{r_0},
\qquad l\ge1.
\end{align}
Note that the vector spherical harmonics are eigenfunctions of the Laplacian:
\begin{align*}
    \Delta_{S^2}d Y^{(l,m)} = l(l+1) dY^{(l,m)}, \qquad \Delta_{S^2}\ast_{S^2} dY^{(l,m)} = l(l+1)\ast_{S^2} dY^{(l,m)}.
    \label{Eq:eigen eq of SVH}
\end{align*}
Here, $\varphi_{(l,m)}(x)$ and $\tilde{\varphi}_{(l,m)}(x)$ are scalar fields on $\Sigma$. The fields $\tilde{\varphi}_{(l,m)}$ and $\varphi_{(l,m)}$ correspond to the four-dimensional longitudinal (pure-gauge) and transverse modes, respectively.
The factor $1/m_l$ is introduced to canonically normalize the scalar fields. Using integration by parts,
\begin{align}
	\int_{S^2} dY^{(l,m)} \wedge \ast_{S^2} dY^{(l',m')} = l (l+1)\delta_{l,l'} \delta_{m,m'}.
\end{align}
Thus, the four-dimensional gauge field decomposes into two-dimensional vector fields $a_{(l,m)}^{(1)} =a_{(l,m)a} dx^a$ and two towers of scalar fields, $\varphi_{(l,m)}$ and $\tilde{\varphi}_{(l,m)}$.
We call $a^{(1)}_{(0,0)}\eqqcolon a^{(1)}$ the (massless) photon, whereas $a^{(1)}_{(l,m)}$ with $l \geq 1$ are called massive KK photons.
Note that there are no massless scalar modes.
Substituting Eqs.~\eqref{Eq:4-dim gauge field A_a decomposition} and \eqref{Eq:4-dim gauge field A_k decomposition} into the Maxwell action \eqref{Eq:Maxwell-action}, we obtain the two-dimensional action for the KK modes, including the St\"uckelberg couplings between $a^{(1)}_{(l,m)}$ and $\tilde{\varphi}_{(l,m)}$:
\begin{align}
	&I_\Maxwell[a^{(1)},a^{(1)}_{(l,m)},\varphi_{(l,m)}, \tilde{\varphi}_{(l,m)}]
	\nn \\
	&= \frac{r_0^2}{4 e^2} \int_\Sigma d^2x \sqrt{\det g_{ab}}~ f_{ab} f^{ab}
	+ \sum_{(l,m)} \frac{r_0^2}{2e^2} \int_{\Sigma} d^2x \sqrt{\det g} \biggl[
        (\del_a \varphi_{(l,m)})^2 
        + m_l^2 \varphi_{(l,m)}^2
	\biggr]
	\nn \\
	& \quad 
	+ \sum_{(l,m)} \frac{r_0^2}{2e^2} \int_\Sigma d^2x \sqrt{\det g_{ab}} \biggl[
		\frac{1}{2}
        f_{(l,m)ab} f_{(l,m)}^{ab}
        + \Bigl(
			m_l a_{(l,m)a} - \del_a \tilde{\varphi}_{(l,m)}
		\Bigr)^2
	\biggr]
	\nn \\
	& \quad 
	- \frac{r_0^2}{e^2} \int_{\del \Sigma} d\tau \sqrt{\det h_{ab}} n_{(2)a} a_b f^{ab}
	- \frac{r_0^2}{e^2} \sum_{(l,m)} \int_{\del \Sigma} d \tau \sqrt{\det h_{ab}} n_{(2)a} 
	\nn \\
	& \qquad
	\times \biggl[
		a_{(l,m)b} 
        f_{(l,m)}^{ab}
		+ \varphi_{(l,m)}\del^a \varphi_{(l,m)}
		- \tilde{\varphi}_{(l,m)} \Bigl(
			m_l a^a_{(l,m)} - \del^a \tilde{\varphi}_{(l,m)}
		\Bigr)
	\biggr],
	\label{Eq:reduced Maxwell action}
\end{align}
where $f^{(2)} := d a^{(1)}$, $f^{(2)}_{(l,m)} := da^{(1)}_{(l,m)}$, and $\sum_{(l,m)} \coloneqq \sum_{l\geq1} \sum_{|m|\leq l}$.
We have also used Eqs.~\eqref{Eq:eigen eq of SVH}, \eqref{Eq:conditions-Y^lm}, and 
\begin{align}
	\int_{S^2} dY^{(l,m)} \wedge d Y^{(l,m)} = 0,
\end{align}
for $l \geq 1$ and
\begin{align}
	d A^{(1)} &= \frac{1}{\sqrt{4\pi}} d a^{(1)} + \sum_{(l,m)} Y^{(l,m)} d a^{(1)}_{(l,m)} - \sum_{(l,m)} \Bigl(
		m_l a^{(1)}_{(l,m)} - d \tilde{\varphi}_{(l,m)}
	\Bigr) \wedge \frac{1}{m_l} d Y^{(l,m)}
	\nn \\
	& \quad 
	+ \sum_{(l,m)} \biggl(
		d \varphi_{(l,m)} \wedge \frac{1}{m_l} \ast_{S^2} d Y^{(l,m)} - r_0^2 m_l \varphi_{(l,m)} \ast_{S^2} Y^{(l,m)}
	\biggr).
\end{align}
The mixing terms on the boundary, such as $m_l a^a_{(l,m)}\partial_a \tilde{\varphi}_{(l,m)} $, can be removed by a gauge-fixing procedure, as seen below.

Owing to the St\"uckelberg couplings, $\tilde{\varphi}_{(l,m)}$ are eaten by $a^{(1)}_{(l,m)}$, providing the corresponding vector fields with masses.
This can be seen from the gauge symmetry.
Under the four-dimensional $\U(1)$ gauge transformation $A^{(1)} \mapsto A^{(1)} + d \Theta$, each mode transforms as 
\begin{align}
	a^{(1)} \mapsto a^{(1)} + d \Theta_{(0,0)},
	\quad 
	a^{(1)}_{(l,m)} \mapsto a^{(1)}_{(l,m)} + d \Theta_{(l,m)},
	\quad 
	\tilde{\varphi}_{(l,m)} \mapsto \tilde{\varphi}_{(l,m)} + m_l \Theta_{(l,m)},
\end{align}
for $l \geq 1$, where the gauge transformation parameter $\Theta$ is expanded as 
\begin{align}
	\Theta(x,y) &=  \sum_{l=0}^\infty \sum_{|m|\leq l} \Theta_{(l,m)}(x) Y^{(l,m)}(y),
	\\
	d \Theta(x,y) &= 
    Y^{(0,0)} (\partial_a \Theta_{(0,0)}(x)) dx^a
    \nn\\
    & \qquad +
    \sum_{(l,m)} \Bigl[
		Y^{(l,m)}(y) (\del_a \Theta_{(l,m)}(x) \bigr) dx^a + \Theta_{(l,m)}(x) \del_k Y^{(l,m)}(y) dy^k
	\Bigr],
\end{align}
where we use $\partial_k Y^{(0,0)}=0$.
The usual two-dimensional $\mathrm{U}(1)$ gauge symmetry associated with $a^{(1)}$ remains intact, and the bulk action is gauge invariant.
Using $\Theta_{(l,m)}$, the scalars $\tilde{\varphi}_{(l,m)}$ can be gauged away and absorbed by the massive KK photons $a^{(1)}_{(l,m)}$, as in the unitary gauge of the four-dimensional Abelian Higgs model.
Here, the two-dimensional boundary conditions for the parameter read
\begin{align}
	n^a_{(2)} \partial_a \Theta_{(0,0)}|_{\partial \Sigma}= n^a_{(2)} \partial_a\Theta_{(l,m)}|_{\partial \Sigma} = 0 \quad (l \geq 1),
	\label{Eq:gauge-transformation-reduction}
\end{align}
since $n^\mu_{(4)} \partial_\mu \Theta=0$ on the boundary in four dimensions, as shown in Eq.~\eqref{Eq:Neumann boundary condition for the four-dim gauge transformation}. 
The boundary condition on the gauge field in four dimensions is imposed as in Eqs.~\eqref{Eq:D-dim Neumann (absolute) boundary condition I} and \eqref{Eq:D-dim Neumann (absolute) boundary condition II}, where Dirichlet conditions are imposed on the normal components at the boundary, and Neumann conditions are imposed on the tangential components.
Hence, we impose
\begin{align}
    n^a_{(2)} \delta a_{a} \big|_{\del \Sigma} = n^a_{(2)} \delta f_{ab} \big|_{\del \Sigma} &=0,
    \label{eq:zero-bc}
    \\
	n^a_{(2)} \del_a \tilde{\varphi}_{(l,m)}\big|_{\del \Sigma} = n^a_{(2)} \del_a \varphi_{(l,m)}\big|_{\del \Sigma} = n^a_{(2)} a_{(l,m)a} \big|_{\del \Sigma} = n^a_{(2)} f_{(l,m)ab} \big|_{\del \Sigma} &= 0
	\quad (l > 0),
    \label{eq:massive-bc}
\end{align}
for two-dimensional fields, where $\delta a_a$ is the quantum fluctuation around the classical background. 
We introduce $\delta f_{ab}:=\partial_a \delta a_b - \partial_b \delta a_a$.
The massless and massive KK photons satisfy absolute boundary conditions, while the massive scalars satisfy Neumann boundary conditions.
These conditions are invariant under $\U(1)$ gauge transformations satisfying Eq.~\eqref{Eq:gauge-transformation-reduction}.
By the same argument as in Eq.~\eqref{Eq:delta_theta_boundary_term} of the four-dimensional theory, the boundary terms in Eq.~\eqref{Eq:reduced Maxwell action} are gauge invariant for field configurations satisfying the above boundary conditions.
For explicit calculations, we choose $n^a_{(2)} \propto \delta^a_\rho$.

\paragraph{Gauge fixing.}

To evaluate the one-loop partition function of the KK modes, we impose a gauge-fixing and derive the corresponding gauge-fixed action.
The KK zero mode $a^{(1)}$ is decomposed into the classical background \eqref{Eq:flux-2d} and a quantum fluctuation $\delta a^{(1)}$ as
\begin{align}
da^{(1)}
=
\frac{\sqrt{\det g_{ab}}}{2!}
\frac{i e^2 Q}{\sqrt{4 \pi}r_0^2}
\varepsilon_{ab}\,dx^a\wedge dx^b
+
d\delta a^{(1)},
\end{align}
and perform the path integral over $\delta a^{(1)}$.
We impose the gauge-fixing conditions
\begin{align}
\nabla_a \delta a^a =0,
\qquad
\nabla_a a^a_{(l,m)} - m_l \tilde{\varphi}_{(l,m)} =0.
\label{Eq:2-dim. gauge-fixing condition}
\end{align}
Note that this condition is obtained from the four-dimensional Lorenz gauge using the metric
\begin{align}
    ds^2 
    = \hat{g}_{\mu\nu}(x) dx^\mu dx^\nu
    = g_{ab}(x) dx^a dx^b + r_0^2 d \Omega^2_{(2)}.
    \label{Eq:hatgmunu}
\end{align}
With this metric, the Lorenz gauge condition reads
\begin{align}
    \hat{g}^{\mu\nu}\hat{\nabla}_\mu A_\nu =
     \frac{1}{\sqrt{4 \pi}} \nabla_a a^a + \sum_{(l,m)} \Bigl[
        \nabla_a a^a_{(l,m)} - m_l \tilde{\varphi}_{(l,m)}
    \Bigr] Y^{(l,m)} =0.
\end{align}
Note that $a_a$ and $a_{(l,m)a}$ can be gauge-fixed independently since $\Theta_{0,0}$ and $\Theta_{(l,m)}$ are independent gauge parameters.
The metric \eqref{Eq:hatgmunu} is obtained from the metric ansatz \eqref{Eq:4-dim metric anzatz} by a Weyl rescaling with a factor $r_0^2/\Phi$. 
Owing to the classical Weyl invariance of the four-dimensional Maxwell action \eqref{Eq:Maxwell-action}, the Lorenz gauge may equivalently be imposed using either metric.
By the standard gauge-fixing procedure, the Faddeev--Popov (FP) ghosts $(b_0,c_0)$ and $(b_{(l,m)}, c_{(l,m)})$ are introduced:
\begin{align}
	I_{\mr{FP} + \mr{GF}} &= \frac{r_0^2}{e^2}\int_\Sigma d^2x \sqrt{\det g_{ab}} \biggl[
		- b_0 \nabla^2 c_0 + \frac{1}{2\xi_0} (\nabla_a \delta a^a)^2
	\biggr]
	\nn \\
	& \quad 
	+ \sum_{(l,m)} \frac{r_0^2}{e^2} \int_\Sigma d^2x \sqrt{\det g_{ab}}\biggl[
		b_{(l,m)} \Bigl(
			- \nabla^2 + m_l^2
		\Bigr) c_{(l,m)} + \frac{1}{2\xi_{(l,m)}} \biggl(
			\nabla_a a^a_{(l,m)} - m_l \tilde{\varphi}_{(l,m)}
		\biggr)^2
	\biggr],
    \label{Eq:FP+GF}
\end{align}
where $\xi_0$ and $\xi_{(l,m)}$ are gauge fixing parameters.
Hereafter, we set $\xi_0=\xi_{(l,m)}=1$ for simplicity in the calculations.
We thus obtain
\begin{equation}
    I_{\text{Maxwell}} +I_{\mr{FP} + \mr{GF}}
    =I_{\text{zero mode}} + 
    \sum_{(l,m)}
    \left[
        I_{(l,m)}^{\text{scalar}} + \tilde{I}_{(l,m)}^{\text{scalar}} + I_{(l,m)}^{\text{vector}}+ I_{(l,m)}^{\text{ghost}}
    \right],
\end{equation}
\begin{align}
    I_{\text{zero mode}} &= 
    \frac{1}{2} \int_{\Sigma} d^2x \sqrt{\det g_{ab}} \frac{e^2 Q^2}{4 \pi r_0^2}
    +\frac{r_0^2}{4e^2} \int_{\Sigma}  d^2x \sqrt{\det g_{ab}}~
    \left[
        \delta f_{ab} \delta f^{ab}
        +4 b_0 \Delta_0 c_0
        + 2 (\nabla_a \delta a^a)^2
    \right]
    \notag\\
    &\qquad-
    \frac{r_0^2}{e^2} \int_{\partial \Sigma} d\tau \sqrt{\det h_{ab}}~  
    n_{(2)a} \delta a_b \delta f^{ab},
    \label{Eq:I_1}
    \\
    I_{(l,m)}^{\text{scalar}}
    &= 
    \frac{r_0^2}{2e^2}
    \int_\Sigma d^2x \sqrt{\det g_{ab}}~
    \left[ 
        (\nabla \varphi_{(l,m)})^2
        +
        m_l^2 \varphi_{(l,m)}^2
    \right]
    -
    \frac{r_0^2}{e^2}
    \int_{\partial\Sigma} d\tau \sqrt{\det h_{ab}}~
    n_{(2)}^a \varphi_{(l,m)} \partial_a \varphi_{(l,m)},
    \label{Eq:I_2}
    \\
    \tilde{I}_{(l,m)}^{\text{scalar}} &=\frac{r_0^2}{2e^2}
    \int_\Sigma d^2x \sqrt{\det g_{ab}}~
    \left[ 
        (\nabla \tilde{\varphi}_{(l,m)})^2
        +
        m_l^2 \tilde{\varphi}_{(l,m)}^2
    \right] -\frac{r_0^2}{e^2}
    \int_{\partial \Sigma} d\tau \sqrt{\det h_{ab}}~
    n_{(2)}^a \tilde{\varphi}_{(l,m)} \partial_a \tilde{\varphi}_{(l,m)},
    \label{Eq:I_3}
    \\
    I_{(l,m)}^{\text{vector}} &=\frac{r_0^2}{4e^2}
    \int_\Sigma d^2 x \sqrt{\det g_{ab}} ~
    \left[ 
        f_{(l,m)ab}f^{ab}_{(l,m)}
        + 2 (\nabla_a a_{(l,m)}^a)^2
        + 2m_l^2 a_{(l,m)a} a_{(l,m)}^a
    \right]
    \notag\\
    &\quad
    -\frac{r_0^2}{e^2}
    \int_{\partial \Sigma}  d\tau \sqrt{\det h_{ab}}~
    n_{(2)a} a_{(l,m)b}f^{ab}_{(l,m)},
    \label{Eq:I_4}
    \\
    I_{(l,m)}^{\text{ghost}}&= \frac{r_0^2}{e^2} \int_{\Sigma} d^2 x \sqrt{\det g_{ab}} ~ 
    b_{(l,m)}
    \left(
        \Delta_0
        + m_l^2
    \right)c_{(l,m)},
    \label{eq:I-ghost}
\end{align}
as the gauge-fixed action.
Here, $\Delta_0:=-\nabla^2$ is the Laplacian for the scalar on $\Sigma$.
Note that the boundary mixing terms in Eq.~\eqref{Eq:reduced Maxwell action} are canceled by the boundary terms generated by integrating the last term in Eq.~\eqref{Eq:FP+GF} by parts.
The FP ghosts satisfy the Neumann boundary conditions from Eq.~\eqref{Eq:gauge-transformation-reduction}
\begin{align}
	n^a_{(2)}\partial_a b_{0}\big|_{\del \Sigma} = n^a_{(2)}\partial_ac_{0}\big|_{\del \Sigma}=n^a_{(2)}\partial_a b_{(l,m)}\big|_{\del \Sigma} = n^a_{(2)}\partial_ac_{(l,m)}\big|_{\del \Sigma} = 0.
	\label{Eq:FP-bc}
\end{align}
For simplicity, we neglect possible contributions from boundary degrees of freedom, including edge modes, throughout this paper.\footnote{
In Maxwell theory, edge-mode contributions to the (entanglement) entropy have been discussed in Refs.~\cite{Donnelly:2014fua,Donnelly:2015hxa,Mukherjee:2023ihb,Ball:2024hqe}.}
Using the boundary conditions \eqref{eq:zero-bc} and \eqref{eq:massive-bc}, we can rewrite the actions \eqref{Eq:I_1}--\eqref{Eq:I_4} as 
\begin{align}
    I_{\text{zero mode}}
    &=\frac{1}{2} \int_{\Sigma} d^2x \sqrt{\det g_{ab}} \frac{e^2 Q^2}{4 \pi r_0^2}
    + \frac{r_0^2}{2e^2} \int_{\Sigma} d^2x \sqrt{\det g_{ab}} 
    \left[ 2b_0 \Delta_0 c_0 +\delta a ^a (\Delta_{1})_{ab} \delta a^b\right],
    \\
    I_{(l,m)}^{\text{scalar}}
    &= 
    \frac{r_0^2}{2e^2}
    \int_\Sigma d^2x \sqrt{\det g_{ab}}~
    \varphi_{(l,m)}
    \left(
        \Delta_0 +m_l^2
    \right)
    \varphi_{(l,m)},
    \label{Eq:KK scalar action}
    \\
    \tilde{I}_{(l,m)}^{\text{scalar}} &=\frac{r_0^2}{2e^2}
    \int_\Sigma d^2x \sqrt{\det g_{ab}}~
    \tilde{\varphi}_{(l,m)}
    \left(
        \Delta_0 +m_l^2
    \right)
    \tilde{\varphi}_{(l,m)},
    \label{Eq:tilde KK scalar action}
    \\
    I_{(l,m)}^{\text{vector}} &=\frac{r_0^2}{2e^2}
    \int_\Sigma d^2 x \sqrt{\det g_{ab}} ~
    a^a_{(l,m)}
    \left(
        (\Delta_1)_{ab} + m_l^2 g_{ab}
    \right)
    a^b_{(l,m)},
    \label{Eq:KK vector action}
\end{align}
where $\Delta_1:=dd^\dagger + d^\dagger d$ is the Laplace operator for one-forms on $\Sigma$ and its component form is
\begin{align}
    (\Delta_{1})_{ab} = -(\nabla^2 g_{ab}+ [\nabla_a, \nabla_b]).
\end{align}
This $2\times2$ matrix differential operator acts on the vector field $\delta a^a$ as
$(\Delta_1 \delta a)_a = (\Delta_1)_{ab} \delta a^b = -\nabla^2 \delta a_a + R_{ab} \delta a^b$.
Note that these actions are independent of the dilaton as a consequence of the classical Weyl invariance of the four-dimensional Maxwell action.

In deriving the two-dimensional actions above, we have used the metric in Eq.~\eqref{Eq:hatgmunu}.
The Weyl transformation from Eq.~\eqref{Eq:4-dim metric anzatz} to Eq.~\eqref{Eq:hatgmunu} involves a dilaton-dependent Weyl factor. 
Although the classical Maxwell action is invariant under this transformation, the functional measure in the path integral is not generally invariant, and an anomaly-induced Wess–Zumino contribution depending on the dilaton can be induced~\cite{Kummer:1997jr,Grumiller:2002nm,Fabbri:2003vy,Easson:2026vte}.
This contribution is not computed and is omitted in what follows.

\subsection{Integrating out massive KK modes}
\label{Sec:Integrating out Kaluza--Kleinphotons}

The gauge-fixed actions for the KK modes were derived in the previous subsection.
We now integrate out the massive KK modes to derive the one-loop effective action for two-dimensional dilaton gravity.
In this analysis, we use the heat-kernel method to extract the local part of the one-loop effective action. 
The contributions from the massless photon and ghost modes are discussed in Sec.~\ref{Sec:Zero modes}. 
The local massless-sector contribution retained in our calculation modifies the two-dimensional black hole entropy but not the effective dilaton potential.

The path integrals over the massive KK modes reduce to functional determinants since the actions are quadratic
\begin{align}
    Z_{\text{massive}}
    &= \prod_{l>0}\prod_{|m|\le l}
    \int [da_{(l,m)}] e^{-I_{(l,m)}^{\text{vector}}}
    \int [d\varphi_{(l,m)}] e^{-I_{(l,m)}^{\text{scalar}}}
    \int [d\tilde{\varphi}_{(l,m)}] e^{-\tilde{I}_{(l,m)}^{\text{scalar}}}
    \int [db_{(l,m)} dc_{(l,m)}] e^{-I_{(l,m)}^{\text{ghost}}}
    \notag
    \\
    &=\prod_{l>0}\prod_{|m|\le l} \frac{1}{\sqrt{{\det}_A[\Delta_1 + m_l^2]}} \times \frac{1}{\sqrt{{\det}_N[\Delta_0 + m_l^2]}} \times \frac{1}{\sqrt{{\det}_N[\Delta_0 + m_l^2]}} \times {\det}_N[\Delta_0 + m_l^2]
    \nn \\
    &=\prod_{l>0}\prod_{|m|\le l}\frac{1}{\sqrt{{\det}_A[\Delta_1 + m_l^2]}},
\end{align}
where we use the boundary conditions \eqref{eq:massive-bc} and \eqref{Eq:FP-bc}.
Here, $\det_N$ and $\det_A$ denote functional determinants evaluated using eigenmodes satisfying Neumann and absolute boundary conditions, respectively.
Note that $\Delta_1$ is a $2\times2$ differential operator. Owing to the mass term, the functional determinant for the massive modes is well defined even in the presence of Laplacian zero modes.
In the last line, the contributions arising from the massive ghosts and the massive KK scalars cancel each other, and only the massive KK photons contribute to the one-loop effective action: 
\begin{align}
    W_{\text{massive}} &:=-\log Z_{\text{massive}}
    = \sum_{(l,m)} W_{(l,m)}, 
    \label{Eq:Wmassive}
\end{align}  
where
\begin{align}
    W_{(l,m)} &:= \frac{1}{2}\log {\det}_A[\Delta_1 + m_l^2].
    \label{Eq:W_lm}
\end{align}
The functional determinant is rewritten using the Schwinger time $s$ with length squared dimension as 
\begin{align}
    W_{(l,m)} = -\frac{1}{2} \int_0^\infty \frac{d s}{s} e^{-s m^2_l} \Tr_A[ e^{- s \Delta_1}],
    \label{eq:Wlm}
\end{align}
where $\Tr_A$ denotes the trace evaluated over eigenmodes satisfying the absolute boundary condition.
We call $\Tr_A [e^{- s \Delta_1}]$ the heat kernel in this paper.
Since evaluating this quantity requires the spectrum of $\Delta_1$, an exact calculation is generally difficult except on special constant-curvature backgrounds\footnote{
See ~\cite{Giombi:2008vd,David:2009xg,Gopakumar:2011qs} for the thermal AdS case. In thermal AdS, the partition function is also calculated by the so-called quasinormal mode method~\cite{Denef:2009kn,Keeler:2014hba}. This is known to be equivalent to the heat-kernel method~\cite{Keeler:2018lza}.
}.
Instead, we consider the asymptotic expansion of the heat kernel in the $s \to 0$ limit~\cite{gilkey2018invariance}
\begin{align}
    \Tr_A [e^{-s \Delta_1}] \approx \sum_{n=0}^\infty c^A_n(\Delta_1) s^{\frac{n}{2} - 1},
    \label{eq:asymptotic-expansion}
\end{align}
so that we evaluate the local quantum corrections to black hole thermodynamics.
The expansion coefficients $c^A_n(\Delta_1)$ are determined solely by local data, such as the curvature. 
Details of the evaluation are summarized in App.~\ref{sec:evaluation-PF}.
Explicit examples are given in Eqs.~\eqref{Eq:c_0w/Dirichlet}--\eqref{Eq:c3A (Delta1)}.
In the present analysis, we use the small-$s$ heat-kernel expansion to evaluate only the leading local part of the one-loop effective action.
Global properties of the spacetime, including the periodicity of the Euclidean time direction and the associated Matsubara contributions, as well as nonlocal terms, are therefore not captured within this approximation. 
Accordingly, the conclusions below concerning the one-loop thermodynamics apply only to the leading local contribution within the derivative expansion.

Using the asymptotic expansion, Eq.~\eqref{eq:Wlm} becomes
\begin{align}
    W_{(l,m)}
    \approx
    -\frac{1}{2}
    \sum_{n=0}^{\infty}
    c_n^A(\Delta_1)
    \int_0^\infty ds\, s^{\frac{n}{2}-2} e^{-m_l^2 s}.
\end{align}
This integral has an ultraviolet (UV) divergence as $s\to 0$ and must therefore be regularized. Using the resulting expression, Eq.~\eqref{Eq:renormalized effective action for the massive case}, we obtain the following derivative (small-$s$) expansion:
\begin{align}
    W^{\text{ren}}_{(l,m)} 
    &\approx - \frac{1}{2} \log \left( \frac{m_l^2}{\mu^2} \right) \left[ m_l^2 c^A_0(\Delta_1) -c^A_2 (\Delta_1) \right] +\frac{1}{2}m_l^2c^A_0 (\Delta_1) + \sqrt{\pi} m_l c^A_1 (\Delta_1) - \frac{1}{2}\sum_{k=3}^\infty \frac{c^A_{k}(\Delta_1)}{m_l^{k-2}} \Gamma\left(\frac{k}{2}-1\right)
    \label{Eq: renomalized effective action of massive KK}
    \\
    &=-\frac{m_l^2}{4\pi}\left[\log \left( \frac{m_l^2}{\mu^2} \right)-1\right]  \int_{\Sigma} d^2x \sqrt{\det g_{ab}}  - \log \left( \frac{m_l^2}{\mu^2} \right)\frac{1}{3}\chi(\Sigma)  +\cdots ,
    \label{Eq: renomalized effective action of massive KK 2}
\end{align}
where $\mu$ is the renormalization scale, and $c_{0,1,2}^A(\Delta_1)$ are given in Eqs.~\eqref{Eq:c0A (Delta1)}--\eqref{Eq:c2A (Delta1)}. 
In this paper, we leave the $\mu$ dependence explicit.
These coefficients contain terms with up to two derivatives in the bulk and one derivative on the boundary.
The omitted terms correspond to higher-curvature contributions, since the heat-kernel coefficients $c_n^A(\Delta_1)$ with $n>2$ are constructed from local curvature invariants in the bulk and on the boundary \cite{gilkey2018invariance}.
Although higher-curvature terms are present in the full one-loop effective action, they are beyond the scope of the present work.\footnote{
The small-$s$ heat-kernel expansion is an asymptotic expansion.
Its truncation as a local derivative expansion is controlled when
$|R_{(2)}|/m_l^2\ll 1$,
$\sqrt{\nabla_a R_{(2)}\nabla^a R_{(2)}}/m_l^3\ll 1$, and
$|K_{(1)}|/m_l\ll 1$,
together with analogous conditions for higher derivatives and boundary
invariants. For the lightest massive KK level, $l=1$, the classical solution
gives
$R_{(2)}/m_1^2=-1+6G_NM/r-6G_Ne^2Q^2/(4\pi r^2)\to -1$
as $r\to\infty$.
Thus, for this level, the truncated local derivative expansion is not
parametrically controlled in the asymptotic region, and we only regard it
as controlled in regions where the above local hierarchy conditions are
satisfied. This limitation concerns the truncation of the local derivative
expansion; the small-$s$ heat-kernel expansion itself remains an asymptotic
expansion as $s\to0$.
}

In the following, we focus on the volume and Euler characteristic terms in Eq.~\eqref{Eq: renomalized effective action of massive KK 2}.
The effective action $W_{\text{massive}}$ involves an infinite sum over all KK modes, as shown in Eq.~\eqref{Eq:Wmassive}, and is therefore divergent. We employ zeta-function regularization to render it finite.
Using Eqs.~\eqref{Eq:KK-sum-1}--\eqref{Eq:KK-sum-3}, the summations over the KK labels are regularized as 
\begin{align}
    \sum_{(l,m)}\log \frac{m_l^2}{\mu^2}
    &= -4\frac{d\zeta}{dz}(-1) + \frac{4}{3} \log (\mu r_0),
    \label{Eq:log m}
    \\
    \sum_{(l,m)} m_l^2 \log \frac{m_l^2}{\mu^2}
    & = \frac{1}{r_0^2}
    \left(  
        -4\frac{d\zeta}{dz}(-3) - 2\frac{d\zeta}{dz}(-1)
        +\frac{2}{15} \log (\mu r_0)
    \right),
    \label{Eq:mlog m}
    \\
    \sum_{(l,m)} m_l^2 
    & = -\frac{1}{15r_0^2},
    \label{Eq:m}
\end{align}
where $\zeta(z)$ is the zeta function defined by 
$\zeta(z) \coloneqq \sum_{n=1}^\infty n^{-z}$, 
and the values of $\frac{d\zeta}{dz}$ are
\begin{align}
    \frac{d\zeta}{dz} (-1) &= \frac{1}{12} - \log \ms{A} = -0.16542 \cdots,
    \qquad
    \frac{d\zeta}{dz}(-3) = 0.00537 \cdots.
\end{align}
The constant $\ms{A}=1.2824\cdots $ is the Glaisher--Kinkelin constant.
Note that apart from the renormalization scale $\mu$, the only dimensionful scale appearing in the above results is $r_0$.
Finally, integrating out the massive KK modes yields the following regularized effective action:
\begin{align}
    W_{\text{massive}} &\approx -S_{\text{massive}} + \frac{1}{2} \int_{\Sigma} d^2x \sqrt{\det g_{ab}} ~\frac{\mathcal{E}}{4\pi r_0^2} ,
    \label{Eq:W massive}
\end{align}
where $S_{\text{massive}}$ and $\mathcal{E}$ are defined by
\begin{align}
    S_{\text{massive}} &\coloneqq 
    \left[
        -\frac{4}{3}\frac{d\zeta}{dz}(-1) + \frac{4}{9} \log (\mu r_0)
    \right]\chi(\Sigma),
    \label{Eq:S0-in-Wmassive}
    \\
    \mathcal{E}&\coloneqq   
        8\frac{d\zeta}{dz}(-3) +4\frac{d\zeta}{dz}(-1)
        -\frac{2}{15}
        -\frac{4}{15} \log (\mu r_0).
    \label{Eq:mathcalE-in-Wmassive}
\end{align}
The term $S_{\text{massive}}$, being proportional to the Euler characteristic $\chi(\Sigma)$, induces only a constant shift in the entropy. 
On the other hand, $\mathcal E$ arises from the one-loop renormalization of the cosmological constant and enters the dilaton potential in the same way as the classical electric-field energy density shown in Eq.~\eqref{Eq:2-dim Maxwell action on-shell}. 
It can therefore be interpreted as a renormalization of the effective charge parameter.

\subsection{Corrections from massless modes}
\label{Sec:Zero modes}

In this subsection, we evaluate the local contribution from the nonzero
modes of the Laplacian in the massless sector. Within the approximation
described below, this contribution is topological and does not modify the
effective dilaton potential.
To this end, we consider the path integral over the massless modes: the $\U(1)$ gauge-field fluctuation $\delta a_a$ and the massless ghosts $b_0$ and $c_0$.
Since the action is quadratic, the Gaussian integrals yield
\begin{align}
    Z_{\text{zero mode}}
    &= \int[d \delta a d b_0 dc_0]e^{-I_{\text{zero mode}}}
    \notag\\
    &= \exp 
    \left[
        -\frac{1}{2} \int_{\Sigma} d^2x \sqrt{\det g_{ab}}~ \frac{e^2 Q^2}{4 \pi r_0^2}
    \right]
    \int[d \delta a] \exp 
    \left[
        -\frac{r_0^2}{2e^2} \int_{\Sigma} d^2x \sqrt{\det g_{ab}} ~\delta a ^a (\Delta_{1})_{ab} \delta a^b
    \right]
    \notag
    \\
    &\times
    \int[d b_0 dc_0] \exp 
    \left[
        -\frac{r_0^2}{2e^2} \int_{\Sigma} d^2x \sqrt{\det g_{ab}} ~b_0 \Delta_0 c_0
    \right]
    \notag
    \\
    &\approx 
    \frac{{\det'}_N \Delta_0}{\sqrt{{\det'}_A\Delta_1}} 
    \exp 
    \left[
        -\frac{1}{2} \int_{\Sigma} d^2x \sqrt{\det g_{ab}}~ \frac{e^2 Q^2}{4 \pi r_0^2}
    \right],
    \label{Eq:zero mode partition function}
\end{align}
where the classical contribution has been included and the boundary conditions \eqref{eq:zero-bc} and \eqref{Eq:FP-bc} have been imposed.
Here, $\det'_N$ and $\det'_A$ denote functional determinants evaluated over the nonzero Laplacian eigenvalues satisfying Neumann and absolute boundary conditions, respectively.
A more complete treatment would require a careful analysis of harmonic forms, the path-integral measure, and the gauge volume, all of which may contribute to the entropy~\cite{Donnelly:2012st,Donnelly:2015hxa}. In this work, however, we restrict our attention to the contribution from nonzero Laplacian eigenvalues together with the classical background term, neglecting these additional effects.

Let us evaluate the one-loop determinant contribution to the effective action
\begin{align}
    W^{\text{1-loop}}_{\text{zero mode}} := \frac{1}{2} \log {\det}'_A \Delta_1 - \log {\det}'_N \Delta_0 =- \frac{1}{2} \int_0^\infty \frac{ds}{s}
    \left[ \Tr_A'[e^{-s\Delta_1}] - 2\Tr_N'[e^{-s\Delta_0}]\right],
\end{align}
where we introduce the Schwinger time $s$ and $\Tr_A'$ and $\Tr_N'$ denote traces over nonzero Laplacian eigenmodes satisfying absolute and Neumann boundary conditions, respectively.
Let $b_1$ and $b_0$ denote the numbers of zero modes of the one-form Laplacian with absolute boundary conditions and the scalar Laplacian with Neumann boundary conditions, respectively.\footnote{
By the de Rham--Hodge theorem (see Ref.~\cite{gilkey2018invariance} for example), these are given by the zeroth and first Betti numbers $b_0=\dim H_N^0(\Sigma),~b_1=\dim H_A^1(\Sigma)$
where
\begin{align*}
    H_A^p(\Sigma)
    &:=
    \left\{
    \omega^{(p)}
    \,\middle|\,
    \Delta_p\omega^{(p)}=0,\ 
    \omega^{(p)} \text{ satisfies the absolute boundary condition}
    \right\},
    \\
    H_N^0(\Sigma)
    &:=
    \left\{
    f
    \,\middle|\,
    \Delta_0 f=0,\ 
    f \text{ satisfies the Neumann boundary condition}
    \right\}.
\end{align*}
}
Then, 
\begin{align}
    \Tr'_A[e^{-s\Delta_1}] = \Tr_A[e^{-s\Delta_1}] - b_1, \quad \Tr'_N[e^{-s\Delta_0}] = \Tr_N[e^{-s\Delta_0}] -b_0.
    \label{Eq:zero-mode-subtraction}
\end{align}
Since the heat kernels admit the asymptotic expansions
\begin{align}
    \Tr_A[e^{-s\Delta_1}] \approx \sum_{n=0}^\infty c_n^A(\Delta_1) s^{\frac{n}{2}-1},
    \quad
    \Tr_N[e^{-s\Delta_0}] \approx \sum_{n=0}^\infty c_n^N(\Delta_0)s^{\frac{n}{2}-1},
\end{align}
the effective action becomes
\begin{align}
    W^{\text{1-loop}}_{\text{zero mode}}
    &\approx- \frac{1}{2} \int_0^\infty ds
    \left[ \sum_{n\neq2} (c^A_n(\Delta_1)-2c^N_n(\Delta_0))s^{\frac{n}{2}-2} + (c^A_2(\Delta_1)-2c^N_2(\Delta_0)-b_1+2b_0) s^{-1}\right].
\end{align}
The Schwinger time integral in this equation diverges both in the UV ($s\to0$) and in the IR ($s\to\infty$), which must be regularized.
The details of the regularization are presented in App.~\ref{sec:evaluation-PF}.
Replacing $c_2(\hat{D})$ with $c^A_2(\Delta_1)-2c^N_2(\Delta_0)-b_1+2b_0$ in Eq.~\eqref{Eq:renormalized effective action}, we find the renormalized action
\begin{align}
    W^{\text{1-loop}}_{\text{zero mode},\mathrm{ren}} =&\log \left(\frac{\mu}{m_{\text{IR}}}\right)
    (\chi(\Sigma) + b_1 - 2b_0) =: -S_{\text{zero mode}} ,
    \label{eq:zero-mode-enropy}
\end{align}
where $m_{\rm IR}$ is the IR cutoff, $\mu$ is the renormalization scale, and $c_2^N(\Delta_0)$ and $c_2^A(\Delta_1)$ are given in Eqs.~\eqref{Eq:c2N (Delta0)} and \eqref{Eq:c2A (Delta1)}, respectively.
With the regularization described in App.~\ref{sec:evaluation-PF} and the
IR power divergences discarded, Eq.~\eqref{eq:zero-mode-enropy} gives only
the logarithmic contribution extracted from the local heat-kernel expansion,
with the zero-mode subtraction in Eq.~\eqref{Eq:zero-mode-subtraction}.
The resulting contribution depends only on the Euler characteristic and
the Betti numbers and is therefore topological, consistent with the structure
of the logarithmic contribution discussed in Ref.~\cite{Barvinsky:1995dp}.
It gives the entropy shift $S_{\mathrm{zero~mode}}$ without modifying the
effective dilaton potential. This result does not include the full
IR-sensitive and nonlocal contributions, the integration over harmonic
modes and the gauge-group volume, the normalization of the path-integral
measure, or contributions from boundary and edge modes. Thus,
$S_{\mathrm{zero~mode}}$ should be understood as a partial contribution
rather than the complete one-loop entropy correction from the massless
sector. The omitted contributions may affect the entropy and thermodynamics
and are left for future investigation.

\subsection{Effective dilaton gravity including KK modes}

Combining the results obtained so far, we arrive at the effective dilaton gravity action including the KK contributions:
\begin{align}
    I_{\text{eff}}[g,\Phi]
    &=I_{\text{EH}} + I_{\text{GHY}} +I_{\rm{Maxwell}}|_{\text{on-shell}}
    + W_{\rm massive} + W^{\text{1-loop}}_{\text{zero mode},\mathrm{ren}}
    \\
    &= - \frac{1}{4 G_N}\int_{\Sigma} d^2x \sqrt{\det g_{ab}}
    \left[ 
        \Phi\left(R_{(2)}+\frac{2}{r_0^2}\right) -\frac{2\Lambda}{r_0^2} \Phi^2
        +\frac{3}{2\Phi} (\nabla \Phi)^2
        -\frac{G_N}{2\pi r_0^2}(e^2Q^2 + \mathcal{E})
    \right]
    \notag\\
    &\qquad-\frac{1}{2G_N}\int_{\partial \Sigma} dx \sqrt{\det h_{ab}}~\Phi K_{(1)}  -S_{\text{massive}}-S_{\text{zero mode}} ,
\end{align}
where $I_{\text{EH}} + I_{\text{GHY}} +I_{\rm{Maxwell}}|_{\text{on-shell}}$  is given in Eq.~\eqref{eq:total-tree-action}.
Recall that higher-curvature terms, Weyl anomaly terms, and the additional contributions from the massless sector discussed in Sec.~\ref{Sec:Zero modes} are not included in this effective action.
After eliminating the dilaton kinetic term via the Weyl transformation \eqref{Eq:Weyl transformation}, the effective action becomes
\begin{align}
    I_\eff[g,\Phi] &= - S_{\mr{massive}} - S_{\text{zero mode}}
    \nn \\
    & \qquad 
    - \frac{1}{4 G_N} \int_{\Sigma} d^2x \sqrt{\det g_{ab}} \biggl[
        \Phi R_{(2)} + V_\eff(\Phi)
    \biggr] - \frac{1}{2 G_N} \int_{\del \Sigma} dx \sqrt{\det h_{ab}} \Phi K_{(1)}.
    \label{Eq: effective dilaton gravity w/ massive KK mode}
\end{align}
Here, the effective dilaton potential is obtained by adding the second term in Eq.~\eqref{Eq:W massive} to \eqref{Eq: effective dilaton potential w/o massive KK mode}:
\begin{align}
	V_\eff(\Phi)&= 2 r_0 \left[ \Phi^{-\frac{1}{2}} -  \Lambda \Phi^{\frac{1}{2}} - \frac{1}{4\pi} G_N (e^2Q^2 +\mathcal{E})\Phi^{-\frac{3}{2}} \right].
	\label{Eq: effective dilaton potential w/ massive KK mode}
\end{align}
The thermodynamic quantities, including only the retained local one-loop contributions, can then be obtained by repeating the analysis of Sec.~\ref{Sec:two-dimensional dilaton gravity and phase transition}:
\begin{align}
    F_{(2)}(\Phi_H,Q) &= E_{(2)}(\Phi_H, Q) - T_{(2)}(\Phi_H, Q)S(\Phi_H),
    \label{Eq:2-dim free energy w/ KK modes}
    \\
    T_{(2)}(\Phi_H, Q) &= \frac{V_\eff(\Phi_H)}{4\pi r_0},
    \\
    E_{(2)}(\Phi_H, Q) &= \frac{1}{G_N} \left(
        \Phi_H^{\frac{1}{2}} - \frac{\Lambda}{3} \Phi_H^{\frac{3}{2}} + \frac{G_N}{4\pi} (e^2Q^2+\mc{E}) \Phi_H^{- \frac{1}{2}} 
    \right),
    \label{Eq:2-dim energy w/ KK modes}
    \\
    S(\Phi_H) &= \frac{\pi \Phi_H}{G_N} + S_{\text{massive}} + S_{\text{zero mode}} .
    \label{Eq:2-dim entropy w/ KK modes}
\end{align}
We summarize the semiclassical and one-loop results in Tab.~\ref{tab:summary}.

Before finishing this section, a few comments are in order.
As noted above, the local one-loop contributions only shift the entropy and the effective charge parameter,
\begin{align}
\frac{\pi\Phi_H}{G_N}\to \frac{\pi\Phi_H}{G_N}+S_{\text{massive}} + S_{\text{zero mode}},
\qquad
e^2Q^2\to e^2Q^2+\mathcal E.
\end{align}
Within this approximation, the phase diagram is not qualitatively altered so long as these corrections remain small.
Since the massive KK photons are neutral, the shift $\mathcal E$ does not represent a renormalization of the electromagnetic coupling $e$, but arises from the one-loop vacuum energy of the massive KK modes. We also note that the free energy diverges for the AdS background with $Q=0$ and $\Phi_H=0$. Since the geometric assumptions underlying the two-dimensional effective theory break down at $\Phi=0$, where the $S^2$ shrinks to zero size, this divergence should be regarded as a signal of the breakdown of the effective description rather than absorbed into a local counterterm.
The contributions omitted in our analysis, including higher-curvature terms, Weyl anomaly terms, and the additional contributions from the massless sector, might modify the phase structure. Moreover, our heat-kernel analysis does not capture nonlocal terms or finite-temperature effects. Boundary degrees of freedom such as edge modes are also not included. These effects may be important for understanding non-extremal black hole thermodynamics within effective dilaton gravity.

\begin{table}[t]
    \centering
    \begin{tabular}{|c||c|c|}\hline
         & Semiclassical & One-loop \\
        \hhline{|=#=#=|}
        Effective dilaton potential $V_\eff$ & \eqref{Eq: effective dilaton potential w/o massive KK mode} & \eqref{Eq: effective dilaton potential w/ massive KK mode}\\
        Free energy $F_{(2)}$ & \eqref{Eq:2-dim free energy w/o KK modes} & \eqref{Eq:2-dim free energy w/ KK modes} \\
        Energy $E_{(2)}$ & \eqref{Eq:2-dim energy w/o KK modes} & \eqref{Eq:2-dim energy w/ KK modes}
        \\
        Black hole entropy $S$ & \eqref{Eq:2-dim entropy w/o KK modes} & \eqref{Eq:2-dim entropy w/ KK modes} \\
         \hline
    \end{tabular}
    \caption{
    Summary of the semiclassical and one-loop results at leading order in the derivative expansion. Here, $S_{\text{massive}}$, $\mathcal E$, and $S_{\text{zero mode}}$ are defined in Eqs.~\eqref{Eq:S0-in-Wmassive}, \eqref{Eq:mathcalE-in-Wmassive}, and \eqref{eq:zero-mode-enropy}, respectively.
    }
    \label{tab:summary}
\end{table}

\section{Conclusions}
\label{Sec:Conclusion}

In this paper, we derived a two-dimensional effective dilaton gravity from four-dimensional spherically symmetric charged black holes and studied its thermodynamics. At the semiclassical level, the resulting effective theory reproduces the phase structure of the original four-dimensional black holes. In particular, it captures the Hawking--Page transition for Schwarzschild--AdS black holes and the small/large black-hole transition for RN--AdS black holes. This shows that the nonlinear dilaton potential obtained by dimensional reduction retains the information necessary to describe non-extremal black-hole thermodynamics, which is not captured by the near-horizon JT limit alone.

We also studied the one-loop corrections to the effective dilaton gravity arising from electromagnetic fluctuations around the spherically symmetric background. 
To this end, we performed the KK reduction of the four-dimensional Maxwell
field on $S^2$, included the full massive KK tower, and evaluated its
leading local contribution using the heat-kernel expansion.
The retained local one-loop contributions appear as constant shifts in the entropy and in the electric-field energy density, or equivalently, the effective charge parameter entering the dilaton potential. The conclusion that the semiclassical phase structure remains qualitatively unchanged applies only within the  approximation described in Sec.~\ref{Sec:Partition function and effective action}, which omits the dilaton-dependent Weyl anomaly terms and includes only a partial contribution from the massless sector. For the KK modes, we have not studied the full determinants.

The results in Eqs.~\eqref{Eq:2-dim free energy w/ KK modes}--\eqref{Eq:2-dim entropy w/ KK modes} depend on the renormalization scale $\mu$ through $S_{\mathrm{massive}}$, $\mathcal{E}$, and $S_{\mathrm{zero~mode}}$. 
The $\mu$-dependence associated with the retained local heat-kernel terms can be absorbed into the corresponding local counterterms. By contrast, the full one-loop effective action generally contains nonlocal terms, such as $\log(\Delta/\mu^2)$, which may generate genuine logarithmic corrections to physical quantities~\cite{Xiao:2021zly}.
Moreover, the higher-curvature terms omitted in our analysis may modify the phase structure. A more complete evaluation of the contributions from the lowest KK modes, including higher-curvature and nonlocal terms, is left for future work. 
The dilaton-dependent Weyl anomaly terms and the additional contributions from the massless sector discussed in Sec.~\ref{Sec:Zero modes} may also affect the entropy and thermodynamics. Evaluating these contributions is necessary for a complete description of the one-loop corrections and is left for future work.

Several issues remain for future work. First, the gauge-fixing dependence of the effective potential \eqref{Eq: effective dilaton potential w/ massive KK mode} should be examined more carefully. Although the one-loop determinant in Maxwell theory on curved backgrounds is known to be gauge independent under suitable choices of UV regularization, path-integral measure, and boundary conditions~\cite{Vassilevich:1997iz,Solodukhin:2012jh}, it remains to be verified that these conditions are satisfied in the present setup. Second, boundary degrees of freedom, such as edge modes, as well as harmonic forms, were neglected in this work. Since such effects can contribute to black-hole entropy~\cite{Donnelly:2014fua,Donnelly:2015hxa,Mukherjee:2023ihb,Ball:2024hqe}, they should be incorporated into a complete treatment of quantum corrections.

Another important issue is the inclusion of finite-temperature effects associated with the thermal cycle, which were not captured by our local heat-kernel analysis. Existing approaches, such as the Matsubara expansion and the thermal quotient method~\cite{David:2009xg,Gopakumar:2011qs,Keeler:2018lza}, generally require detailed spectral information. While finite-temperature effects can be computed explicitly in special constant-curvature backgrounds~\cite{Giombi:2008vd,David:2009xg,Gopakumar:2011qs}, no systematic method is currently available for more general geometries, including those considered here. Developing methods to estimate such effects without solving the full spectral problem remains an important direction for future work.

Finally, we have focused on the KK modes of the electromagnetic field while treating the gravitational sector semiclassically. Incorporating gravitational KK modes and their one-loop determinants would be an important step toward a more complete two-dimensional description of quantum corrections to non-extremal black-hole thermodynamics.

\section*{Acknowledgments}
\noindent
The authors thank Luca V. Iliesiu, Masamichi Miyaji, and Seiji Terashima for valuable discussions.
This work is supported by JST Grant No. JPMJPF2221 (YA), by JSPS KAKENHI Grant No. JP22K03601 (TH), and by JST SPRING, Japan Grant No. JPMJSP2123 (YM).

\appendix

\section{Notation}
\label{sec:notation}

In this appendix, we summarize our notation for geometry and related topics in a Euclidean spacetime theory.

\subsection{Differential form}

We write a $p$-form as
\begin{align}
    \omega^{(p)} = \frac{1}{p!} \omega_{\mu_1 \dots \mu_p} dx^{\mu_1} \wedge \cdots \wedge dx^{\mu_p}.
\end{align}
However, we drop the subscript $(0)$ for a zero-form for simplicity. 
The exterior derivative acts on $\omega^{(p)}$ as 
\begin{align}
    d \omega^{(p)} = \frac{1}{p!} \del_\mu \omega_{\nu_1 \cdots \nu_p} dx^\mu \wedge d x^{\nu_1} \wedge \cdots \wedge d x^{\nu_p},
\end{align}
which maps a $p$-form to the $(p+1)$-form.
The interior product by a vector field $v = v^\mu \del_\mu$ acts on a $p$-form $\omega^{(p)}$ as 
\begin{align}
    \iota_v \omega^{(p)} := \frac{1}{(p-1)!} v^\nu \omega_{\nu\mu_1\cdots\mu_{p-1}} dx^{\mu_1} \wedge \cdots \wedge dx^{\mu_{p-1}}.
\end{align}
This maps a $p$-form to the $(p-1)$-form.

The $D$-dimensional antisymmetric symbol $\varepsilon_{\mu_1\dots\mu_D}$ is normalized as 
\begin{align}
    \varepsilon_{0,1,\dots, D-1}  = +1.
\end{align}
The Hodge star of a $p$-form $\omega^{(p)}$ with respect to the $D$-dimensional metric $g_{\mu\nu}$ is defined by
\begin{align}
    \ast \omega^{(p)} \coloneqq \frac{\sqrt{\det g_{\mu\nu} }}{p!(D-p)!}\omega^{\nu_1 \cdots \nu_p} \varepsilon_{\nu_1 \cdots \nu_p \mu_1 \cdots \mu_{D-p}} dx^{\mu_1} \wedge \cdots \wedge d x^{\mu_{D-p}},
\end{align}
and acts on the unity and the $D$-dimensional volume factor as 
\begin{align}
    \ast 1 = \sqrt{\det g_{\mu\nu}} ~dx^0 \wedge dx^1 \wedge \cdots \wedge dx^{D-1},
    \qquad 
    \ast \left(\sqrt{\det g_{\mu\nu}}~ dx^0 \wedge dx^1 \wedge \cdots \wedge dx^{D-1} \right)= 1.
\end{align}
This Hodge star satisfies 
\begin{align}
    \ast \ast \omega^{(p)} = (-1)^{p(D-p)} \omega^{(p)},
    \qquad 
    \ast^{-1} \omega^{(p)} = (-1)^{p(D-p)} \ast \omega^{(p)}.
\end{align}

If we introduce the inner product of $p$-forms as 
\begin{align}
    (\omega^{(p)}, \zeta^{(p)}) \coloneqq \int \omega^{(p)} \wedge \ast \zeta^{(p)},
\end{align}
we define the conjugate of the differential operator $d$ by
\begin{align}
    (d \omega^{(p-1)}, \zeta^{(p)}) = (\omega^{(p-1)}, d^\dagger \zeta^{(p)}).
\end{align}
By explicitly calculating and dropping the surface term
\begin{align}
    (d \omega^{(p-1)}, \zeta^{(p)}) &= \int d \omega^{(p-1)} \wedge \ast \zeta^{(p)} = (-1)^p \int \omega^{(p-1)} \wedge d \ast \zeta^{(p)} 
    \nn \\
    &= \int \omega^{(p-1)} \wedge  \ast \bigl( (-1)^{D(p+1)+1}\ast d \ast \zeta^{(p)} \bigr),
\end{align}
we obtain
\begin{align}
    d^\dagger \omega^{(p)} = (-1)^{D(p+1) +1} \ast d \ast \omega^{(p)},
\end{align}
which maps a $p$-form to a $(p-1)$-form similarly to the interior product.
This operator is also called a co-differential operator.
Using this operator, we write the covariant Laplacian operator for $p$-form fields as 
\begin{align}
    \Delta_p 
    := d^\dagger d + d d ^\dagger
    = (-1)^{D(p+1) +1}
    \left( (-1)^D\ast d \ast d +  d\ast d \ast \right).
\end{align}
In particular, the action on a scalar field (zero-form field) $\Psi$ is expressed as 
\begin{align}
    d^\dagger d \Psi= -\ast d \ast d \Psi = - \frac{1}{\sqrt{\det g_{\mu\nu}}}\del_\mu \bigl( g^{\mu\nu} \sqrt{\det g_{\mu\nu}} \del_\nu \Psi \bigr),
\end{align}
which is called the Laplace operator of the metric $g_{\mu\nu}$.

\subsection{Connection and curvatures}

We show our conventions for geometric quantities used in this paper:
\begin{align}
	\Gamma^\mu_{\nu\rho} &= \frac{1}{2} g^{\mu\sigma}(\del_\nu g_{\sigma \rho} + \del_\rho g_{\nu\sigma} - \del_\sigma g_{\nu\rho}),
	\\
	R_{\mu\nu}{}^\rho{}_\sigma &= \del_\mu \Gamma^\rho_{\nu\sigma} - \del_\nu \Gamma^\rho_{\mu\sigma} + \Gamma^\rho_{\mu\lambda}\Gamma^\lambda_{\nu \sigma} - \Gamma^\rho_{\nu\lambda} \Gamma^\lambda_{\mu \sigma},
	\\
	R_{\mu\nu} &= R_{\rho\mu}{}^\rho{}_\nu
	= \del_\rho \Gamma^\rho_{\mu\nu} - \del_\mu \Gamma^\rho_{\rho\nu} + \Gamma^\rho_{\rho\lambda} \Gamma^\lambda_{\mu\nu} - \Gamma^\rho_{\mu \lambda} \Gamma^\lambda_{\rho \nu},
	\\
	R&= g^{\mu\nu} R_{\mu\nu}.
\end{align}

\section{Counterterms}
\label{Sec:Counterterms}

In this section, we show that the two-dimensional counterterm introduced in Sec.~\ref{Sec:two-dimensional dilaton gravity and phase transition} agrees with the counterterm obtained by the dimensional reduction of the four-dimensional one.

First, we consider the asymptotically flat case with $\Lambda=0$. In this case, the Gibbons--Hawking--York term evaluated in the flat reference background is conventionally used as a counterterm to remove the large-radius divergence of the spherically symmetric gravitational on-shell action~\cite{Gibbons:1976ue}
\begin{align}
    I_c^{\text{flat}} := \frac{1}{8\pi G_N} \int_{\partial \mathcal{M}} d^3 x \sqrt{\det h_{\mu\nu}}~K_{(3)}^{\text{flat}}.
\end{align}
Here, $K_{(3)}^{\text{flat}}$ denotes the extrinsic curvature computed with respect to the flat metric $\eta_{\mu\nu}$ given by
\begin{align}
    K_{(3)}^{\text{flat}} &= \frac{2}{r_\infty},
\end{align}
where $r = r_\infty$ represents the location of the boundary, and we take the limit $r_\infty \to \infty$.
Setting $\Phi (r)=r^{2}$ and using the metric ansatz Eq.~\eqref{Eq:4-dim metric anzatz} and integrating over $S^2$, we obtain
\begin{align}
    I_c^{\text{flat}}
    = \frac{1}{2 G_N} \int_{\partial \Sigma} d\tau \sqrt{\det h_{ab}} ~\frac{2\Phi}{r_0}.
\end{align}
Performing the Weyl transformation~\eqref{Eq:Weyl transformation}, the counterterm reads
\begin{align}
    I_c^{\text{flat}}=\frac{1}{2G_N} \int_{\partial \Sigma} d\tau \sqrt{\det h_{ab}} ~2r_0^{\frac{1}{2}}\Phi^{\frac{1}{4}}.
\end{align}
The same quantity can also be computed from the general formula in terms of the prepotential $W$. For the effective potential \eqref{Eq: effective dilaton potential w/o massive KK mode} with $\Lambda=0$, the corresponding prepotential is given by
\begin{align}
    W(\Phi) = 4r_0\left[\Phi^{\frac{1}{2}}
    +\frac{G_Ne^2Q^2}{4\pi}\Phi^{-\frac{1}{2}}\right].
\end{align}
Using this, the general formula for the two-dimensional counterterm \eqref{Eq:2-dim counterterm} yields
\begin{align}
    I_c
    = \frac{1}{2G_N}\int_{\partial\Sigma} d\tau \sqrt{\det h_{ab}}\,\sqrt{W(\Phi)}
    = \frac{1}{2G_N}\int_{\partial\Sigma} d\tau \sqrt{\det h_{ab}}\,2r_0^{1/2}\Phi^{1/4},
\end{align}
where we have taken the asymptotic limit $\Phi\to\infty$ on the boundary.
This agrees precisely with the counterterm obtained by dimensional reduction of the four-dimensional theory.

Next, we consider the asymptotically AdS case with $\Lambda <0$.
In this case, the counterterm is given by~\cite{Balasubramanian:1999re}
\begin{align}
    I_c^{\text{AdS}}
    =
    \frac{1}{8\pi G_N}
    \int_{\partial \mathcal{M}} d^{3}x \sqrt{\det h_{\mu\nu}}
    \left[
        2\sqrt{\frac{|\Lambda|}{3}}
        +\frac{1}{2}\sqrt{\frac{3}{|\Lambda|}}\,R^{\text{bdy}}
    \right],
    \label{Eq:AdS_counterterm}
\end{align}
where $R^{\text{bdy}}$ is the scalar curvature of the boundary metric.
Since the boundary $\partial\mathcal M$ is $S^1\times S^2$, with $S^2$ of radius $r_\infty$ and $S^1$ of circumference $\beta_{(4)}$, the boundary curvature receives contributions only from the $S^2$ factor. Hence,
\begin{align}
    R^{\text{bdy}}
    =
    \frac{2}{r_\infty^2}.
\end{align}
As in the flat case, by performing the Weyl transformation~\eqref{Eq:Weyl transformation} and integrating over $S^{2}$, we obtain
\begin{align}
    I^{\text{AdS}}_c = \frac{1}{2G_N} \int_{\partial \Sigma} d\tau \sqrt{\det h_{ab}}
    ~r_0^{\frac{1}{2}}
    \left[ 
        2 \sqrt{\frac{|\Lambda|}{3}}  \Phi^{\frac{3}{4}} + \sqrt{\frac{3}{|\Lambda|}} \Phi^{-\frac{1}{4}}
    \right],
\end{align}
where we used $\Phi(x)=r^2$.
Similarly, the same quantity can be obtained from the general formula.
The prepotential $W$ of the effective potential in Eq.~\eqref{Eq: effective dilaton potential w/o massive KK mode} is given by
\begin{align}
    W(\Phi) = 4r_0\left[\Phi^{\frac{1}{2}}
    + \frac{|\Lambda|}{3} \Phi^{\frac{3}{2}}
    +\frac{G_Ne^2Q^2}{4\pi}\Phi^{-\frac{1}{2}}\right].
\end{align}
Then, the two-dimensional counterterm in Eq.~\eqref{Eq:2-dim counterterm} for $\Lambda<0$ reads
\begin{align}
    I_c = \frac{1}{2G_N}\int_{\partial \Sigma} d\tau \sqrt{\det h_{ab}} \sqrt{W(\Phi)}
    =\frac{1}{2G_N}\int_{\partial \Sigma} d\tau \sqrt{\det h_{ab}}~r_0^{\frac{1}{2}}
    \left[ 
        2\sqrt{\frac{|\Lambda|}{3}} \Phi^{\frac{3}{4}} + \sqrt{\frac{3}{|\Lambda|}} \Phi^{-\frac{1}{4}}
    \right],
\end{align}
where we have taken the limit $\Phi\to\infty$ on the boundary. Thus, the dimensionally reduced counterterm agrees with the two-dimensional counterterm \eqref{Eq:2-dim counterterm} for $\Lambda<0$.

\section{Free energy computation in four dimensions}
\label{Sec:free energy}

In this section, we evaluate the on-shell Euclidean action of a four-dimensional RN--AdS black hole with mass $M$ and charge $Q$, including the counterterm \eqref{Eq:AdS_counterterm}. The boundary $\partial\mathcal{M}$ is located at the cutoff surface $r=r_\infty$.
We use the Euclidean metric
\begin{align}
    ds^2
    =
    f(r)d\tau^2+\frac{dr^2}{f(r)}+r^2d\Omega_{(2)}^2,
    \qquad
    f(r)
    =
    1-\frac{2G_NM}{r}
    +\frac{G_Ne^2Q^2}{4\pi r^2}
    -\frac{\Lambda}{3}r^2.
\end{align}
The horizon radius $r_H$ is defined by $f(r_H)=0$, which gives
\begin{align}
    M
    =
    \frac{r_H}{2G_N}
    \left(
        1
         -\frac{\Lambda}{3}r_H^2
        +\frac{G_Ne^2Q^2}{4\pi r_H^2}
    \right).
    \label{Eq:M and r_H in AdS}
\end{align}
The Euclidean action in the canonical ensemble is
\begin{align}
    I_{\rm ren}
    =
    I_{\rm EH}
    +I_{\rm GHY}
    +I_{\rm Maxwell}
    +I_{\rm c}.
\end{align}
Since the Maxwell stress tensor is traceless, the Einstein equation gives $R=4\Lambda$.
Therefore the gravitational bulk contribution is
\begin{align}
    I_{\rm EH}
    =
    \frac{|\Lambda|}{8\pi G_N }
    \int_{\mathcal M}d^4x\sqrt{\det g_{\mu\nu}}
    =
    \frac{|\Lambda|\beta_{(4)}}{6G_N}
    \left(
        r_\infty^3-r_H^3
    \right).
\end{align}
The unit normal vector $n_{(4)}^\mu \partial_\mu$	and the extrinsic curvature K of the hypersurface $r=r_\infty$ are given by
\begin{align}
    n_{(4)}^\mu\partial_\mu = \sqrt{f(r_\infty)}\partial_r, \quad
    K := \nabla_\mu n^\mu = \frac{f'(r_\infty)}{2\sqrt{f(r_\infty)}}+ \frac{2}{r_\infty}\sqrt{f(r_\infty)}.
\end{align}
The Gibbons--Hawking term is then evaluated at $r=r_\infty$ as
\begin{align}
    I_{\rm GHY}
    =
    -\frac{\beta_{(4)}}{2G_N}
    \left[
        \frac{r_\infty^2}{2}f'(r_\infty)
        +2r_\infty f(r_\infty)
    \right]
    =
    -\frac{\beta_{(4)}}{2G_N}
    \left[
        |\Lambda| r_\infty^3
        +2r_\infty -3G_N M
    \right]
    +\mathcal{O}(r_\infty^{-1})
    .
\end{align}
The counterterm \eqref{Eq:AdS_counterterm} contribution reads
\begin{align}
    I_{\rm c}
    =
    \frac{\beta_{(4)}}{2G_N}
    r_\infty^2\sqrt{f(r_\infty)}
    \left(
        2\sqrt{\frac{|\Lambda|}{3}}
        +\sqrt{\frac{3}{|\Lambda|}}\frac{1}{r_\infty^2}
    \right)
    =\frac{\beta_{(4)}}{2G_N} \left[ \frac{2|\Lambda|}{3}r_\infty^3 + 2r_\infty - 2G_NM \right]
    .
\end{align}
Combining these terms, we find
\begin{align}
    I_{\rm EH}
    +I_{\rm GHY}
    +I_{\rm c}
    &=
    \frac{\beta_{(4)}}{2G_N}
    \left(
       G_N M 
       +\frac{\Lambda}{3}r_H^3
    \right)
    +O(r_\infty^{-1}) 
    \notag
    \\
    &=
    \frac{\beta_{(4)}}{4G_N}
    \left(
        r_H
       +\frac{\Lambda}{3}r_H^3
        +\frac{G_Ne^2Q^2}{4\pi r_H}
    \right)
    +O(r_\infty^{-1}),
\end{align}
where we use Eq.~\eqref{Eq:M and r_H in AdS} in the second line.
The on-shell Maxwell action for the RN--AdS solution~\eqref{Eq:4-dim gauge field} is evaluated in the same way as Eq.~\eqref{Eq:2-dim Maxwell action on-shell}
\begin{align}
    I_\Maxwell &=  \frac{1}{4 e^2} \int_\mc{M} d^4 x \sqrt{\det g_{\mu\nu}} F_{\mu\nu} F^{\mu\nu} - \frac{1}{e^2} \int_{\del\mc{M}} d^3 x \sqrt{\det h_{\mu\nu}} n_{(4)\mu}  A_\nu F^{\mu\nu}
    \nn \\
    &=
    \frac{\beta_{(4)} e^2Q^2}{8\pi}
    \left(
        \frac{1}{r_H}
        -\frac{1}{r_\infty}
    \right).
\end{align}
Taking the limit $r_\infty\to\infty$, the total on-shell action becomes
\begin{align}
    I_{\rm ren}|_{\rm on-shell}
    =
    \frac{\beta_{(4)}}{4G_N}
    \left(
        r_H
        +\frac{\Lambda}{3}r_H^3
        +\frac{3G_Ne^2Q^2}{4\pi r_H}
    \right).
\end{align}
Therefore the four-dimensional free energy in the fixed-charge ensemble is
\begin{align}
    F_{(4)}(r_H)
    =
    T_{(4)}I_{\rm ren}|_{\rm on-shell}
    =
    \frac{1}{4G_N}
    \left(
        r_H
        +\frac{\Lambda}{3}r_H^3
        +\frac{3G_Ne^2Q^2}{4\pi r_H}
    \right).
    \label{eq:free-energy-in-4d}
\end{align}
In particular, the free energy of pure AdS is obtained by setting $Q=0$ and $r_H=0$ in the above expressions:
\begin{align}
    F_{(4)} =0.
    \label{eq:Pure-AdS-free-energy-in-4d}
\end{align}

\section{Evaluation of partition function}
\label{sec:evaluation-PF}

In this section, we review the heat-kernel method  for evaluating partition functions.
See Refs.~\cite{gilkey2018invariance,Hawking:1976ja,Vassilevich:2003xt,Gilkey:2004dm} for details.

\subsection{Heat-kernel method}

We consider the computation of the one-loop effective action for a quantum field theory in curved spacetime. The partition function on a fixed background geometry with metric $g_{ab}$ is
\begin{align}
Z(g)=\int[d\varphi]\,e^{-I[\varphi;g]}.
\end{align}
Here, $I[\varphi;g]$ denotes the action of a bosonic field
$\varphi=\bar\varphi+\delta\varphi$
on a curved spacetime with metric $g_{ab}$, where $\bar\varphi$ is a classical solution satisfying
$(\delta I[\varphi;g]/\delta\varphi)\big|_{\varphi=\bar\varphi}=0$.
Integrating over the quadratic fluctuations $\delta\varphi$, one obtains
\begin{align}
Z(g)
=
\sum_{\bar\varphi}
\det\!\left(
\left.
\frac{\delta^2I[\varphi;g]}
{\delta\varphi^2}
\right|_{\varphi=\bar\varphi}
\right)^{-1/2}
e^{-I[\bar\varphi;g]}.
\end{align}
For a fixed classical solution $\bar\varphi$, the one-loop contribution to the effective action is
\begin{align}
-\log Z\big|_{\rm 1\mbox{-}loop}
=
\frac12
\log\det\!\left(
\left.
\frac{\delta^2I[\varphi;g]}
{\delta\varphi^2}
\right|_{\varphi=\bar\varphi}
\right)
=
\frac12
\Tr\log\!\left(
\left.
\frac{\delta^2I[\varphi;g]}
{\delta\varphi^2}
\right|_{\varphi=\bar\varphi}
\right).
\end{align}

Let us now consider the computation of the one-loop effective action associated with a non-negative second-order differential operator $\hat D$:
\begin{align}
    W \coloneqq \frac12 \Tr \log \hat D .
\end{align}
For each positive eigenvalue $\lambda$ of $\hat D$, the identity
\begin{align}
\log \lambda
=
-\int_0^\infty \frac{ds}{s}\,e^{-s\lambda},
\label{Eq:log_lambda}
\end{align}
holds up to a $\lambda$-independent divergent constant.\footnote{
To show this, one may differentiate both sides of Eq.~\eqref{Eq:log_lambda}.
The left-hand side gives $1/\lambda$, while the right-hand side gives
\begin{align*}
-\frac{d}{d\lambda}
\int_0^\infty \frac{ds}{s}\,e^{-s\lambda}
=
\int_0^\infty ds\,e^{-s\lambda}
=
\frac{1}{\lambda}.
\end{align*}
Hence, the two sides differ only by a $\lambda$-independent constant.
}
Using this identity, the effective action can be written in terms of the Schwinger proper time as
\begin{align}
W
=
-\frac{1}{2}
\int_0^\infty \frac{ds}{s}\,
\Tr\!\left[e^{-s\hat D}\right].
\label{Eq:W & heat kernel}
\end{align}
The resulting expression is also defined only up to a $\hat D$-independent divergent constant, which can be absorbed into the overall normalization of the path integral.
Since $\hat D$ has mass dimension $2$ and the combination $s\hat D$ is dimensionless, the Schwinger proper time $s$ has dimension $-2$.
It is known that the heat-kernel trace admits the asymptotic expansion~\cite{Gilkey:2004dm}
\begin{align}
    \Tr [e^{- s \hat{D}}]
    \approx
    \sum_{n=0}^\infty
    c_n(\hat{D})\, s^{\frac{n}{2}-1},
    \qquad
    (s\to0),
    \label{Eq:asymptotic expansion of Tr}
\end{align}
from which
\begin{align}
    W
    \approx
    -\frac12
    \sum_{n=0}^\infty
    c_n(\hat{D})
    \int_0^\infty \frac{ds}{s}
    s^{\frac{n}{2}-1}
    \label{Eq:W-asymptotic expansion}
\end{align}
follows.
The coefficients $c_n(\hat D)$ depend on the boundary conditions imposed on the fields.
For later use, we list several coefficients below~\cite{Gilkey:2004dm}.
\begin{itemize}
	\item Dirichlet boundary condition
	\begin{align}
		c_0^D(\Delta_0) &= \frac{1}{4\pi} \int_\Sigma d^2x \sqrt{\det g_{ab}},
        \label{Eq:c_0w/Dirichlet}
		\\
		c_1^D(\Delta_0) &= -\frac{1}{4} \frac{1}{\sqrt{4\pi}} \int_{\del \Sigma} d\tau \sqrt{\det h_{ab}},
		\\
		c_2^D(\Delta_0) &= \frac{1}{24\pi} \biggl[
			\int_\Sigma d^2x \sqrt{\det g_{ab}} R_{(2)} + 2 \int_{\del \Sigma} d\tau \sqrt{\det h_{ab}} K_{(1)} 
		\biggr] = \frac{1}{6} \chi(\Sigma),
		\\
		c^D_3(\Delta_0) &= - \frac{1}{384} \frac{1}{\sqrt{4\pi}} \int_{\del \Sigma} d\tau \sqrt{\det h_{ab}} [20 R_{(2)} - 3 K_{(1)}^2 ].
	\end{align}

	\item Neumann boundary condition
	\begin{align}
		c_0^N(\Delta_0) &= \frac{1}{4\pi} \int_\Sigma d^2x \sqrt{\det g_{ab}},
        \label{Eq:c0N (Delta0)}
		\\
		c_1^N(\Delta_0) &= \frac{1}{4} \frac{1}{\sqrt{4\pi}} \int_{\del \Sigma} d \tau \sqrt{\det h_{ab}},
		\\
		c_2^N(\Delta_0) &= \frac{1}{24\pi} \biggl[
			\int_\Sigma d^2x \sqrt{\det g_{ab}} R_{(2)} + 2 \int_{\del \Sigma} d\tau \sqrt{\det h_{ab}} K_{(1)}
		\biggr] = \frac{1}{6} \chi(\Sigma),
        \label{Eq:c2N (Delta0)}
		\\
		c_3^N(\Delta_0)&= \frac{1}{384} \frac{1}{\sqrt{4\pi}} \int_{\del \Sigma} d\tau \sqrt{\det h_{ab}} \bigl[
			20 R_{(2)} + 15 K_{(1)}^2
		\bigr].
	\end{align}
	\item Absolute boundary condition
	\begin{align}
		c_0^A(\Delta_1) &=\frac{1}{2\pi} \int_\Sigma d^2x\sqrt{\det g_{ab}},
        \label{Eq:c0A (Delta1)}
		\\
		c_1^A(\Delta_1) &=0,
        \label{Eq:c1A (Delta1)}
		\\
		c_2^A(\Delta_1) &= -\frac{1}{6 \pi} \biggl[
			\int_\Sigma d^2x \sqrt{\det g_{ab}} R_{(2)} + 2 \int_{\del \Sigma} d\tau \sqrt{\det h_{ab}} K_{(1)}
		\biggr] = -\frac{2}{3} \chi(\Sigma),
        \label{Eq:c2A (Delta1)}
		\\
		c^A_3(\Delta_1) &= \frac{3}{64} \frac{1}{\sqrt{4\pi}} \int_{\del \Sigma} d\tau \sqrt{\det h_{ab}} K_{(1)}^2.
        \label{Eq:c3A (Delta1)}
	\end{align}
\end{itemize}
The integral in Eq.~\eqref{Eq:W-asymptotic expansion} diverges at both the upper and lower limits:
\begin{align}
\int_0^{\infty} \frac{ds}{s} s^{\frac{n}{2}-1}
=
\begin{dcases}
\displaystyle
\lim_{s\to0}
\frac{2}{n-2}
\frac{1}{s^{1-\frac{n}{2}}}
& (n=0,1),
\\[1ex]
\displaystyle
\lim_{s\to\infty}\log s
-\lim_{s\to0}\log s
& (n=2),
\\[1ex]
\displaystyle
\lim_{s\to\infty}
\frac{2}{n-2}
s^{\frac{n}{2}-1}
& (n>2).
\end{dcases}
\label{Eq:s-integral}
\end{align}
Since the Schwinger proper time $s$ has dimension $-2$, the divergences at $s\to0$ and $s\to\infty$ correspond to UV and IR divergences, respectively.
The IR divergence can be regulated by introducing an IR cutoff $m_{\rm IR}$ and replacing
$\hat D\to \hat D+m_{\rm IR}^2$.
This modifies Eq.~\eqref{Eq:W & heat kernel} by an extra factor $e^{-s m_{\rm IR}^2}$, which exponentially suppresses the large-$s$ region.
If $\hat D$ already contains a mass term, e.g., $\hat D=\Delta+m^2$, no additional IR regulator is required.
The UV divergence is regulated by introducing a UV regulator $\epsilon$ and replacing
\begin{align}
    \int_0^\infty \frac{ds}{s}
    \;\to\;
    \tilde{\mu}^{2\epsilon}
    \int_0^\infty \frac{ds}{s^{1-\epsilon}},
\end{align}
where $\tilde{\mu}$ is an auxiliary mass parameter introduced to keep the integral dimensionless.
The divergence at $s\to0$ is regulated by the factor $s^\epsilon$.
The regularized effective action becomes
\begin{align}
    W_{\epsilon,m_{\rm IR}}
    &:=
    -\frac12
    \tilde{\mu}^{2\epsilon}
    \int_0^\infty
    \frac{ds}{s^{1-\epsilon}}
    \Tr\!\left[e^{-s(\hat D+m_{\rm IR}^2)}\right]
    =
    -\frac12
    \sum_{n=0}^{\infty}
    c_n(\hat D)\,
    \tilde{\mu}^{2\epsilon}
    \int_0^\infty
    ds\,
    s^{\frac n2-2+\epsilon}
    e^{-s m_{\rm IR}^2}.
    \label{Eq:W-epsilon-m}
\end{align}
Performing the $s$-integration, we obtain
\begin{align}
    f_n(\epsilon)
    :=
    \tilde{\mu}^{2\epsilon}
    \int_0^{\infty}
    ds\,
    s^{\frac{n}{2}-2+\epsilon}
    e^{-s m_{\rm IR}^2}
    =
    \tilde{\mu}^{2\epsilon}
    m_{\rm IR}^{-n+2-2\epsilon}
    \Gamma\!\left(\frac{n}{2}-1+\epsilon\right),
\end{align}
which is well defined for $0<\epsilon<1$.
The gamma function admits the expansion around $\epsilon=0$
\begin{align}
    \Gamma\!\left(\frac{n}{2}-1+\epsilon\right)
    =
    \begin{dcases}
        -\frac{1}{\epsilon}
        +\gamma_E-1
        +\mathcal O(\epsilon),
        & n=0,
        \\
        \frac{1}{\epsilon}
        -\gamma_E
        +\mathcal O(\epsilon),
        & n=2,
        \\
        \Gamma\!\left(\frac{n}{2}-1\right)
        +\mathcal O(\epsilon),
        & n\neq0,2,
    \end{dcases}
\end{align}
where $\gamma_E$ denotes the Euler--Mascheroni constant.
Expanding $f_n(\epsilon)$ around $\epsilon=0$, we find
\begin{align}
    f_n(\epsilon)
    =
    \begin{dcases}
        -\frac{m_{\rm IR}^2}{\epsilon}
        +m_{\rm IR}^2
        \log\!\left(\frac{m_{\rm IR}^2}{\mu^2}\right)
        -m_{\rm IR}^2
        +\mathcal O(\epsilon),
        & n=0,
        \\
        -2\sqrt{\pi}\,m_{\rm IR}
        +\mathcal O(\epsilon),
        & n=1,
        \\
        \frac{1}{\epsilon}
        -\log\!\left(\frac{m_{\rm IR}^2}{\mu^2}\right)
        +\mathcal O(\epsilon),
        & n=2,
        \\
        \frac{\Gamma\!\left(\frac{n}{2}-1\right)}
        {m_{\rm IR}^{\,n-2}}
        +\mathcal O(\epsilon),
        & n>2,
    \end{dcases}
    \label{Eq:f-n-epsilon}
\end{align}
where we have used $\Gamma(-1/2)=-2\sqrt{\pi}$.
Following the standard convention, we introduce the renormalization scale
\begin{align}
    \mu := e^{-\gamma_E/2}\tilde{\mu}.
\end{align}
By subtracting the UV pole $1/\epsilon$ in Eq.~\eqref{Eq:f-n-epsilon} and taking the limit $\epsilon\to0$, we obtain the UV-renormalized effective action
\begin{align}
    W_{\mathrm{ren}}
    =&
    -\frac{1}{2}\log \left(\frac{m_{\rm IR}^2}{\mu^2}\right)
    \left[m_{\rm IR}^2 c_0 (\hat{D}) - c_2(\hat{D})\right]
    +\frac{1}{2}m_{\rm IR}^2c_0 (\hat{D})
    +\sqrt{\pi} m_{\rm IR} c_1 (\hat{D})
    \nonumber\\
    &
    -\frac{1}{2}\sum_{n=3}^\infty
    \frac{c_{n}(\hat{D})}{m_{\rm IR}^{n-2}}
    \Gamma\left(\frac{n}{2}-1\right).
    \label{Eq:W-ren-mIR}
\end{align}
Taking the limit $m_{\rm IR}\to0$, the terms proportional to $m_{\rm IR}^2 c_0(\hat D)$ and $m_{\rm IR} c_1(\hat D)$ vanish, whereas the terms associated with $c_n(\hat D)$ for $n>2$ become power divergent. Since such IR power divergences are not determined universally within a local heat-kernel analysis, we discard them and retain only the logarithmic contribution.
This yields
\begin{align}
    W_{\mathrm{ren}}
    =
    \frac{1}{2}
    \log\!\left(\frac{m_{\rm IR}^2}{\mu^2}\right)
    c_2(\hat D).
    \label{Eq:renormalized effective action}
\end{align}
Note that the coefficients $c_n(\hat D)$ with $n>2$ contain higher-curvature invariants~\cite{gilkey2018invariance,Vassilevich:2003xt,Gilkey:2004dm}.

Next, let us consider the case of $\hat{D}=\Delta+m^2$.
As mentioned above, we do not need to introduce the IR regulator because the IR divergence is already suppressed by $e^{-sm^2}$:
\begin{align}
    W= - \frac{1}{2} \int_0^\infty \frac{ds}{s} \Tr [e^{-s (\Delta+m^2)}]
    =-\frac{1}{2}\sum_{n=0}^\infty c_n (\Delta)\int_0^{\infty}  s^{\frac{n}{2} -2 } e^{-sm^2}ds,
\end{align}
where we used the asymptotic expansion for $\Tr[e^{-s\Delta}]$.
The UV divergence is regularized in the same way as discussed above.
\begin{align}
    W_{\epsilon}
    :=-\frac{1}{2}\sum_{n=0}^\infty c_n (\Delta)\tilde{\mu}^{2\epsilon}\int_0^{\infty}  s^{\frac{n}{2} -2 +\epsilon} e^{-sm^2}ds.
    \label{Eq:W-epsilon}
\end{align}
Repeating the same computation to derive Eq.~\eqref{Eq:W-ren-mIR}, we find the following asymptotic expansion of the renormalized effective action:
\begin{align}
    W_{\mathrm{ren}} =  
    -\frac{1}{2}\log \left(\frac{m^2}{\mu^2}\right)
    \left[m^2 c_0 (\Delta) - c_2(\Delta)\right]
    +\frac{1}{2}m^2c_0 (\Delta)
    +\sqrt{\pi} m c_1 (\Delta)
    -\frac{1}{2}\sum_{n=3}^\infty
    \frac{c_{n}(\Delta)}{m^{n-2}} \Gamma\left(\frac{n}{2}-1\right).
    \label{Eq:renormalized effective action for the massive case}
\end{align}
Thus, the heat-kernel expansion yields the large-mass expansion of the effective action.

The UV regularization corresponds to the zeta-function regularization.
The generalized zeta-function regularization is defined as 
\begin{align}
    \zeta(z;\hat{D}) := \Tr[\hat{D}^{-z}]
\end{align}
Using the Gamma function, this function can be written as the Mellin transform of $\Tr[e^{-s\hat{D}}]$:
\begin{align}
    \zeta(z;\hat{D}) = \frac{1}{\Gamma(z)} \int_0^\infty \frac{ds}{s^{1-z}} \Tr[e^{-s \hat{D}}].
\end{align}
Note that the UV-regularized effective action is given by 
\begin{align}
    W_\epsilon \coloneqq - \frac{1}{2} \tilde{\mu}^{2\epsilon} \int_0^\infty \frac{ds}{s^{1-\epsilon}} \Tr[e^{- s \hat{D}}]
    =-\frac{1}{2} \tilde{\mu}^{2\epsilon} \Gamma(\epsilon) \zeta(\epsilon;\hat{D}).
\end{align}
Here, we ignore the IR regularization for simplicity.
Around $\epsilon=0$, this is expanded as 
\begin{align}
    W_\epsilon = - \frac{1}{2} \biggl(
        \frac{1}{\epsilon} - \gamma_E + 2 \log\tilde{\mu} 
    \biggr) \zeta(0; \hat{D}) - \frac{1}{2} \frac{d\zeta}{dz}(0,\hat{D}) + \mc{O}(\epsilon).
\end{align}
Subtracting this pole $1/\epsilon$, we can write the renormalized effective action using the zeta function 
\begin{align}
    W_{\mr{ren}} = - \frac{1}{2} \biggl[
        \log(\mu^2) \zeta(0;\hat{D}) + \frac{d\zeta}{dz}(0;\hat{D})
    \biggr].
\end{align}
For details of zeta-function regularization, see, for example, Refs.~\cite{Hawking:1976ja,Vassilevich:2003xt}.

\subsection{Regularization of the infinite Kaluza--Klein mode sums}
\label{Sec:Regularization of the infinite Kaluza--Klein mode sums}

The infinite sums in Eqs.~\eqref{Eq:log m}--\eqref{Eq:m} are regularized using zeta-function regularization:
\begin{align}
    \sum_{l=1}^\infty (2l+1) \log \frac{l(l+1)}{\mu^2 r_0^2}
    &= 
    4 \sum_{l=1}^\infty l \log l
    -2\log (\mu r_0)
    \sum_{l=1}^\infty 
    \left[ 2 l + 1\right]
    \notag\\
    &=-4 \frac{d\zeta}{dz}(-1)
    -2\log (\mu r_0)
    \left[ 2 \zeta(-1)+ \zeta(0)\right]
    \notag\\
    &=-4 \frac{d\zeta}{dz}(-1) + \frac{4}{3} \log (\mu r_0) ,
    \label{Eq:KK-sum-1}
    \\
    \sum_{l=1}^\infty (2l+1)l(l+1) \log \frac{l(l+1)}{\mu^2 r_0^2}
    &=  
    \sum_{l=1}^\infty 
    \left[
        4l^3 \log l + 2l\log l
    \right]
    -
    2 \log (\mu r_0) \sum_{l=1}^\infty
    \left[ 2 l^3 + 3 l^2 +  l \right]
    \notag\\
    &= 
    \left[
        -4\frac{d\zeta}{dz}(-3) - 2\frac{d\zeta}{dz}(-1)
    \right]
    -2 \log (\mu r_0)
    \left[ 2 \zeta(-3) + 3 \zeta(-2) +  \zeta(-1) \right]
    \notag\\
    &=
    -4\frac{d\zeta}{dz}(-3) - 2\frac{d\zeta}{dz}(-1)
    +\frac{2}{15} \log (\mu r_0),
    \label{Eq:KK-sum-2}
    \\
    \sum_{l=1}^\infty (2l+1)l(l+1)
    &= 2\zeta(-3) + 3\zeta(-2) + \zeta(-1)= -\frac{1}{15}
    \label{Eq:KK-sum-3}
\end{align}
where we use the definition of the zeta function and its derivative
\begin{align}
    \zeta(z) = \sum_{n=1}^\infty n^{-z},
    \qquad 
    \frac{d\zeta}{dz}(z) = - \sum_{n=1}^\infty n^{-z} \log n,
\end{align}
and special values of the zeta function
\begin{align}
    \zeta(0)= -\frac{1}{2}, \quad \zeta(-1) = - \frac{1}{12}, \quad \zeta(-2)=0, \quad \zeta(-3)= \frac{1}{120}.
\end{align}

\newcommand{\arxivfont}{\rmfamily}
\bibliographystyle{yautphysm}
\bibliography{ref}

\end{document}